\documentclass[a4paper,12pt]{article}
\pdfoutput=1
\usepackage{color,colortbl}
\usepackage[square, comma, sort&compress, numbers]{natbib}
\usepackage{tipa,bbm}
\usepackage{epsf,amsmath,amssymb,mathrsfs}
\usepackage{graphicx,subfigure}
\usepackage{booktabs,multirow,makecell}
\usepackage{threeparttable}
\newlength{\dinwidth}
\newlength{\dinmargin}
\newcommand{\half}{\frac{1}{2}}

\begin{document}
\title{\bf  Fits of Weak Annihilation and Hard Spectator Scattering Corrections in $B_{u,d}\to VV$ Decays}
\author{Qin Chang$^{a,b}$, Xiao-Nan Li$^{a}$, Jun-Feng Sun$^{a}$ and Yue-Ling Yang$^{a}$\\
{ $^a$\small Institute of Particle and Nuclear Physics, Henan Normal University, Henan 453007, P.~R. China}\\
{ $^b$\small State Key Laboratory of Theoretical Physics, Institute of Theoretical Physics,}\\[-0.2cm]
{     \small Chinese Academy of Sciences, P.~R. China}}
\date{}
\maketitle
\begin{abstract}
In this paper, the contributions of weak annihilation and hard spectator scattering in $B\to \rho K^*$, $ K^* \bar{K}^*$, $\phi K^*$, $\rho\rho$ and $\phi\phi$ decays are investigated within the framework of QCD factorization. Using the experimental data available, we perform $\chi^2$ analyses of end-point parameters in four cases based on the topology-dependent and polarization-dependent parameterization schemes. The fitted results indicate that: (i) In the topology-dependent scheme, the relation $(\rho_A^{i},\phi_A^{i})\neq (\rho_A^{f},\phi_A^{f})$ gotten through $B\to PP$ and $PV$ decays is favored by the penguin-dominated $B\to VV$ decays at 95\% C. L.; (ii) The large hard spectator scattering corrections and/or the simplification $ (\rho_H,\phi_H)= (\rho_A^{i},\phi_A^{i})$ are challenged by ${\cal B}(\bar{B}^0\to \rho^0 \rho^0)$, even though they are allowed by  $B\to PP$ and $PV$ decays and helpful for resolving  ``$\pi\pi$ puzzle"; (iii)  In the polarization-dependent scheme, the relation $(\rho_A^{L},\phi_A^{L})\neq (\rho_A^{T},\phi_A^{T})$ is always required. Moreover, we have updated  the theoretical results for $B\to VV$ decays with the best-fit values of end-point parameters.  A few observables, such as the ones of pure annihilation $B_d\to \phi \phi$ decay, are also identified for probing the annihilation corrections. 
\end{abstract}
\noindent{{\bf PACS numbers:} 13.25.Hw, 14.40.Nd, 12.39.St}
\section{Introduction}
The non-leptonic charmless two-body $B$ meson decays provide a festival ground for testing the flavor pictures of Standard Model~(SM) and probing the possible hints of new physics~(NP). Theoretically, in order to obtain the reliable prediction, one of the main roles is to  evaluate the short-distance QCD corrections to hadronic matrix elements of $B$ meson decays.  In this respect,  the QCD factorization (QCDF) approach \cite{Beneke:1999br,Beneke:2000ry}, the perturbative QCD (pQCD) approach \cite{Keum:2000ph,Keum:2000wi} and the soft-collinear effective theory (SCET)~\cite{scet1,scet2,scet3,scet4} are explored and widely used to calculate the amplitudes of $B$ meson decays. 
 
In the ${\cal O}(\alpha_s)$ corrections, although the weak annihilation~(WA) amplitudes are formally $\Lambda_{\rm QCD}/m_b$ power-suppressed, they are generally  nontrivial, especially for the flavor-changing-neutral-current~(FCNC) dominated and pure annihilation B decays. Furthermore, because of the possible strong phase provided by the WA amplitude, the WA contribution also play an indispensable role for evaluating the charge-parity~(CP) asymmetry. Unfortunately, in the collinear factorization approach, the calculation of WA corrections always suffers from the divergence at the end-point of convolution integrals of meson's light-cone distribution amplitudes~(LCDA). In the SCET, the annihilation diagrams are factorizable and real to the leading power of ${\cal O}(\alpha_{s}(m_{b})\Lambda_{QCD}/m_{b})$~\cite{msf,scetAnni}.  In the QCDF, the end-point divergence are usually parameterized by the phenomenological parameter $X_{A}$ defined as \cite{Beneke:2003zv} 
\begin{eqnarray}
\label{Xa}
\int_{0}^{1}\frac{dx}{x}\to X_{A}(\rho_A,\phi_A)=(1+\rho_{A}e^{i\phi_{A}})\ln\frac{m_{b}}{\Lambda_{h}}\,,
\end{eqnarray}
in which $\Lambda_{h}=0.5{\rm GeV}$, $\rho_A$ and $\phi_A$ are phenomenological parameters and responsible for the strength and the possible strong phase of WA correction near the end-point, respectively. In addition, for the hard spectator scattering~(HSS) contributions, the calculation of twist-3 distribution amplitudes also suffers from end-point divergence, which is usually dealt with the same parameterization scheme as Eq.~\eqref{Xa} and labeled by $X_{H}(\rho_H\,,\phi_H)$. 
 
So far, the values of $(\rho_A,\phi_A)$ are utterly unknown from the first principles of QCD dynamics, and thus can only be obtained through the experimental data. Originally, a conservative choice of $\rho_A\sim 1$ with an arbitrary strong interaction phase $\phi_A$ is introduced ( in practice, for the specific final states PP, PV, VP and VV, the different values of $(\rho_A,\phi_A)$ are suggested to fit the data, see Ref.~\cite{Beneke:2003zv} for detail). Meanwhile, the values of $\rho_A$ and $\phi_A$ are treated as universal inputs for different annihilation topologies~\cite{Beneke:2003zv,Beneke:2006hg,Cheng:2009cn,Cheng:2009mu,Cheng:2008gxa}. However,  in 2012, the measurements of the pure annihilation $B_s\to\pi^+\pi^-$ decay, ${\cal B}(B_s\to\pi^+\pi^-)=(0.57\pm0.15\pm0.10)\times 10^{-6}$~(CDF)~\cite{Aaltonen:2011jv} and $(0.95^{+0.21}_{-0.17}\pm0.13)\times 10^{-6}$~(LHCb)~\cite{Aaij:2012as}, present a challenge to the traditional QCDF estimation of the WA contributions, which results in a small prediction $(0.26^{+0.00+0.10}_{-0.00-0.09})\times 10^{-6}$~\cite{Cheng:2009mu}. In the pQCD approach, the possible un-negligible large WA contributions are noticed first in Refs.~\cite{Keum:2000ph,Keum:2000wi,Lu:2000em,Li:2005vu}. Moreover, the prediction of ${\cal B}(B_s\to\pi^+\pi^-)$ with the same central value as the data is presented~\cite{Ali:2007ff,Xiao:2011tx}. 

Recently, motivated by the possible large WA contributions observed by CDF and LHCb collaborations, some researches have been done within both the SM and the NP scenarios, for instance Refs.~\cite{Xiao:2011tx,Chang:2012xv,Gronau:2012gs,Cheng:2014rfa,Li:2015xna,Bobeth:2014rra,Zhu:2011mm,Wang:2013fya}. Especially, some theoretical studies within the QCDF framework are renewed. In Ref.~\cite{Bobeth:2014rra}, the global fits of  WA parameters $X_A(\rho_A,\phi_A)$ are performed. It is found that,  for the decays related by $(u\leftrightarrow d)$ quark exchange,  a universal and relative large WA parameter is supported by the data except for the $B\to\pi K$ system, which exhibits the well-known ``$\Delta A_{CP}(\pi K)$ puzzle", and some tensions in $B\to\phi K^*$ decays. In Refs.~\cite{Zhu:2011mm,Wang:2013fya}, after carefully studying the flavor dependence of the WA parameter $X_A$ on the initial states in $B\to PP$ system, the authors present a ``new treatment"~(a topology-dependent scheme) for the end-point parameters. It is suggested that $X_A$ should be divided into two independent complex parameters $X_A^{i}$ and $X_A^{f}$, which correspond to non-factorizable and factorizable topologies ( gluon emission from the initial and final states, respectively), respectively. Meanwhile, the flavor dependence of $X_A^{i}$ on the initial states, $B_d$ and $B_s$, should be carefully considered. Moreover, the global fits of the end-point parameters in $B\to PP$ and $B\to PV$ decays have confirmed such ``new treatment", except for that the flavor symmetry breaking effect of WA parameters is hard to be distinguished due to the experimental errors and theoretical uncertainties~\cite{Chang:2014yma,Sun:2014tfa}. Numerically, with the simplification $X_H=X_A^{i}$, the best-fit results~\cite{Chang:2014yma,Sun:2014tfa}
 \begin{eqnarray}
    \left\{ \begin{array}{l}
  ({\rho}_{A}^{i},{\phi}_{A}^{i}[^{\circ}]) = 
  (2.98^{+1.12}_{-0.86},-105^{+34}_{-24}),  \\
  ({\rho}_{A}^{f},{\phi}_{A}^{f}[^{\circ}]) = 
  (1.18^{+0.20}_{-0.23},-40^{+11}_{-8}),
    \end{array} \right. \quad {\rm for~PP~final~states}
    \label{PPSA}
   \end{eqnarray}
 \begin{eqnarray}
    \left\{ \begin{array}{l}
  ({\rho}_{A}^{i},{\phi}_{A}^{i}[^{\circ}]) = 
 (2.87^{+0.66}_{-1.95},-145^{+14}_{-21}),\\
  ({\rho}_{A}^{f},{\phi}_{A}^{f}[^{\circ}]) =
  (0.91^{+0.12}_{-0.13},-37^{+10}_{-9}),
   \end{array} \right.\quad {\rm for~PV~final~states}
   \label{PVSA}
   \end{eqnarray}
 are suggested. With such values, all of the QCDF results for charmless $B\to PP$ and $PV$ decays, especially for $B\to\pi \pi$ and $\pi K$ decays, can accommodate the current measurements. 

Even though the topology-dependent scheme for the HSS and WA contributions has been tested in $B\to PP$ and $PV$ decays and presents a good agreement with data, it is also worth further testing whether such scheme persist still in $B\to VV$ decays, which involve more observables, such as polarization fractions and  relative amplitude phases, and thus would present much stronger constraints on the HSS and WA contributions. Moreover,  in recent years, many measurements of $B\to VV$ decays are updated at higher precision~\cite{HFAG}. So, it is also worth reexamining the agreement between QCDF's prediction and experimental data, and investigating the effects of HSS and WA corrections on $B\to VV$ decays, especially some puzzles and tensions therein. 

Our paper is organized as follows. After a brief review of the WA  corrections in $B\to VV$ decays in section 2, we present our numerical analyses and discussions in section 3. Our main conclusions are summarized in section 4.  

\section{Brief Review of WA Corrections}
In the SM, the effective weak Hamiltonian responsible for $b\to p$ transition is written as~\cite{Buchalla:1995vs,Buras:1998raa}
%%%%%%%%%%%%%%%%%%%%%%%%%%%%%%%%%%%%%%%%%%%%%%%%%
\begin{eqnarray}\label{eq:eff}
 {\cal H}_{\rm eff} &=& \frac{G_F}{\sqrt{2}} \biggl[V_{ub}
 V_{up}^* \left(C_1 O_1^u + C_2 O_2^u \right) + V_{cb} V_{cp}^* \left(C_1
 O_1^c + C_2 O_2^c \right) - V_{tb} V_{tp}^*\, \big(\sum_{i = 3}^{10}
 C_i O_i \big. \biggl. \nonumber\\
 && \biggl. \big. + C_{7\gamma} O_{7\gamma} + C_{8g} O_{8g}\big)\biggl] +
 {\rm h.c.},
\end{eqnarray}
%%%%%%%%%%%%%%%%%%%%%%%%%%%%%%%%%%%%%%%%%%%%%%%%%
where $V_{qb} V_{qp}^{\ast}$~$(p=d,s)$ are products of the Cabibbo-Kobayashi-Maskawa~(CKM) matrix elements, $C_{i}$ are the Wilson coefficients, and $O_i$ are the relevant four-quark operators. The essential theoretical problem for obtaining the amplitude of $B\to M_1M_2$ decay is the evaluation of the hadronic matrix elements of the local operators in Eq.~\eqref{eq:eff}. Based on the collinear factorization scheme and color transparency hypothesis, the QCDF approach is developed to deal with the hadronic matrix elements~\cite{Beneke:1999br,Beneke:2000ry}. 

 Up to power corrections of order ${\Lambda}_{QCD}/m_{b}$,
  the factorization formula for $B$ decaying into two light meson
  is given by~\cite{Beneke:1999br,Beneke:2000ry}
  \begin{eqnarray}
  {\langle}M_{1}M_{2}{\vert}O_{i}{\vert}B{\rangle}
  &=&
  \sum\limits_{j}\Big\{ F_{j}^{B{\to}M_{1}}
  {\int}dz\,T^{I}_{ij}(z){\Phi}_{M_{2}}(z)
   +(M_{1}{\leftrightarrow}M_{2})\Big\}
  \nonumber \\ & &
   +{\int}dx\,dy\,dz\,T^{II}_{i}(x,y,z)
  {\Phi}_{B}(x){\Phi}_{M_{1}}(y){\Phi}_{M_{2}}(z)
  \label{eq:ff}.
  \end{eqnarray}
Here, $F_{j}^{B{\to}M_1}$ denotes the  form factor of $B$ ${\to}$ $M_1$ transition,  and ${\Phi}_{X}(z)$ is the light-cone wave function for the two-particle  Fock state of the participating meson $X$, both of which are nonperturbative inputs. $T^{I}(z)$ and $T^{II}(x,y,z)$ denote hard scattering kernels, which could be systematically calculated order by order with the perturbation theory in principle. The QCDF framework for  $B\to VV$ decays has been fully developed in  Refs.~\cite{Beneke:2006hg,Cheng:2008gxa,Bell:2009fm,Beneke:2009ek,Bell:2015koa}.

For the WA contributions, the convolution integrals in $B\to VV$ decays exhibit not only the logarithmic infrared divergence regulated by Eq.~(\ref{Xa}) but also the linear infrared divergence appeared in the transverse building blocks $A_{1,2}^{i-}$, which is different from the case of $B\to PP,\,PV$ decays. With the treatment similar to $X_{A}$ in Eq.~(\ref{Xa}), the linear divergence is usually extracted into unknown complex quantity $X_{L}$ defined as~\cite{Beneke:2006hg}  
\begin{equation}\label{XL}
\int_{0}^{1}\frac{dx}{x^2}\to X_{L}(\rho_A,\phi_A)=(1+\rho_{A}e^{i\phi_{A}})\frac{m_{b}}{\Lambda_h}\,.
\end{equation}
In such a parameterization scheme, even though the predictive power of QCDF is partly weakened due to  the incalculable WA parameters, such scheme provides a feasible way to evaluate the effects of WA corrections in a phenomenological view point. Traditionally, the end-point parameters $X_{A\,,L}^{i,f}(\rho_A^{i,f},\phi_A^{i,f})$ are assumed to be universal for different WA topologies of $B\to VV$ decays, and ones take the values $\rho_A^{f}=\rho_A^{i}\sim 0.7$  and $\phi_A^{f}=\phi_A^{i}\sim -50^{\circ}$~\cite{Beneke:2006hg,Cheng:2008gxa} as input. In this paper, in order to test the proposal of  Refs.~\cite{Zhu:2011mm,Wang:2013fya} mentioned in introduction, $(\rho_A^{f},\phi_A^{f})$ and $(\rho_A^{i},\phi_A^{i})$ are treated as independent parameters, and responsible for the end-point corrections of factorizable and non-factorizable WA topologies, respectively.   

After evaluating the convolution integral with the asymptotic light-cone distribution amplitudes, one can get the basic building blocks of WA amplitudes, which are explicitly written as~\cite{Beneke:2006hg,XQLi}
\begin{eqnarray}
A_{1}^{i,0}&\simeq& A_{2}^{i,0}\simeq 18\pi\alpha_{s}\left[(X_{A}^{i}-4+\frac{\pi^2}{3})+r_{\chi}^{V_{1}}r_{\chi}^{V_{2}}(X_{A}^{i}-2)^2\right]\,,\\
\label{a3iz}
A_{3}^{i,0}&\simeq&
18\pi\alpha_{s}(r_{\chi}^{V_{1}}-r_{\chi}^{V_{2}})\left[-(X_{A}^{i})^2+2X_{A}^{i}-4+\frac{\pi^2}{3}\right]\,,\\
A_{3}^{f,0}&\simeq&
18\pi\alpha_{s}(r_{\chi}^{V_{1}}+r_{\chi}^{V_{2}})(2X_{A}^{f}-1)(2-X_{A}^{f})\,,
\end{eqnarray}
for the non-vanishing longitudinal contributions, and 
\begin{eqnarray}
A_{1}^{i,+}&\simeq& A_{2}^{i,+}\simeq 18\pi\alpha_{s}\frac{m_{1}m_{2}}{m_{B}^2}\left[2(X_{A}^{i})^2-3X_{A}^{i}+6-\frac{2\pi^2}{3}\right]\,,\\
A_{1}^{i,-}&\simeq& A_{2}^{i,-}\simeq
18\pi\alpha_{s}\frac{m_{1}m_{2}}{m_{B}^2}\left(\half X_{L}^{i}+\frac{5}{2}-\frac{\pi^2}{3}\right),\\
\label{a3im}
A_{3}^{i,-}&\simeq&
18\pi\alpha_{s}\left(\frac{m_{1}}{m_{2}}r_{\chi}^{V_{2}}-\frac{m_{2}}{m_{1}}r_{\chi}^{V_{1}}\right)\left[(X_{A}^{i})^2-2X_{A}^{i}+2\right]\,,\\
A_{3}^{f,-}&\simeq&
18\pi\alpha_{s}\left(\frac{m_{1}}{m_{2}}r_{\chi}^{V_{2}}+\frac{m_{2}}{m_{1}}r_{\chi}^{V_{1}}\right)\left[2(X_{A}^{f})^2-5X_{A}^{f}+3\right]\,,
\end{eqnarray}
for the transverse ones. Generally, $A_3^{i,0}$ and $A_3^{i,-}$ given by Eqs.~\eqref{a3iz} and \eqref{a3im} are very small and therefore negligible for the case of light final states. One may refer to Refs.~\cite{Beneke:2006hg,Cheng:2009mu,Cheng:2008gxa} for the further explanation. 

The decays modes considered in this paper include the penguin-dominated $B\to \rho K^*$ decays induced by $b\to s \bar{q} q$~($q=u,d$) transition and $B\to \phi K^*$ decays induced by $b\to s \bar{s} s$ transition, the tree-dominated $B\to \rho \rho$ decays induced by $b\to d \bar{q} q$ transition, and the penguin- and/or annihilation-dominated $B\to K^* \bar{K}^*$ and $\phi\phi$ decays induced by $b\to d \bar{s} s$ transition. The explicit expressions of their amplitudes are summarized in Appendix A. As is known, the penguin-dominated and color-suppressed tree dominated decays are very sensitive to the WA and the HSS contributions, respectively. So, it is expected that their precisely measured observables could present strong constraints on the end-point parameters. The pure annihilation decays, such as $\bar{B}^0\to \bar{K}^{*-} K^{*+}$ and $\phi\phi$ decay modes, are much more suitable for probing the annihilation contributions without the interference effects. However, there is no available experimental result by now. So, we leave them as our predictions, which will be tested by the forthcoming measurements at LHC and super-KEKb.    

\section{Numerical Analyses and Discussions}

\begin{table}[t]
\caption{\small  The values of input parameters: Wolfenstein parameters, pole and running quark masses, decay constants, form factors and Gegenbauer moments. For the other inputs, such as masses and lifetimes of mesons, we take their central values given by PDG~\cite{PDG}.}
 \label{ppvalue}
\begin{footnotesize}
 \vspace{0.1cm}
 \small
\doublerulesep 0.1pt \tabcolsep 0.1in
\begin{tabular}{c}
\Xhline{2pt}
  $\bar{\rho}=0.145_{-0.007}^{+0.013}$,~~ $\bar{\eta}=0.343_{-0.012}^{+0.011}$,~~
  $A=0.810_{-0.024}^{+0.018}$,~~ $\lambda=0.22548_{-0.00034}^{+0.00068}$ ~\cite{CKMfitter},\\\Xhline{1pt}
  $m_{c}=1.67\pm0.07$ GeV,~~  $m_{b}=4.78\pm0.06$ GeV,~~  $m_{t}=173.21\pm0.87$ GeV,\\
  $\frac{\bar{m}_{s}(\mu)}{\bar{m}_{u,d}(\mu)}=27.5\pm1.0$,~~  ${\bar{m}_{s}(2 {\rm GeV})}=95\pm5$ MeV,~~  ${\bar{m}_{b}(m_{b})}=4.18\pm0.03$ GeV ~\cite{PDG},\\\Xhline{1pt}
  $f_{B_{u,d}}=190.6\pm4.7$ MeV,~~$f_{\rho}=216\pm3$ MeV,~~$f_{\rho}^{\perp}=165\pm9$ MeV,\\
  $f_{K^{*}}=220\pm5$ MeV,~~$f_{K^{*}}^{\perp}=185\pm10$ MeV,
  $f_{\phi}=215\pm5$ MeV,~~$f_{\phi}^{\perp}=186\pm9$ MeV~\cite{lattice,Ball:2006eu},\\\Xhline{1pt}
  $A_{0}^{B\to\rho}=0.303\pm0.029$ GeV,~~$A_{1}^{B\to\rho}=0.242\pm0.023$ GeV,~~
  $V^{B_{u}\to\rho}=0.323\pm0.030$ GeV,~~\\
  $A_{0}^{B\to K^{*}}=0.374\pm0.034$ GeV,~~$A_{1}^{B\to K^{*}}=0.292\pm0.028$ GeV,~~
  $V^{B\to K^{*}}=0.411\pm0.033$ GeV ~\cite{Ball:2004rg},\\
  \Xhline{1pt}
  $a_{1}^{\parallel,\perp}(\rho)^{\mu=\text{2GeV}}=0$,~
  $a_{1}^{\parallel,\perp}(\phi)^{\mu=\text{2GeV}}=0$,~
   $a_{1}^{\parallel,\perp}(K^*)^{\mu=\text{2GeV}}=0.02$,~\\ 
   $a_{2}^{\parallel,\perp}(\rho)^{\mu=\text{2GeV}}=0.10$,~
  $a_{2}^{\parallel,\perp}(\phi)^{\mu=\text{2GeV}}=0.13$,~   
  $a_{2}^{\parallel,\perp}(K^*)^{\mu=\text{2GeV}}=0.08$~\cite{Ball:2007rt}.\\
\Xhline{2pt}
\end{tabular}
\end{footnotesize}
\end{table}

In this paper, the independent observables, including CP-averaged branching fraction, CP asymmetries, polarization fractions and relative amplitude phases, are evaluated. For these observables, we take the same definition and convention as Ref.~\cite{Beneke:2006hg}. The available experimental results averaged by HFAG~\cite{HFAG} are listed in the ``Exp." columns of Tables \ref{tab:rhok}, \ref{tab:kk}, \ref{tab:phik} and \ref{tab:rhorho}~(the recent measurement of $\bar{B}^{0}\to\rho^{+}\rho^{-}$ decay reported by Belle~\cite{Vanhoefer:2015ijw} agrees well with the previous results, and hasn't been included in the HFAG's average), which are employed in the coming fits of WA parameters. In addition, the values of input parameters used in our evaluations are listed in the Table~\ref{ppvalue}. 

 In the $\chi^2$ analyses, with the same statistical $\chi^2$ approach as the one given in the appendix of Refs.~\cite{Chang:2014rla,Hofer:2010ee}, we firstly scan randomly the points of end-point parameters in the conservative ranges and evaluate the $\chi^2$ value of each point. Then, we find out the $\chi_{\rm min}^2$ and get the allowed spaces (points) at $68\%\,,95 \%$ C.L.. If more than one separate spaces are found, we pick each of them out and further deal with them respectively with their local minima of the $\chi^2$ ({\it e.g.} the 4 solutions in the coming Fig.~\ref{KKrhoK}).  

With aforementioned theoretical strategy and inputs, we now proceed to present our numerical results and discussions, which are divided into four cases for different purposes:

\begin{figure}[t]
\begin{center}
 \subfigure[]{\includegraphics[width=8cm]{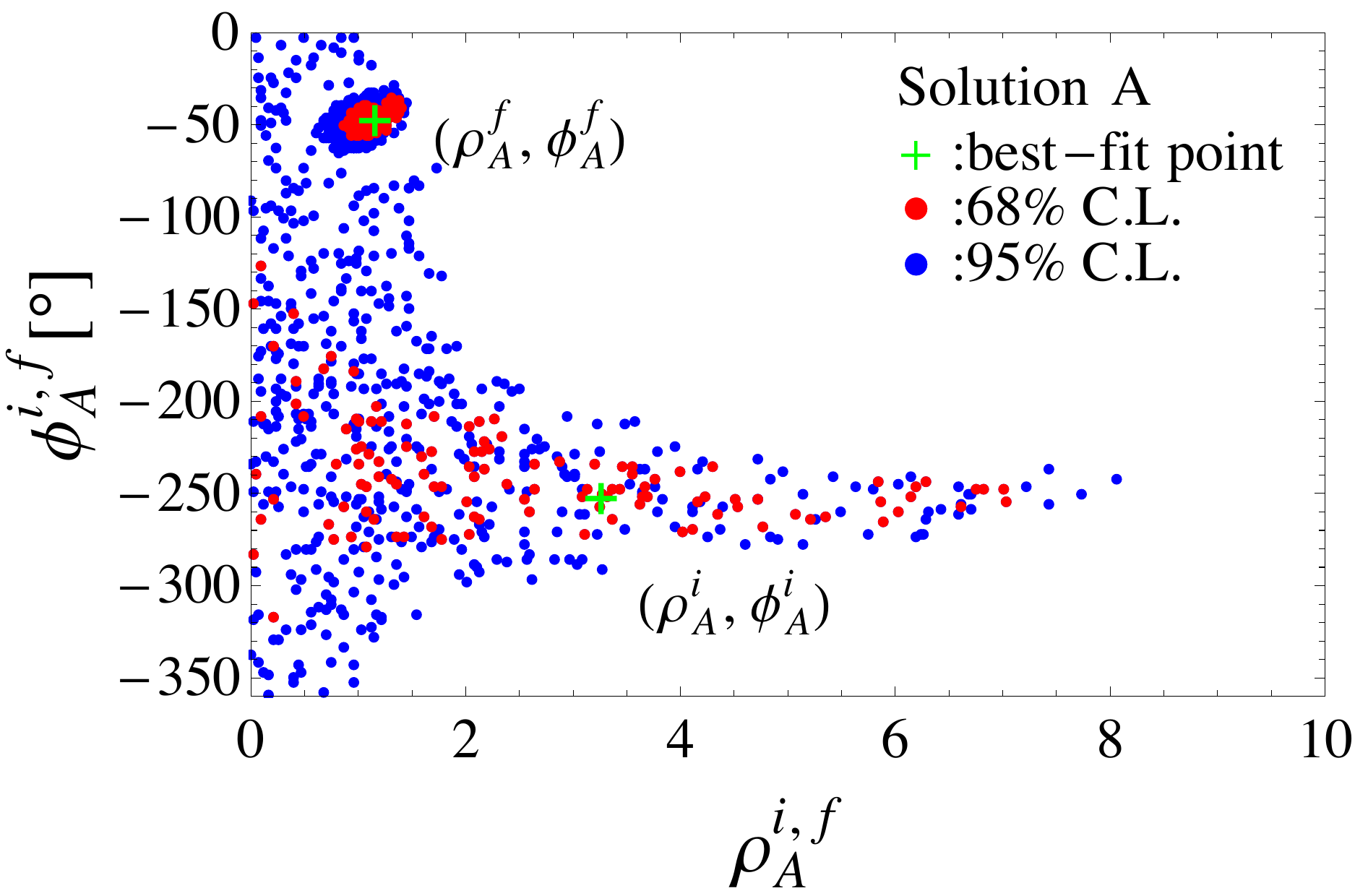}}\quad
 \subfigure[]{\includegraphics[width=8cm]{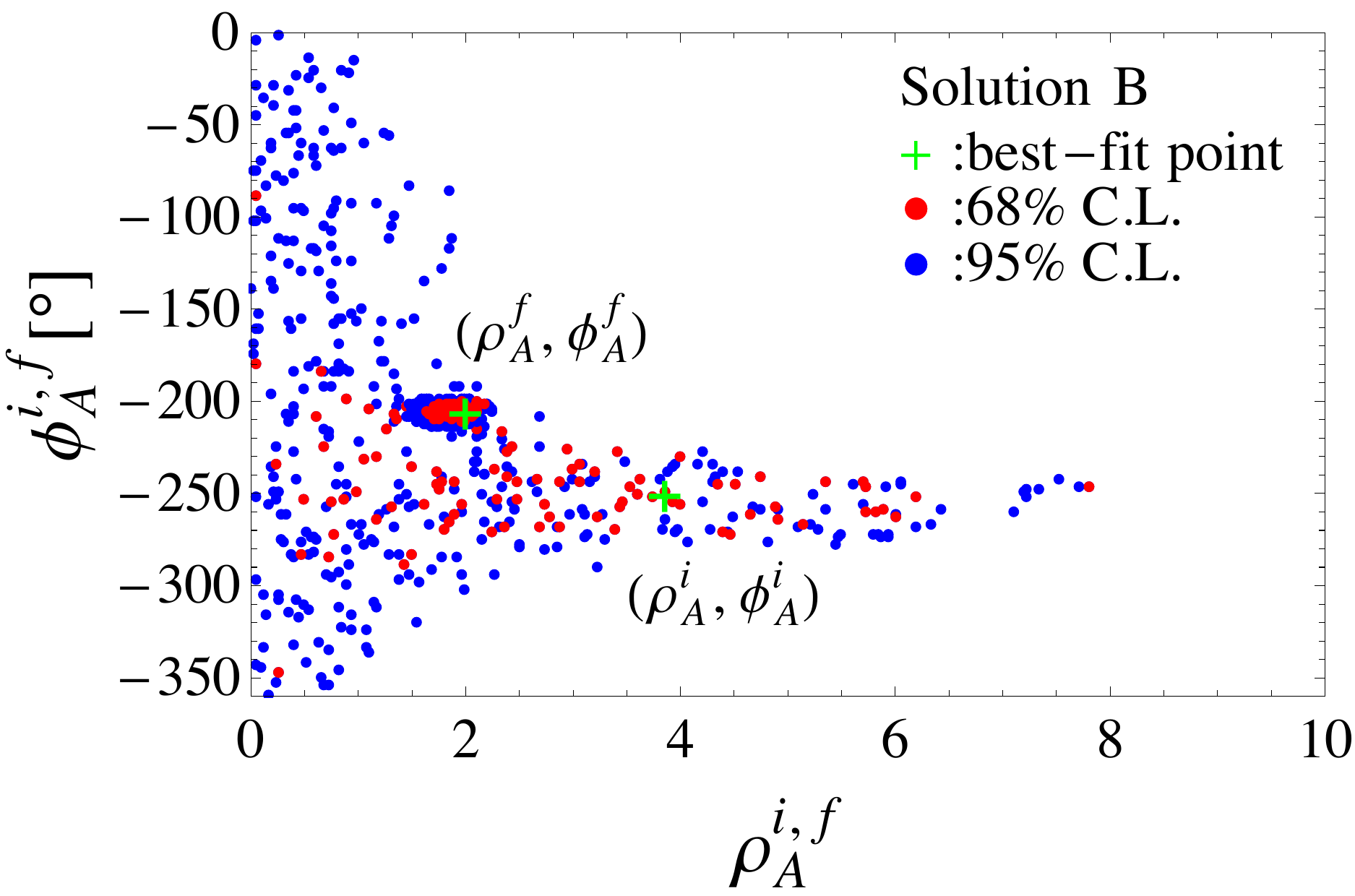}}\\
 \subfigure[]{\includegraphics[width=8cm]{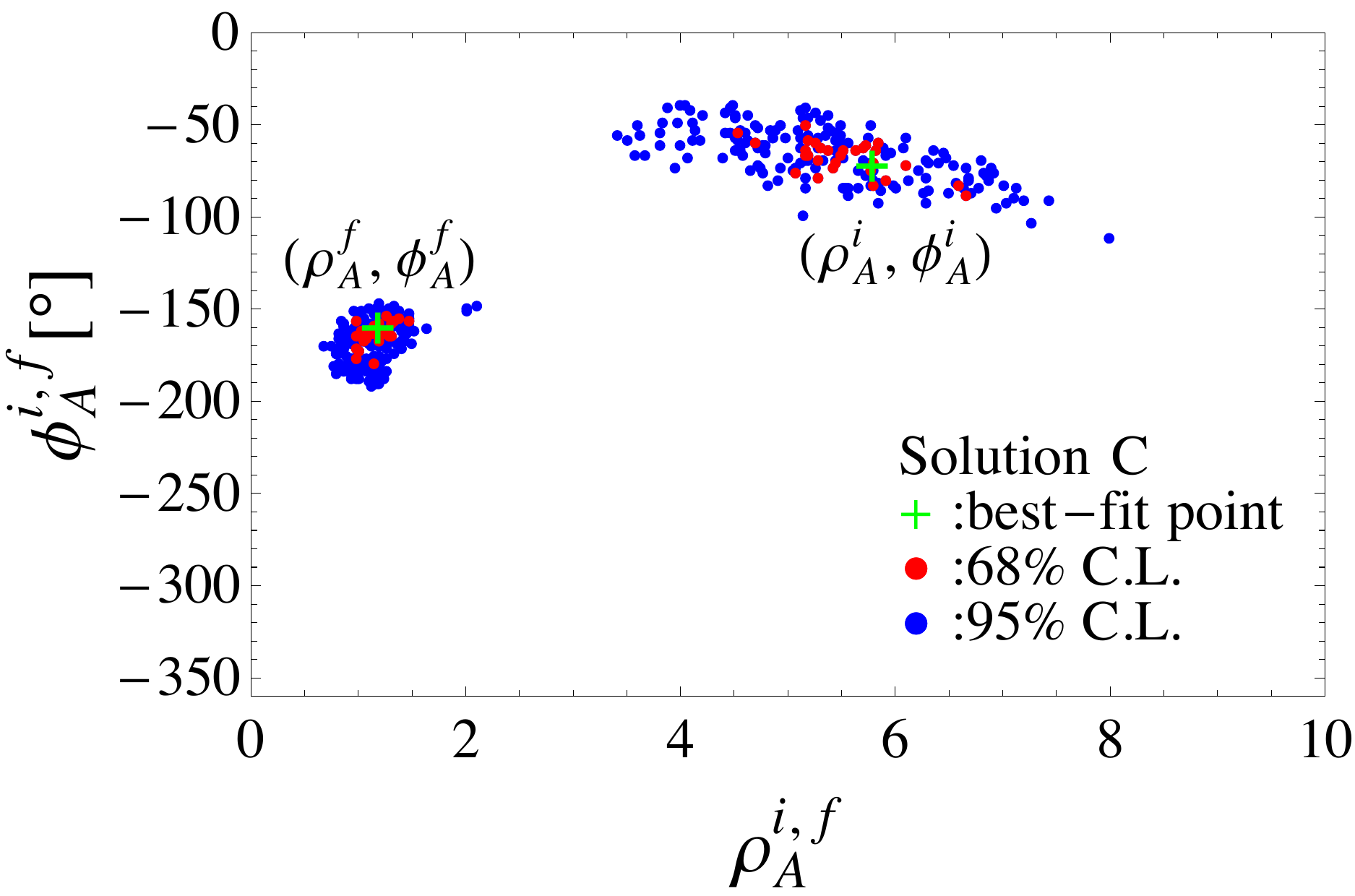}}\quad
 \subfigure[]{\includegraphics[width=8cm]{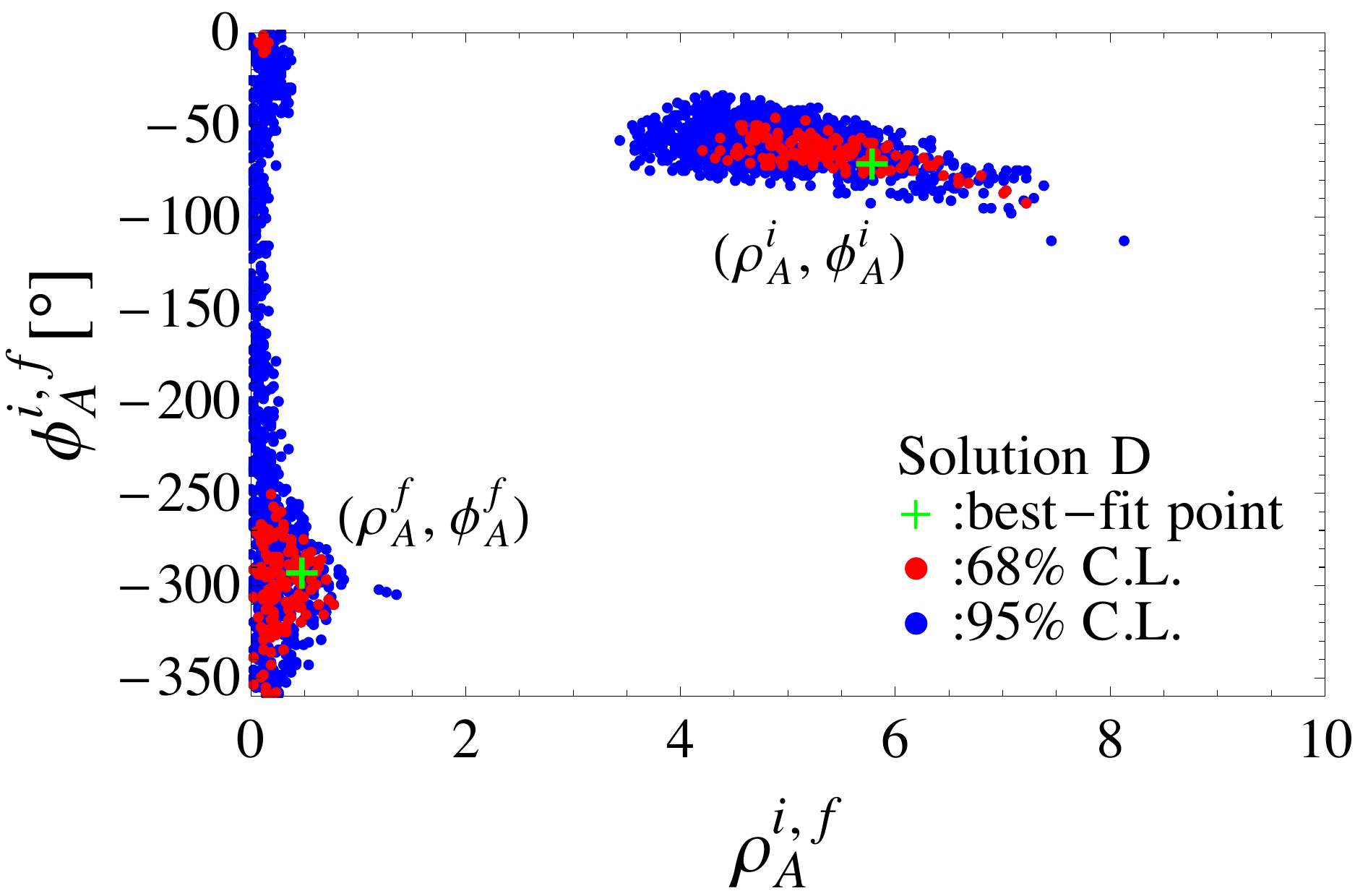}}
\caption{\label{KKrhoK} \small The allowed regions of WA parameters at 68\% and 95\% C. L. with the combined constraints from  $B\to \rho K^*, \bar{K}^*K^*$ decays. The best-fit points of solutions A, B, C and D correspond to $\chi_{\rm min}^2=5.0$, $5.1$, $5.9$ and $6.0$, respectively. One may also see Fig.~\ref{KKrhoK2} plotted in the complex plane.} 
\end{center}
\end{figure}

\subsection{Case I}
For case I, in order to test the topology-dependent scheme, $(\rho_{A}^{i,f},\phi_{A}^{i,f})$ are treated as free parameters. Meanwhile, the simplification $(\rho_{H},\phi_{H})=(\rho_{A}^{i},\phi_{A}^{i})$ allowed by $B\to PP$ and $PV$ decays~\cite{Chang:2014yma,Sun:2014tfa} is assumed. The combined constraints from $B_{u,d}\to \rho K^*,\bar{K}^*K^*$ decays, where 16 observables (see Tables \ref{tab:rhok} and \ref{tab:kk}) are well measured, are considered in the fit.   

For the $B\to \rho K^{*}$ decays, the tree contributions $\alpha_{1,2}$ are strongly suppressed by the CKM factor $|V_{us}^*V_{ub}|\sim {\cal O}(\lambda^4)$, whereas the QCD penguin contribution $\alpha_{4}$  is proportional to $|V_{cs}^*V_{cb}|\sim {\cal O}(\lambda^2)$ and thus dominates the amplitudes. Therefore, the WA contributions with the same CKM factor $|V_{cs}^*V_{cb}|$ as $\alpha_{4}$ would be important for these decays. In their amplitudes given by Eqs.~(\ref{eq:rhok1}-\ref{eq:rhok4}), the main WA contribution is derived from the effective WA coefficient $\beta_3$, which is dominated by the building block $A_3^f$ accompanied by $N_cC_6$. So, $B\to \rho K^{*}$ decays would present strict constraints on $(\rho_{A}^{f},\phi_{A}^{f})$. 

Moreover, the measured penguin-dominated $B^{-}\to K^{*-}K^{*0}$ and $\bar B^{0}\to \bar K^{*0}K^{*0}$ decays,  which amplitudes are given by Eqs.~\eqref{eq:kk1} and \eqref{eq:kk3}, would provide further constraints on $(\rho_{A}^{f},\phi_{A}^{f})$. Besides,  due to the existence of $\beta_{2,4}$, which are relevant to $A_{1,2}^i$ only, $B^{-}\to K^{*-}K^{*0}$ and $\bar B^{0}\to \bar K^{*0}K^{*0}$ decays also may provide some constraints on $(\rho_{A}^{i},\phi_{A}^{i})$. 

Under the combined constraints from $B_{u,d}\to \rho K^*,\bar{K}^*K^*$ decays, the allowed spaces of end-point parameters $(\rho_{A}^{i,f},\phi_{A}^{i,f})$ are shown in Fig.~\ref{KKrhoK}. It could be found that: (i) as expected, the parameters $(\rho_{A}^{f},\phi_{A}^{f})$ are strictly bounded into four separate regions, which are named solutions A-D. However, the constraint on $(\rho_{A}^{i},\phi_{A}^{i})$ is very loose. In addition, the two different spaces of  $(\rho_{A}^{f},\phi_{A}^{f})$  in solutions A and B, as well as the ones in solutions C and D, correspond to almost the same  $(\rho_{A}^{i},\phi_{A}^{i})$ space. In fact, the two solutions (solutions A and B, or solutions C and D) result in the similar WA corrections. Such situation also exists in the $B\to PP$ and $PV$ decays~\cite{Chang:2014yma,Sun:2014tfa}. (ii) The relation $(\rho_{A}^{f},\phi_{A}^{f})\neq (\rho_{A}^{i},\phi_{A}^{i})$ is always required at 68\% C. L., except for the solution B  shown by Fig.~\ref{KKrhoK}~(b) due to the loose constraints on $(\rho_{A}^{i},\phi_{A}^{i})$. (iii) Corresponding to the best-fit points of 4 solutions, the  best-fit values of end-point parameters are 
\begin{equation}
\label{scaseI}
 (\rho_A,\phi_A[^\circ])^{i,\,f}=
\left\{ \begin{array}{l}
(3.27,-251),~(1.17,-45)\,;\quad \text{solution A}\\
(3.86,-250),~(2.00,-205)\,;\quad \text{solution B}\\
(5.80,-70),~(1.19,-158)\,;\quad \text{solution C}\\
(5.79,-69),~(0.48,-291)\,.\quad \text{solution D}
\end{array} \right.
\end{equation}
Interestingly, the result $(\rho_A^{f},\phi_A^{f}[^\circ])=(1.17,-45)$ of solution A is very similar to the results gotten in $B\to PP$ and $PV$ decays given by Eqs.~(\ref{PPSA}) and (\ref{PVSA}). Furthermore, the $(\rho_A^{f},\phi_A^{f}$ of solution B in Eq.~\eqref{scaseI} is also very similar to the other results in $B\to PP$ and $PV$ decays~(solution B given in Refs. \cite{Chang:2014yma} and \cite{Sun:2014tfa}). A more clear comparison will be present in the next case.

In the past years, the penguin-dominated $B\to \phi K^*$ decays have attracted much attention due to the well-known ``polarization anomaly". One may refer to Ref.~\cite{Kagan:2004uw} and the most recent studies in QCDF and pQCD approaches~\cite{Bobeth:2014rra,Zou:2015iwa} for detail. For $B\to \phi K^*$ decays, the complete angular analyses are available, which would present much stricter requirement for the WA contributions. In Ref.~\cite{Bobeth:2014rra}, with the traditional ansatz that the end-point parameters are universal for all of the annihilation topologies, it is found that the current measurements for the observables of $B\to \phi K^*$ decays are hardly to be accommodated simultaneously. So, in the next case, we would like to test whether the possible disagreement could be moderated by the topology-dependent scheme.     

\begin{figure}[t]
\begin{center}
 \subfigure[]{\includegraphics[width=8cm]{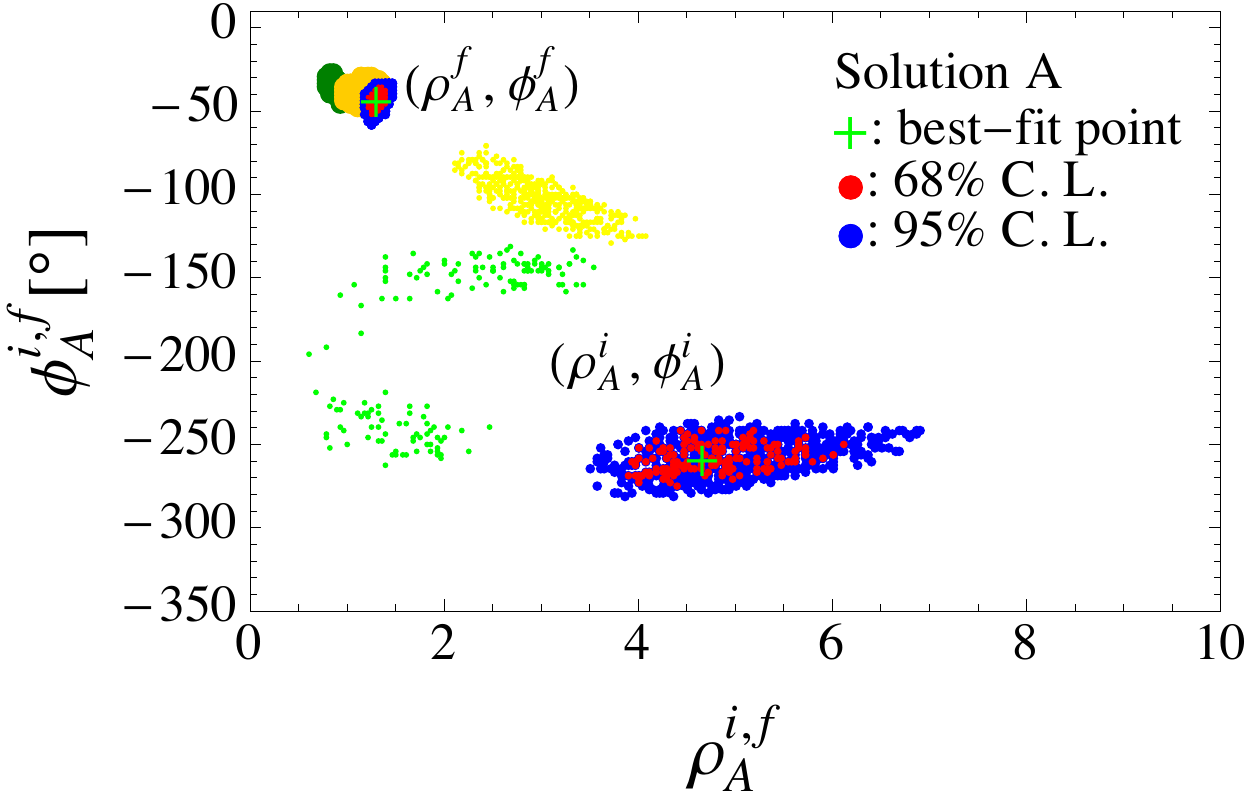}}\quad
 \subfigure[]{\includegraphics[width=8cm]{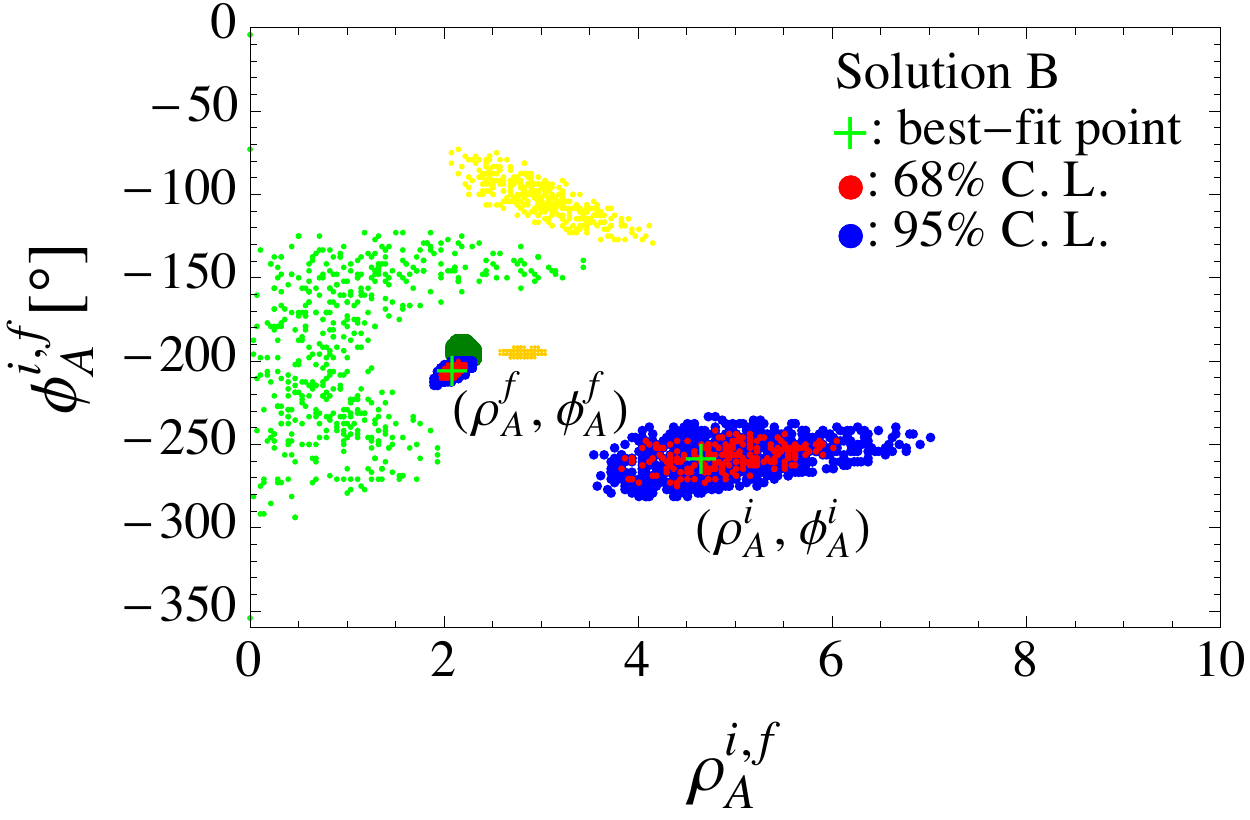}}
\caption{\label{KKrhoKphiK} \small The allowed regions of end-point parameters with the constraints from  $B_{u,d}\to \rho K^*, \bar{K}^*K^*$ and $\phi K^*$ decays. The best-fit points of solutions A and B correspond to $\chi_{\rm min}^2=11.1$. For comparison, the fitted results of $(\rho_{A}^{i(f)},\phi_{A}^{i(f)})$ in $B\to PP$ and $PV$ decays are also shown by light~(dark) yellow and green pointed regions, respectively.  One may also see Fig.~\ref{KKrhoKphiK2} plotted in the complex plane. }
\end{center}
\end{figure}

\begin{figure}[t]
\begin{center}
 \subfigure[]{\includegraphics[width=5.5cm]{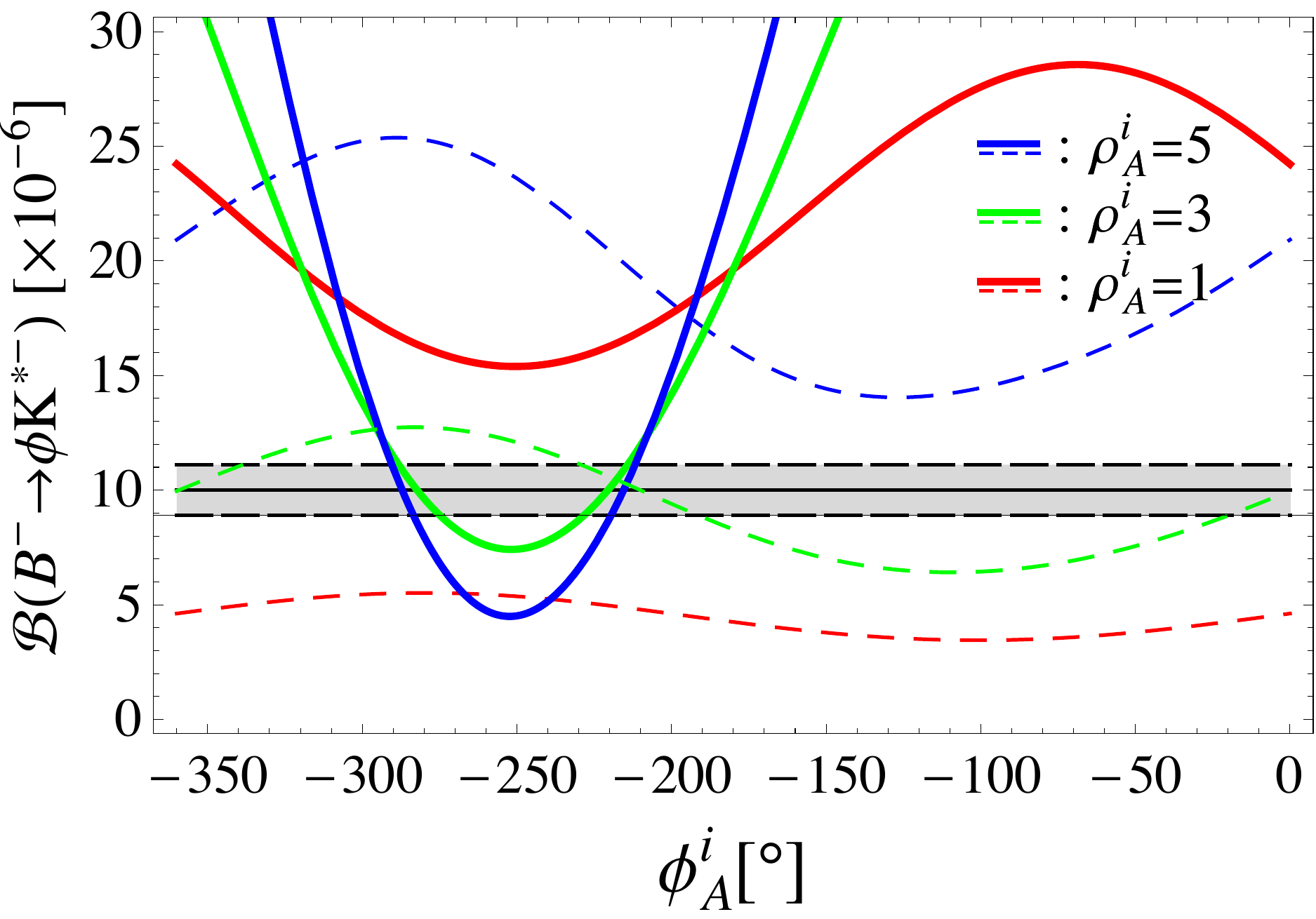}}
 \subfigure[]{\includegraphics[width=5.6cm]{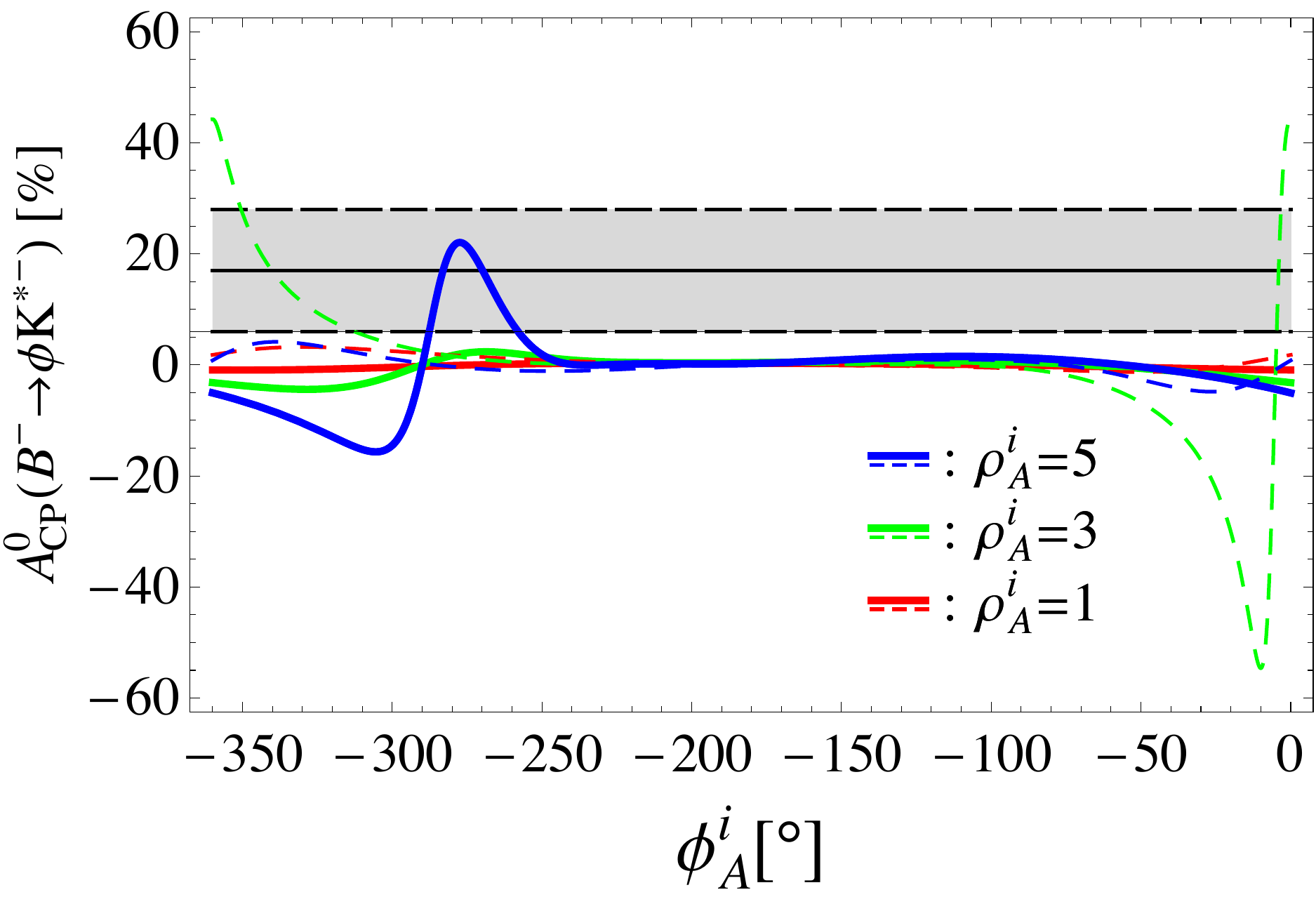}}
  \subfigure[]{\includegraphics[width=5.6cm]{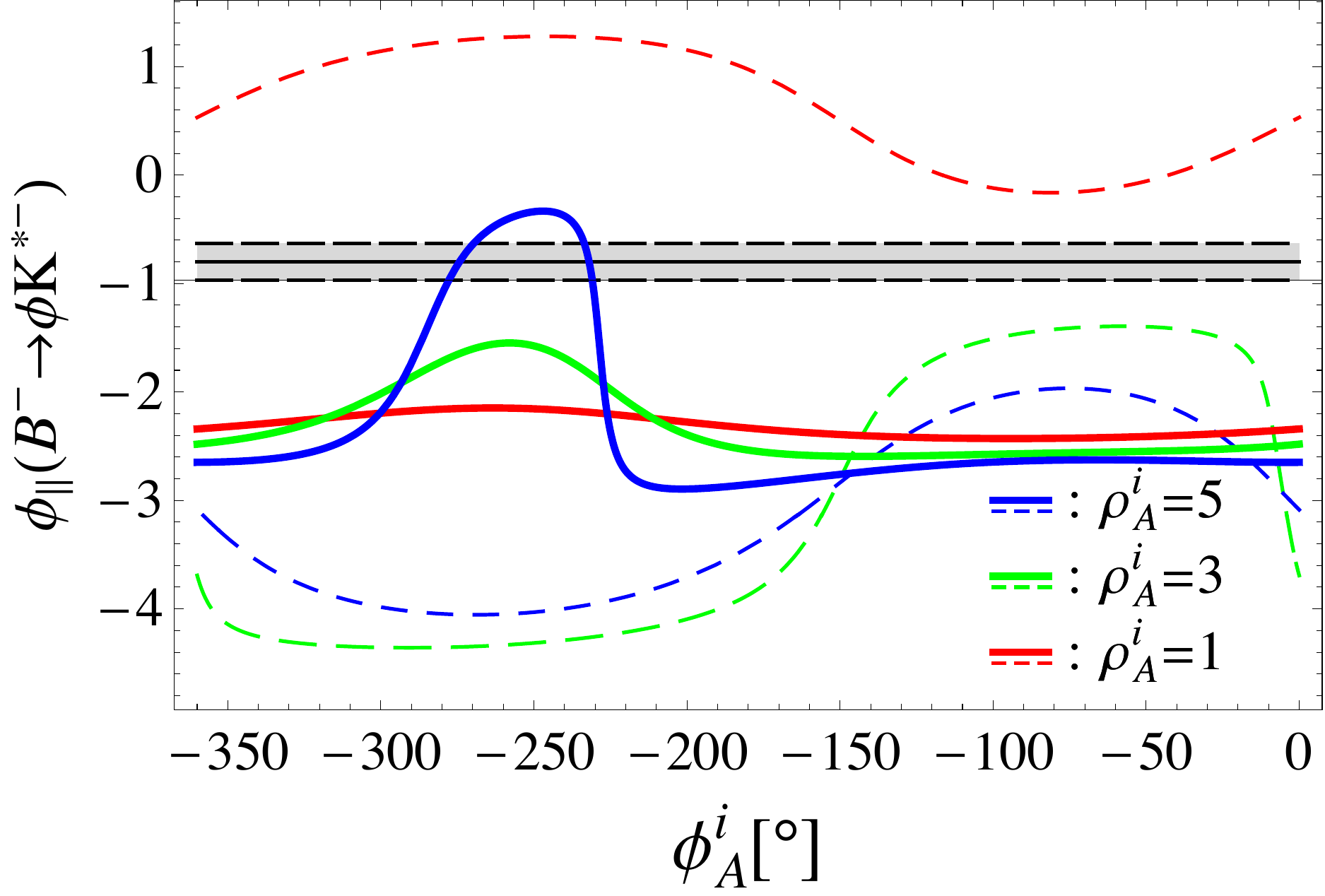}}\\
  \subfigure[]{\includegraphics[width=5.5cm]{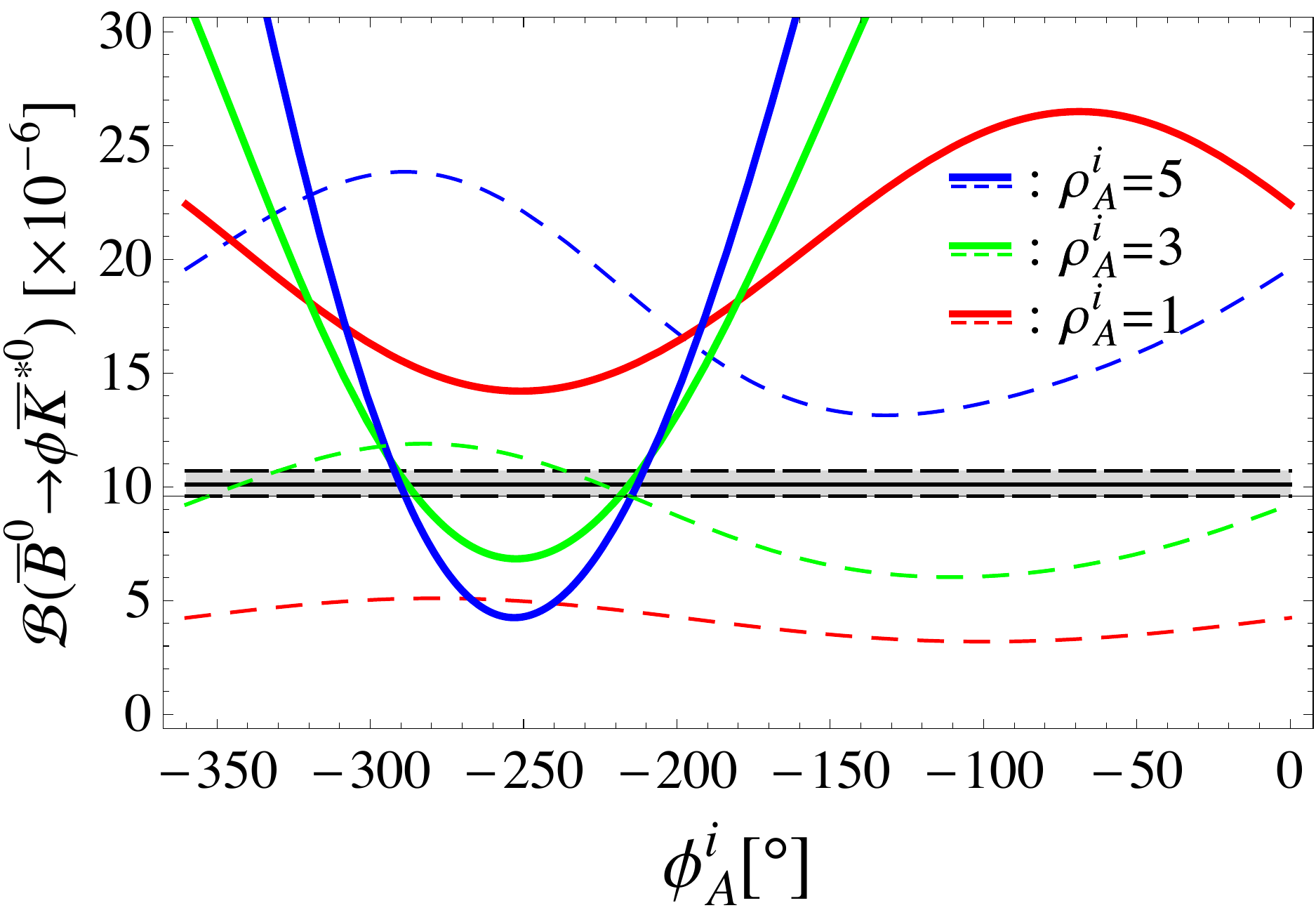}}
 \subfigure[]{\includegraphics[width=5.6cm]{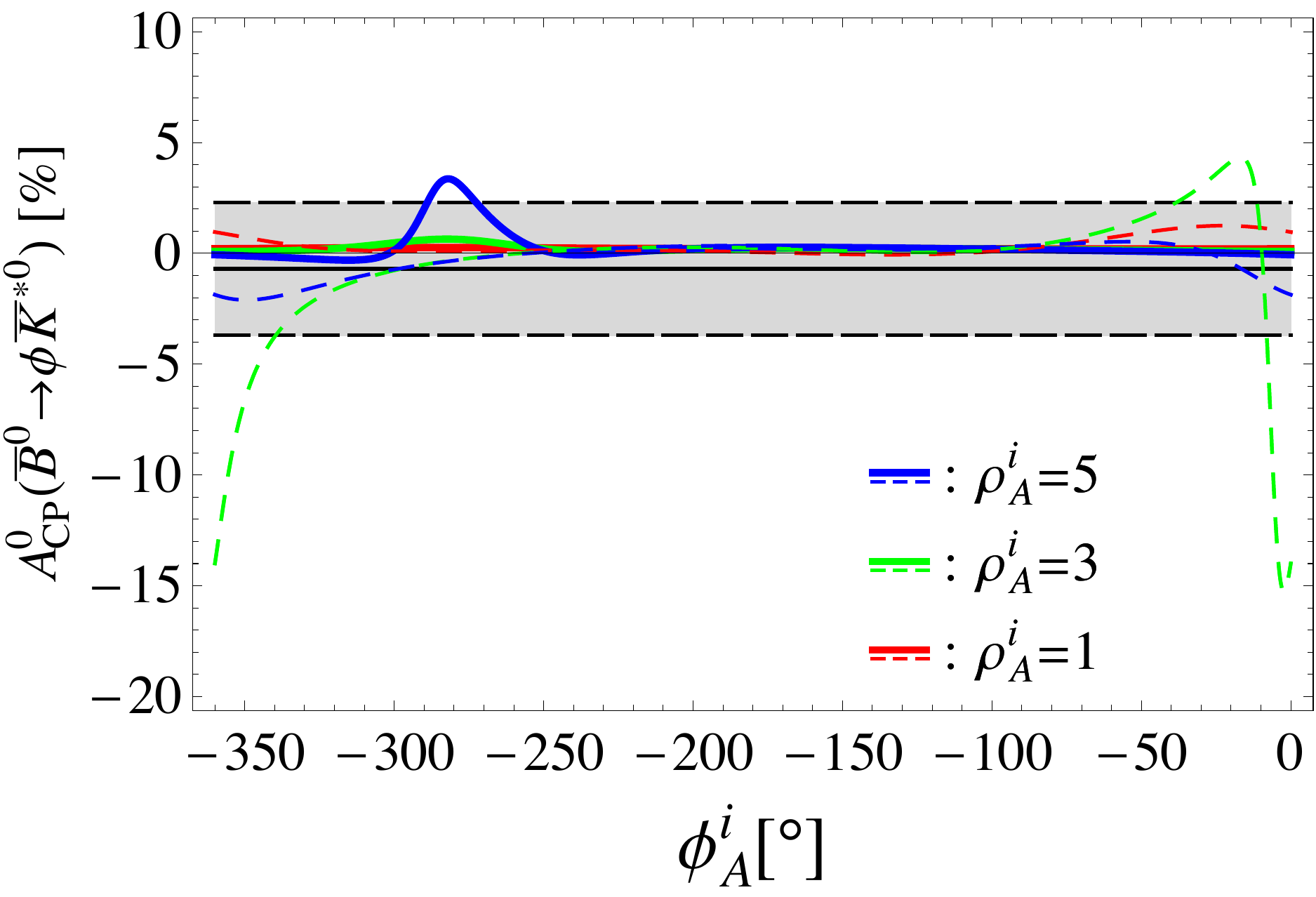}}
  \subfigure[]{\includegraphics[width=5.6cm]{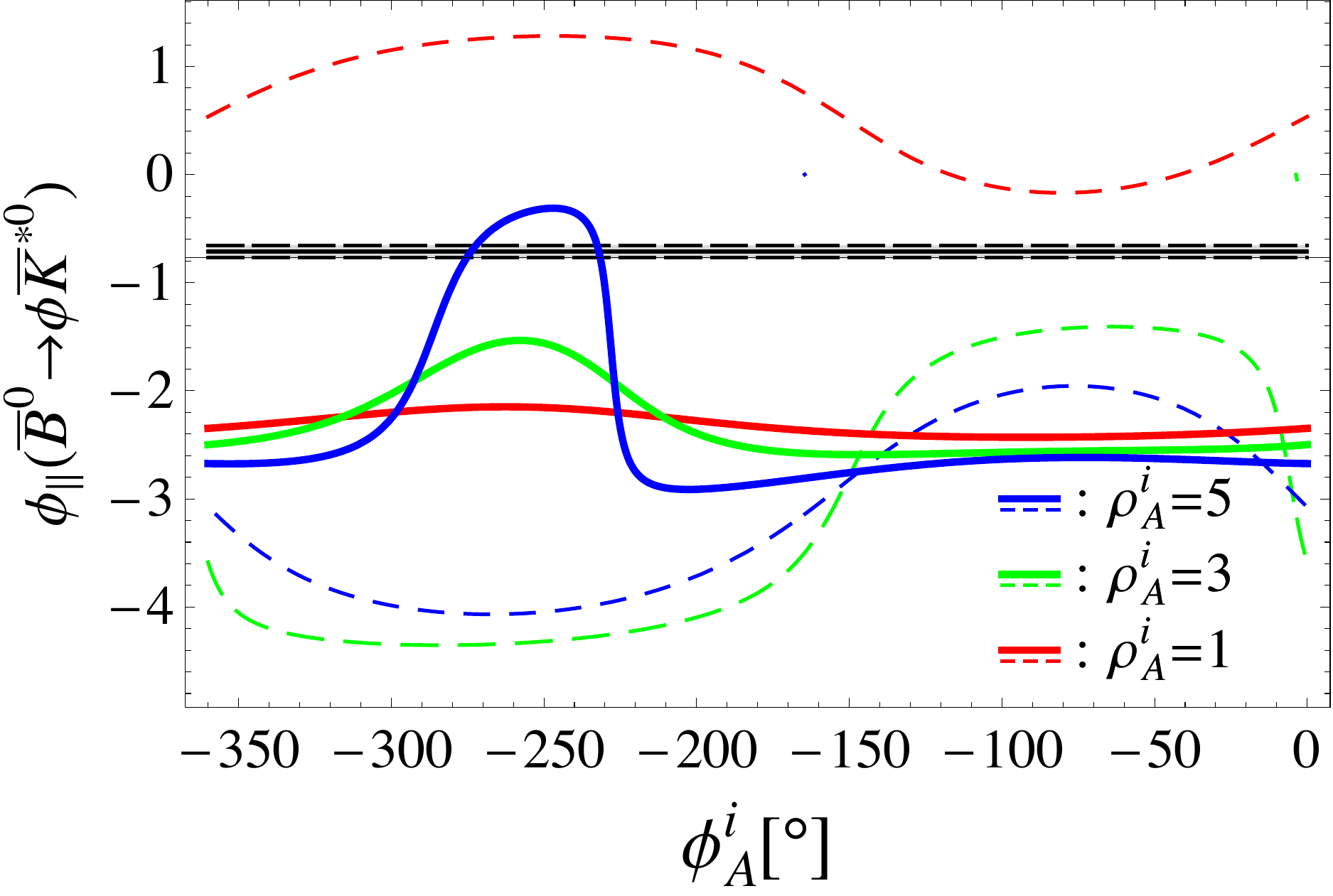}}
\caption{\label{depphiK} \small The dependences of some observables of $B\to \phi K^*$ decays on the parameter $\phi_A^i$ with different values of $\rho_A^i$ and settled $(\rho_{A}^{f},\phi_{A}^{f})$ of solutions A~(solid lines) and C~(dashed lines) in Eq.~(\ref{scaseI}). The shaded bands are experimental results with $1\sigma$ error. }
\end{center}
\end{figure}

\subsection{Case II}
In this case, we take the same ansatz as case I except to take the constraints from $B_{u,d}\to \phi K^*$ decays into account. In the fit, all of the available observables are considered except for $\phi_\bot({\phi K^*})$ because $\phi_\bot\simeq \phi_{\parallel}$ is hold in the QCDF and also supported by the current measurements within errors. With the constraints from 32 measured observables, our fitted results are shown by Fig.~\ref{KKrhoKphiK}.  The value of $\chi_{\rm min}^2$ in this case is much larger than the ones in case I because $B_{u,d}\to \phi K^*$ decay modes and relevant observables are considered as constraint conditions. In addition, to clarify the effects of WA contributions on $B_{u,d}\to \phi K^*$ decays, the dependences of some observables on the end-point parameters are plotted in Fig.~\ref{depphiK}. 

From Fig.~\ref{KKrhoKphiK}, it could be found that: (i) comparing with the solutions A and B of case I, the allowed spaces of end-point parameters are further restricted by $B_{u,d}\to \phi K^*$ decays.  Especially, the regions of $(\rho_{A}^{i},\phi_{A}^{i})$ are strictly bounded around $(4,-250^{\circ})$, which is mainly required by $A_{CP}^0(B^-\to \phi K^{*-})$, $\phi_{\parallel}(B_{u,d}\to \phi K^*)$ and also allowed by the other observables as Fig.~\ref{depphiK}~(solid lines) shows. However, the solutions C and D in case I are entirely excluded by $B\to \phi K^*$, which could be easily understood from Fig.~\ref{depphiK}~(dashed lines); (ii) There is no overlap between the spaces of $(\rho_{A}^{i},\phi_{A}^{i})$ and $(\rho_{A}^{f},\phi_{A}^{f})$ at 95\% C. L., which means that the relation $(\rho_{A}^{f},\phi_{A}^{f})\neq (\rho_{A}^{i},\phi_{A}^{i})$ found in $B\to PP\,,PV$ decays is also required by $B\to VV$ decays; (iii) More interestingly, comparing with the previous fitted results gotten through $B\to PP\,,PV$ decays, it could be found that the allowed spaces of $(\rho_{A}^{f},\phi_{A}^{f})$ in $B\to PP\,,PV$ and $VV$ decays are very close to each other, which implies possible universal $X_A^f(\rho_{A}^{f},\phi_{A}^{f})$ for all of the decay modes. However, no significant relationship could be found for $(\rho_{A}^{i},\phi_{A}^{i})$.

Corresponding to the spaces of  end-point parameters in Fig.~\ref{KKrhoKphiK}, the numerical results are 
 \begin{equation}
\label{scaseII}
 (\rho_A,\phi_A[^\circ])^{i,\,f}=
\left\{ \begin{array}{l}
(4.66_{-0.76}^{+1.47},-259_{-16}^{+18}),~(1.30_{-0.08}^{+0.11},-44_{-8}^{+10})\,,\quad \text{solution A}\\
(4.66_{-0.84}^{+1.39},-259_{-15}^{+17}),~(2.08_{-0.12}^{+0.14},-206_{-5}^{+5})\,,\quad \text{solution B}
\end{array} \right.
\end{equation}
 in which the two solutions for $ (\rho_A^{i},\phi_A^{i})$ are in fact the same. With solution A as input, we then present the updated QCDF's results for $B\to VV$ decays in the ``case II" columns of Tables  \ref{tab:rhok}, \ref{tab:kk}, \ref{tab:phik}, \ref{tab:rhorho} and \ref{tab:phiphi}, in which the data~\cite{HFAG} and the previous theoretical results~\cite{Cheng:2009cn,Cheng:2008gxa,Beneke:2006hg} are also listed for comparison. One may find that most of the updated results are in good agreements with the data within the errors and uncertainties except for an unexpected large ${\cal B}(\bar{B}^0\to\rho^0\rho^0)$. 
 
The $\bar{B}^0\to\rho^0\rho^0$ decay is dominated by color-suppressed tree contribution $\alpha_2$, and thus very sensitive to the HSS corrections. Therefore, the unexpected large theoretical result of ${\cal B}(\bar{B}^0\to\rho^0\rho^0)$ is mainly caused by the large $\rho_A^i$ and the simplification $(\rho_H,\phi_H)=(\rho_{A}^{i},\phi_{A}^{i})$, which are favored by $B\to PP$ and $PV$ decays~\cite{Chang:2014yma,Sun:2014tfa}. So, the current data of ${\cal B}(\bar{B}^0\to\rho^0\rho^0)$ presents a challenge to the large $\rho_H$ and/or the simplification $(\rho_H,\phi_H)=(\rho_{A}^{i},\phi_{A}^{i})$. In addition, recalling the situation in $B\to \pi\pi$ decays, a large HSS correction with $\rho_H\sim 3$ plays an important role for resolving the ``$\pi\pi$ puzzle" and is allowed by the other $B\to PP$ and $PV$ decays~\cite{Chang:2014rla}.  So, any hypothesis for resolving the ``$\pi\pi$ puzzle" through enhancing the HSS corrections should be carefully tested whether it is also allowed by  $\bar{B}^0\to\rho^0\rho^0$ decay. 

It should be noted that the $B\to VV$ decays are relevant to not only the longitudinal  building blocks but also the transverse ones, the latter of which do not contribute to $B\to PP$ and $PV$ decays. In cases I and II, the analyses are based on the findings in $B\to PP$ and $PV$ decays and the ansatz that end-point parameters are universal in longitudinal and transverse building blocks, even though the latter is not essential. In the following cases, we will pay attention to such issue. 

\begin{figure}[t]
\begin{center}
 \subfigure[]{\includegraphics[width=8cm]{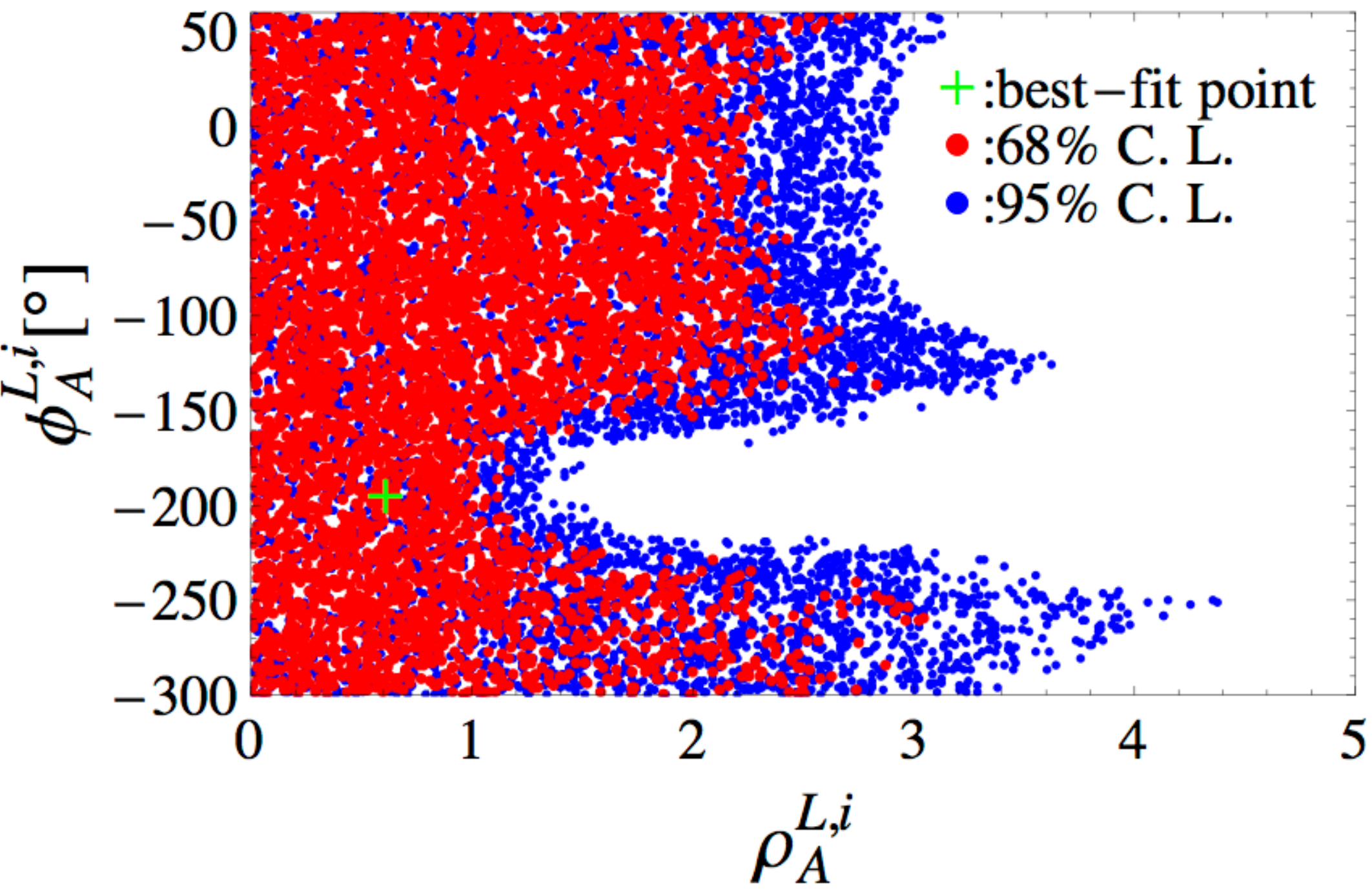}}
 \subfigure[]{\includegraphics[width=8cm]{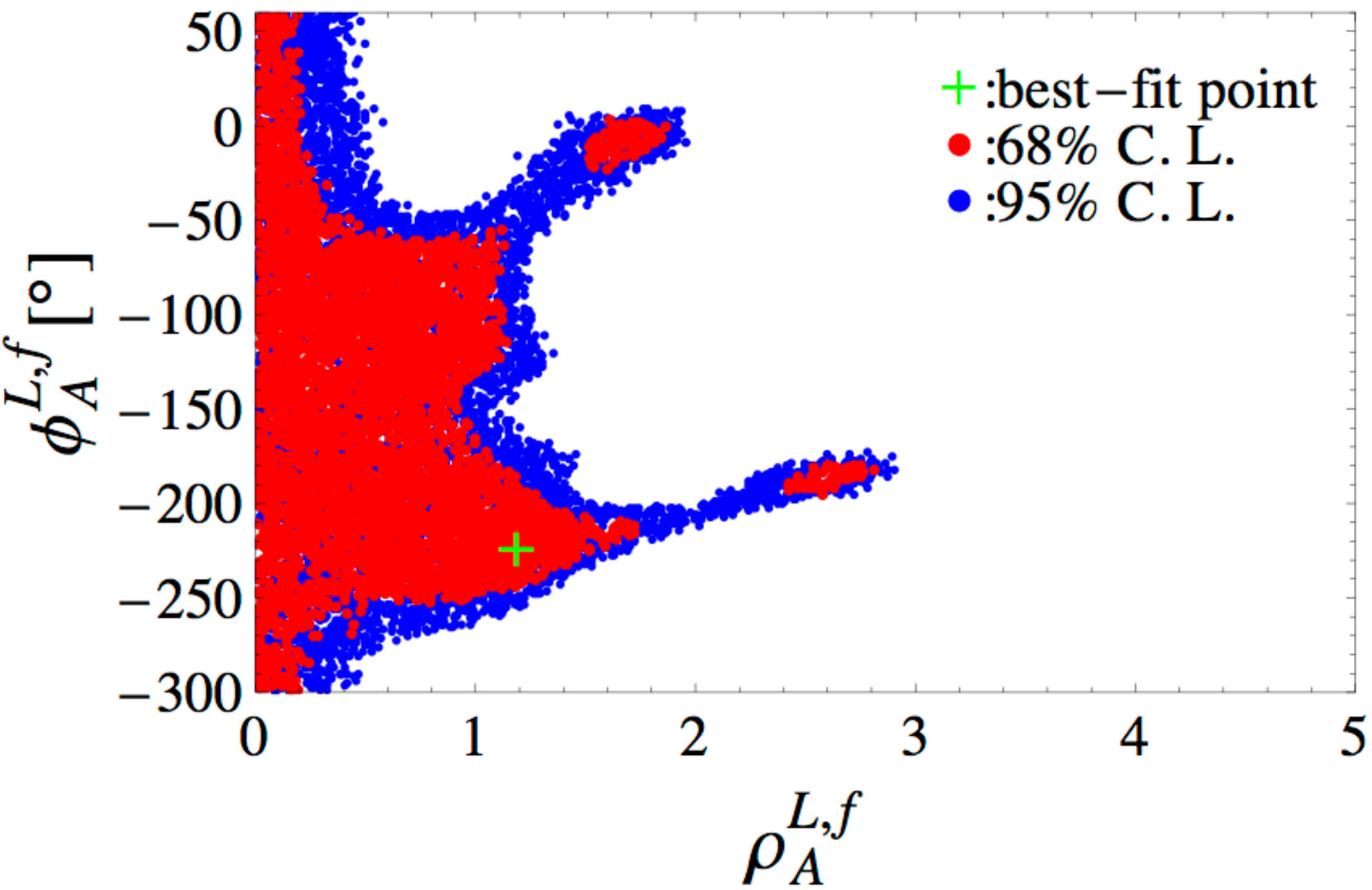}}\\
  \subfigure[]{\includegraphics[width=8cm]{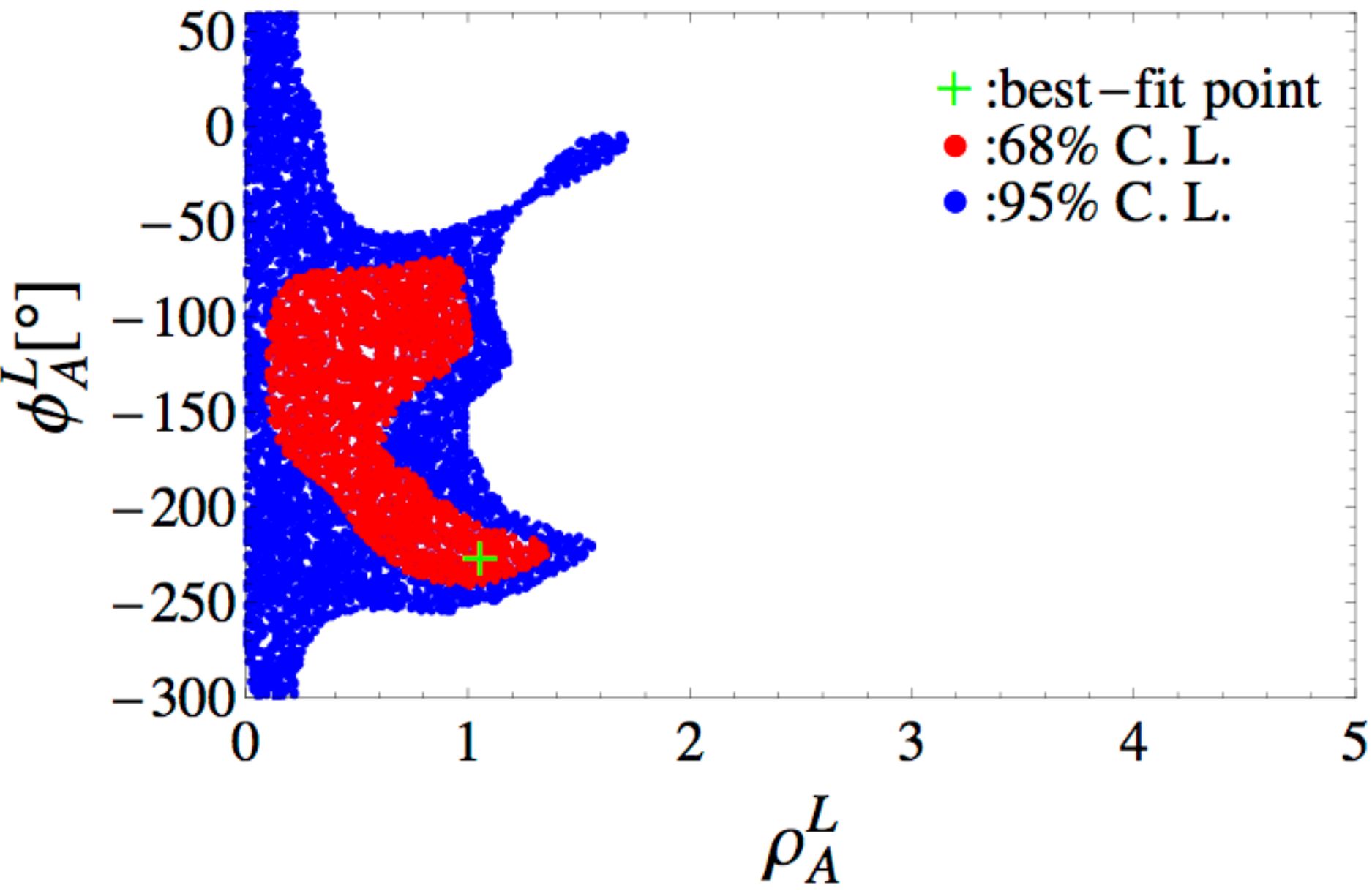}}
\caption{\label{case3} \small The allowed regions of the longitudinal end-point parameters with the constraints from longitudinal-polarization-dominated decay modes. For Fig.~(c), the simplification $(\rho_A^{L,i},\phi_A^{L,i})=(\rho_A^{L,f},\phi_A^{L,f})\equiv(\rho_A^{L},\phi_A^{L})$ is taken. One may also see Fig.~\ref{case32} plotted in the complex plane.}
\end{center}
\end{figure}

\subsection{Case III}
For case III, in order to extract the end-point contributions in the longitudinal building blocks, we pick out the measured longitudinal-polarization-dominated decay modes as constraint conditions, which include $B^-\to \rho^- \bar{K}^{*0}$, $K^{*0}K^{*-}$, $\rho^0 \rho^-$ and $\bar{B}^0\to K^{*0}\bar{K}^{*0}$, $\rho^+\rho^-$ decays. Taking $\rho_A^{T}=0$, the allowed spaces of longitudinal end-point parameters are shown by Fig.~\ref{case3}, in which Figs.~(a,b) and Fig.~(c) correspond to the cases without and with the simplification $(\rho_A^{L,i},\phi_A^{L,i})=(\rho_A^{L,f},\phi_A^{L,f})\equiv(\rho_A^{L},\phi_A^{L})$, respectively. 

From Figs.~\ref{case3}~(a) and (b), one may find that: (i) the large $\rho_A^{L,i}$ is excluded, which is mainly caused by the constraints from $B\to \rho\rho$ decays; (ii) Even though the fit of longitudinal end-point parameters through longitudinal-polarization-dominated decays is an ideal strategy, there is no well-bounded space could be found due to the lack of data and the large theoretical uncertainties, which prevent us to test whether $(\rho_A^{L},\phi_A^{L})$ are topology-dependent. Moreover, as Fig.~\ref{case3}~(c) shows, even if the simplification $(\rho_A^{L,i},\phi_A^{L,i})=(\rho_A^{L,f},\phi_A^{L,f})$ is taken, the spaces of end-point parameters are still hardly to be well restricted. So, the refined experimental measurements are required for a definite conclusion.  

\begin{figure}[t]
\begin{center}
 \subfigure[]{\includegraphics[width=8cm]{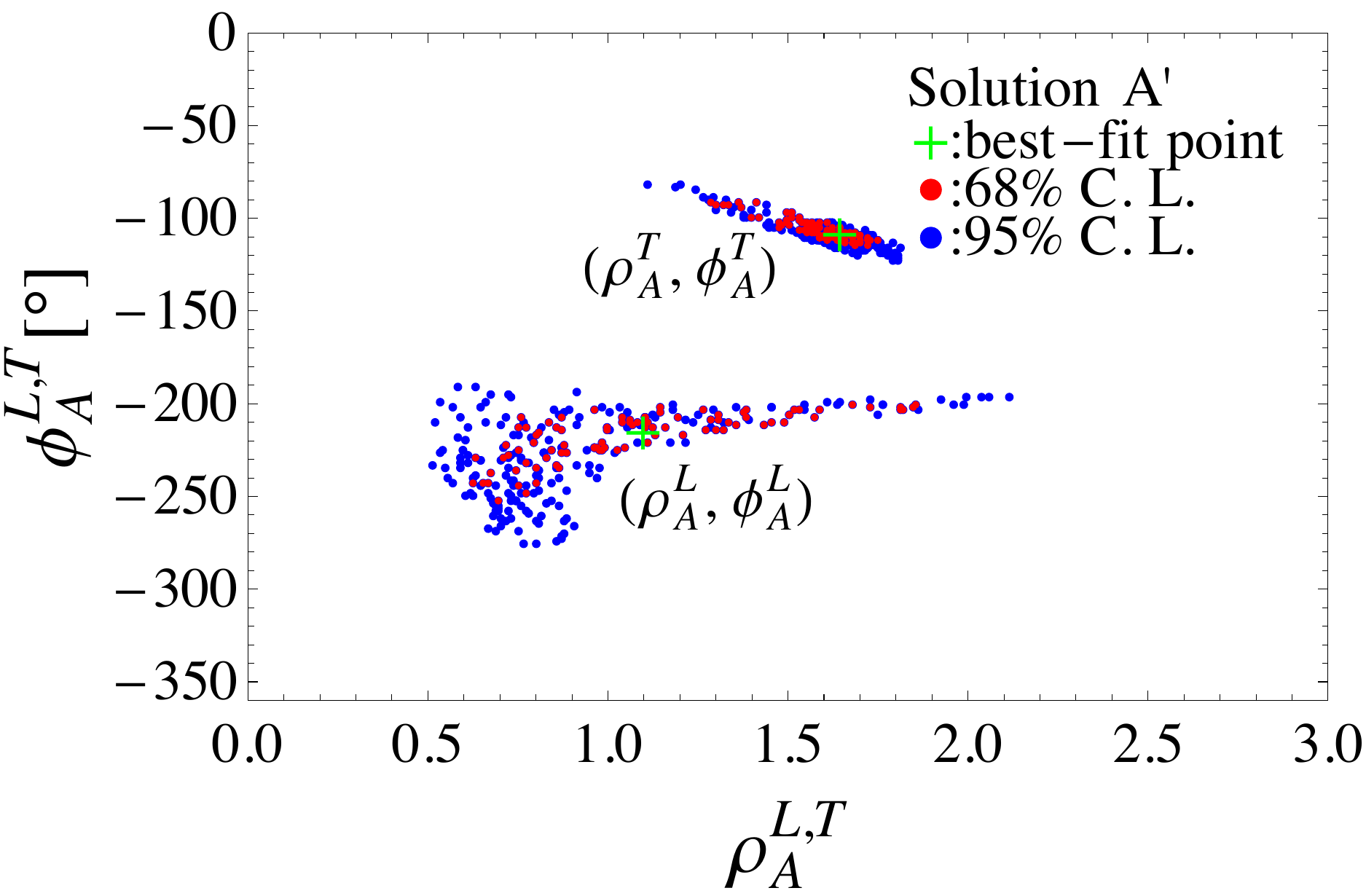}}\quad
 \subfigure[]{\includegraphics[width=8cm]{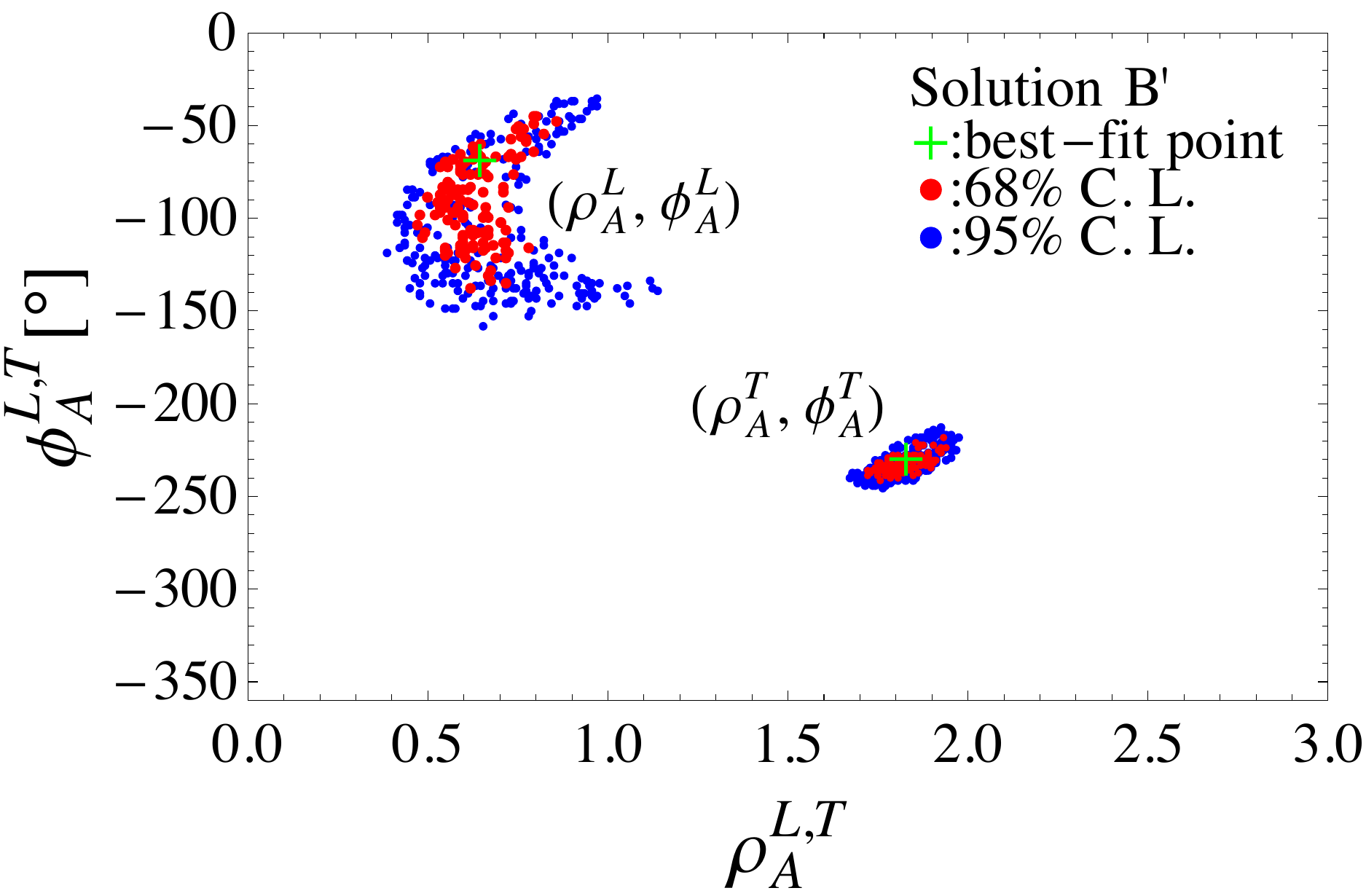}}\\
  \subfigure[]{\includegraphics[width=8cm]{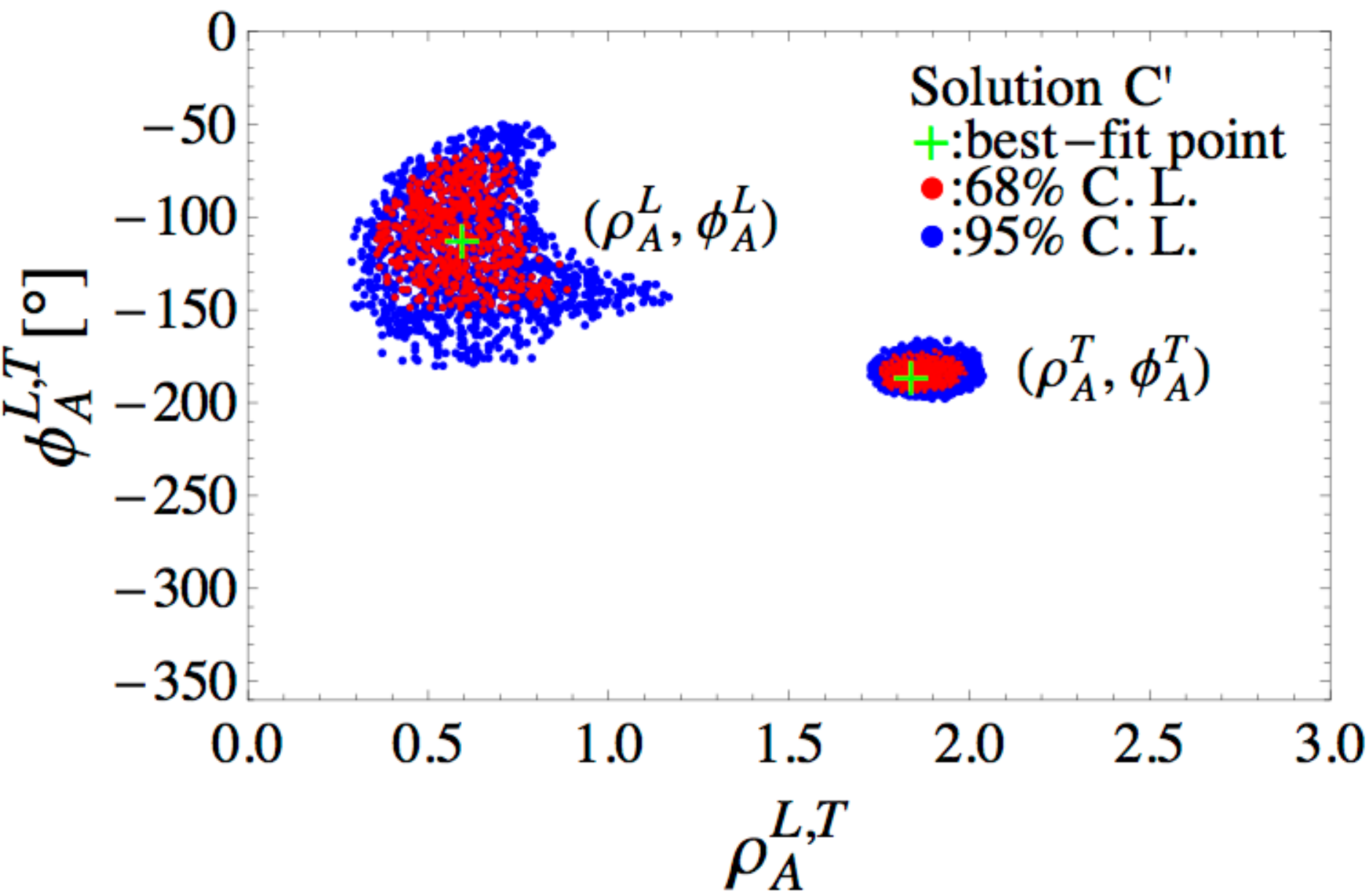}}\quad
    \subfigure[]{\includegraphics[width=8cm]{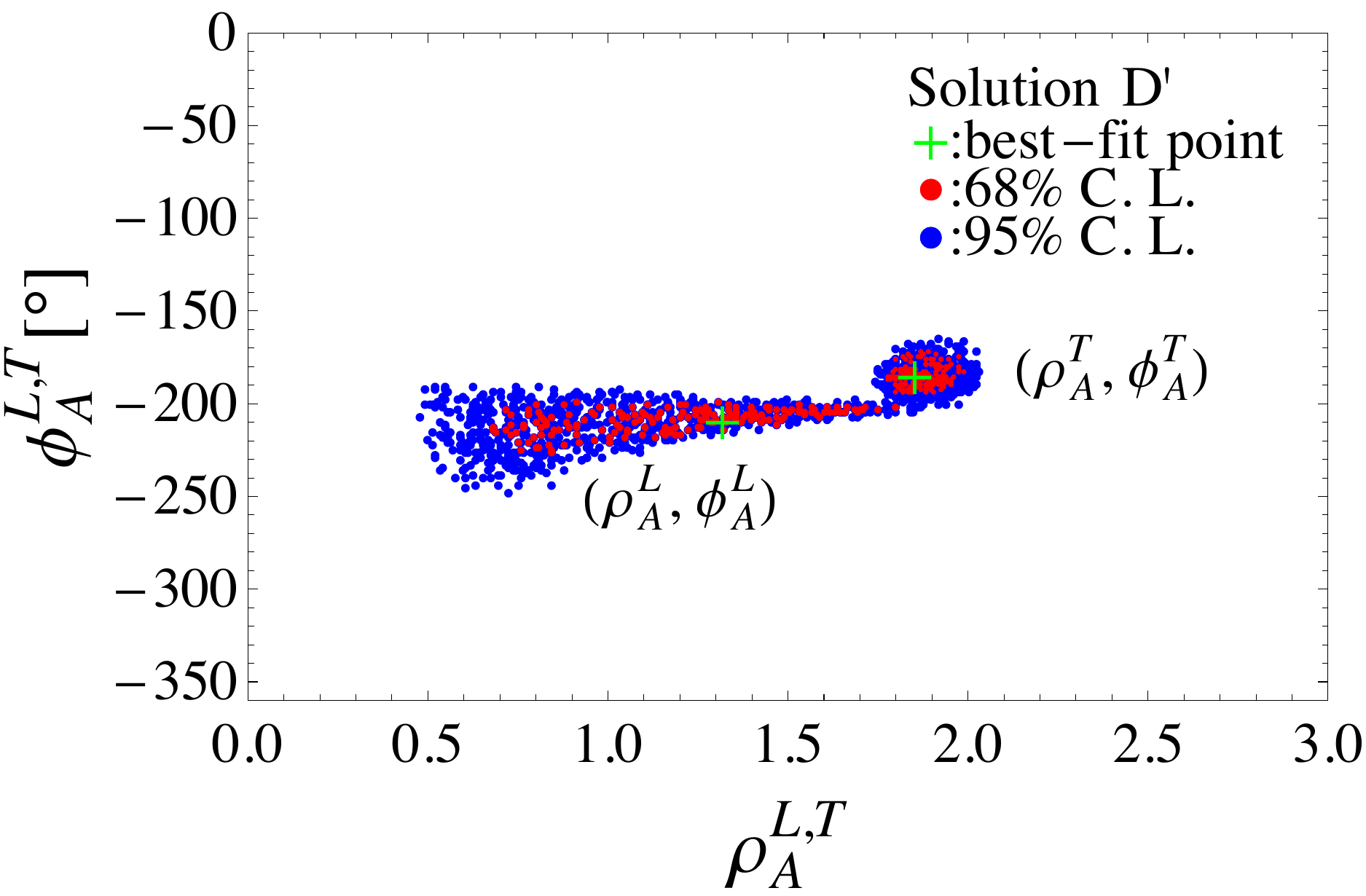}}
\caption{\label{case41} \small The allowed regions of the longitudinal and transverse end-point parameters with the constraints from $B_{u,d}\to \rho K^*, K^*\bar{K}^*$ and $\phi K^*$ decays. The best-fit points of solutions $\rm A'$-$\rm D'$ correspond to $\chi^2_{\rm min}=10.1$, $10.0$, $11.5$ and $12.1$, respectively. One may also see Fig.~\ref{case412} plotted in the complex plane.}
\end{center}
\end{figure}

\subsection{Case IV}
 For case IV,  we assume that the end-point parameters are topology-independent but  non-universal for  longitudinal and transverse building blocks~(a polarization-dependent scheme). The free parameters are $(\rho_A^{L},\phi_A^{L})$ and $(\rho_A^{T},\phi_A^{T})$. 

With the same constraint condition as case II, where 32 observables relevant to the penguin-dominated $B\to \rho K^*, K^*\bar{K}^*,\phi K^*$ decays are considered, the allowed spaces of end-point parameters $(\rho_A^{L,T},\phi_A^{L,T})$ are strictly restricted. Explicitly, the allowed spaces consist of 4 separate parts, named solutions $\rm A'$, $\rm B'$, $\rm C'$ and $\rm D'$, shown by Fig.~\ref{case41}. It could be found that: (i) the values of $\rho_A^T$ are a little larger than the ones of $\rho_A^L$ due to the requirement of large transverse polarization fractions of $B\to \phi K^*$ decays; (ii) There is no significant overlap between the allowed regions of $(\rho_A^{L},\phi_A^{L})$ and $(\rho_A^{T},\phi_A^{T})$ at 95\% C. L.,  which implies that $(\rho_A^{L},\phi_A^{L})\neq (\rho_A^{T},\phi_A^{T})$; (iii) Numerically, the best-fit values of four solutions are
\begin{equation}
\label{scaseIV}
 (\rho_A,\phi_A[^\circ])^{L,\,T}=
\left\{ \begin{array}{l}
(1.10,-213),~(1.65,-106)\,,\quad \text{solution $\rm A'$} \\
(0.64,-66),~(1.83,-227)\,,\quad \text{solution $\rm B'$}\\
(0.60,-110),~(1.84,-184)\,,\quad \text{solution $\rm C'$}\\
(1.32,-208),~(1.86,-183)\,,\quad \text{solution $\rm D'$}
\end{array} \right.
\end{equation}
which are much smaller than the end-point parameters of cases I and II, and thus theoretically more acceptable due to the power counting rules.  Moreover, in the view of minimum $\chi^2$, the solution $\rm A'$ in this case is much more favored by the data than the results of case II because $\chi^2_{\rm min}[{\rm solution~A'}]<\chi^2_{\rm min}[{\rm case~II}]$.  Such findings imply that the end-point contributions are possibly not only topology-independent but also polarization-dependent.  

\begin{figure}[t]
\begin{center}
 \subfigure[]{\includegraphics[width=5.5cm]{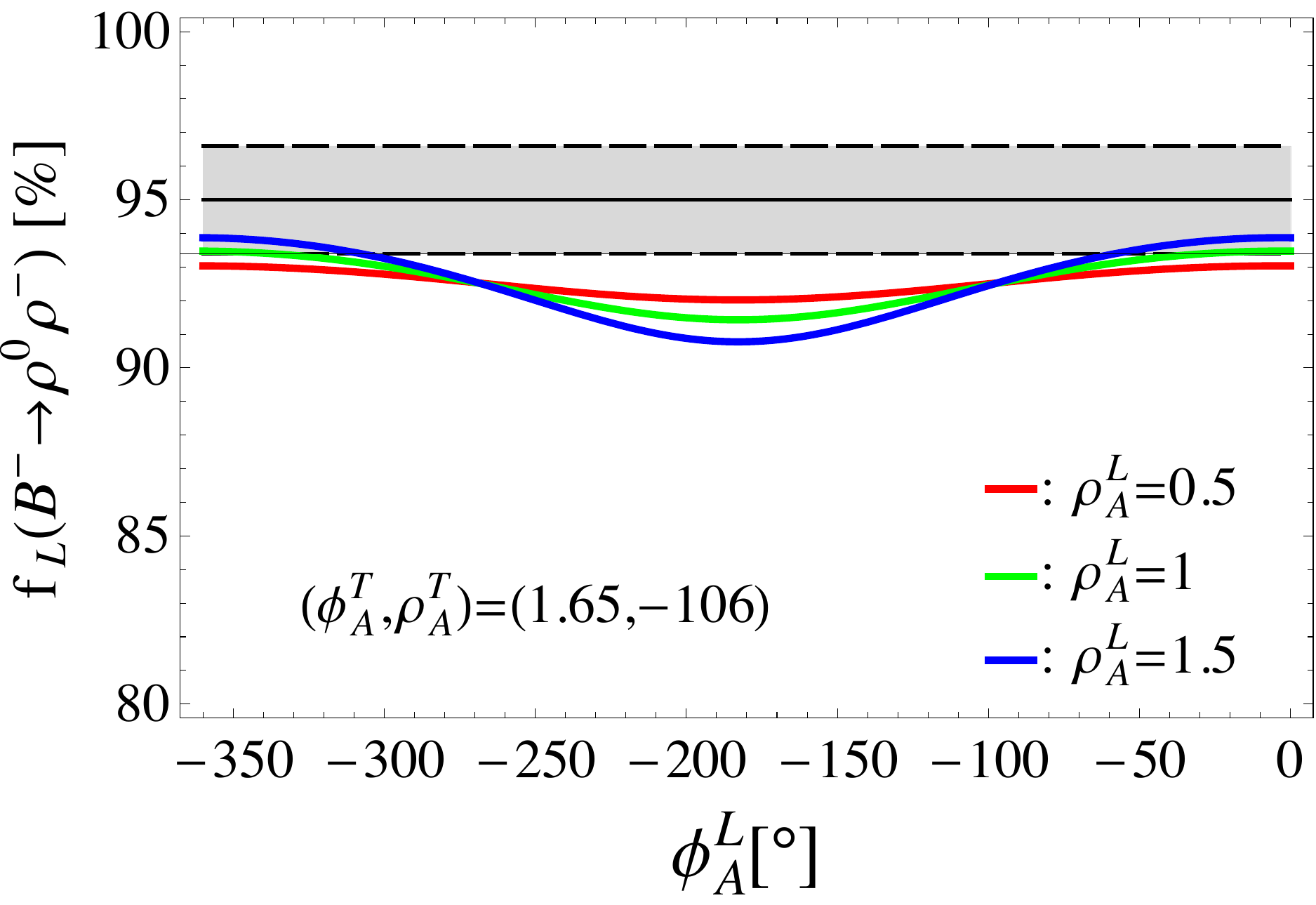}}
 \subfigure[]{\includegraphics[width=5.5cm]{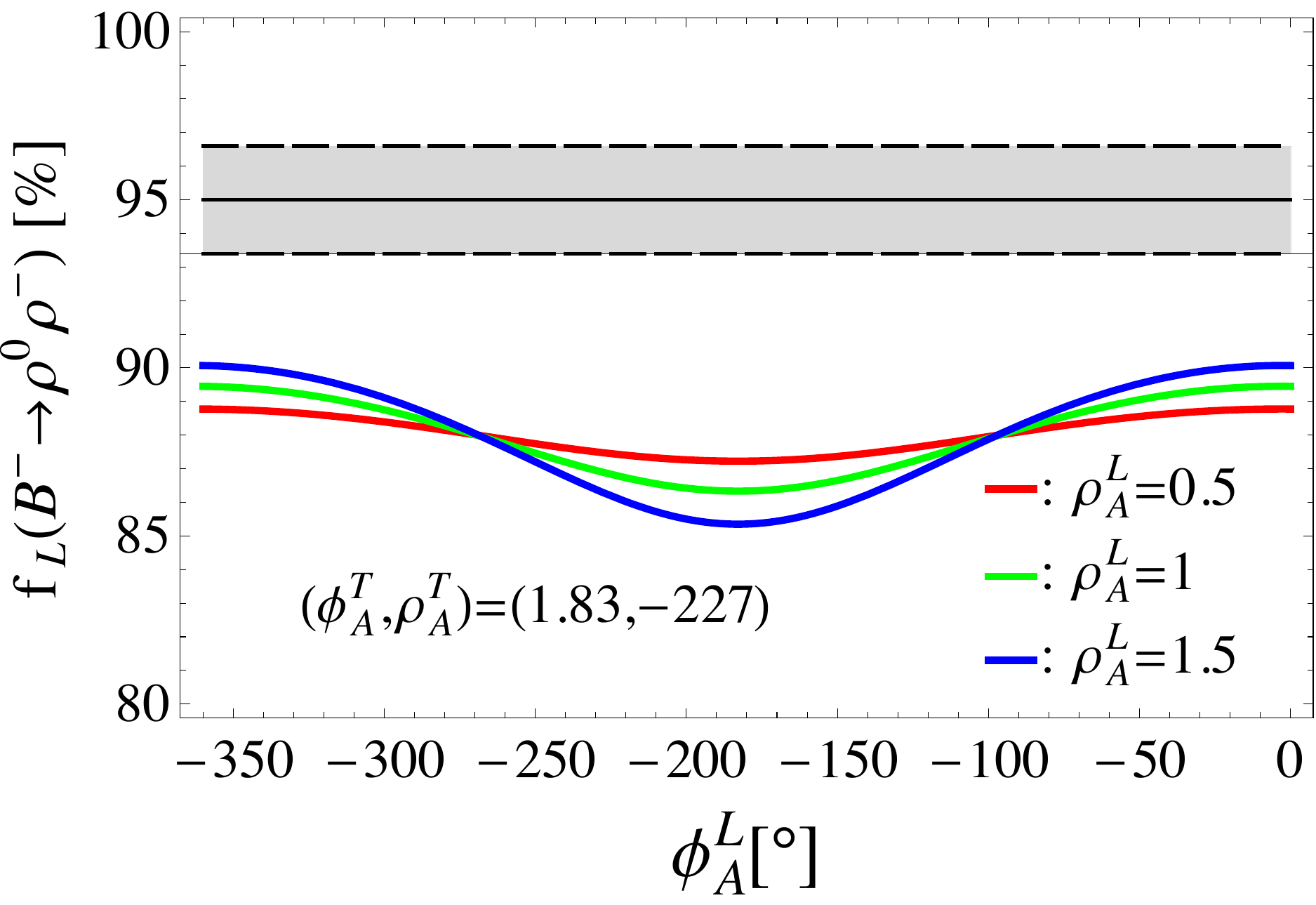}}
  \subfigure[]{\includegraphics[width=5.5cm]{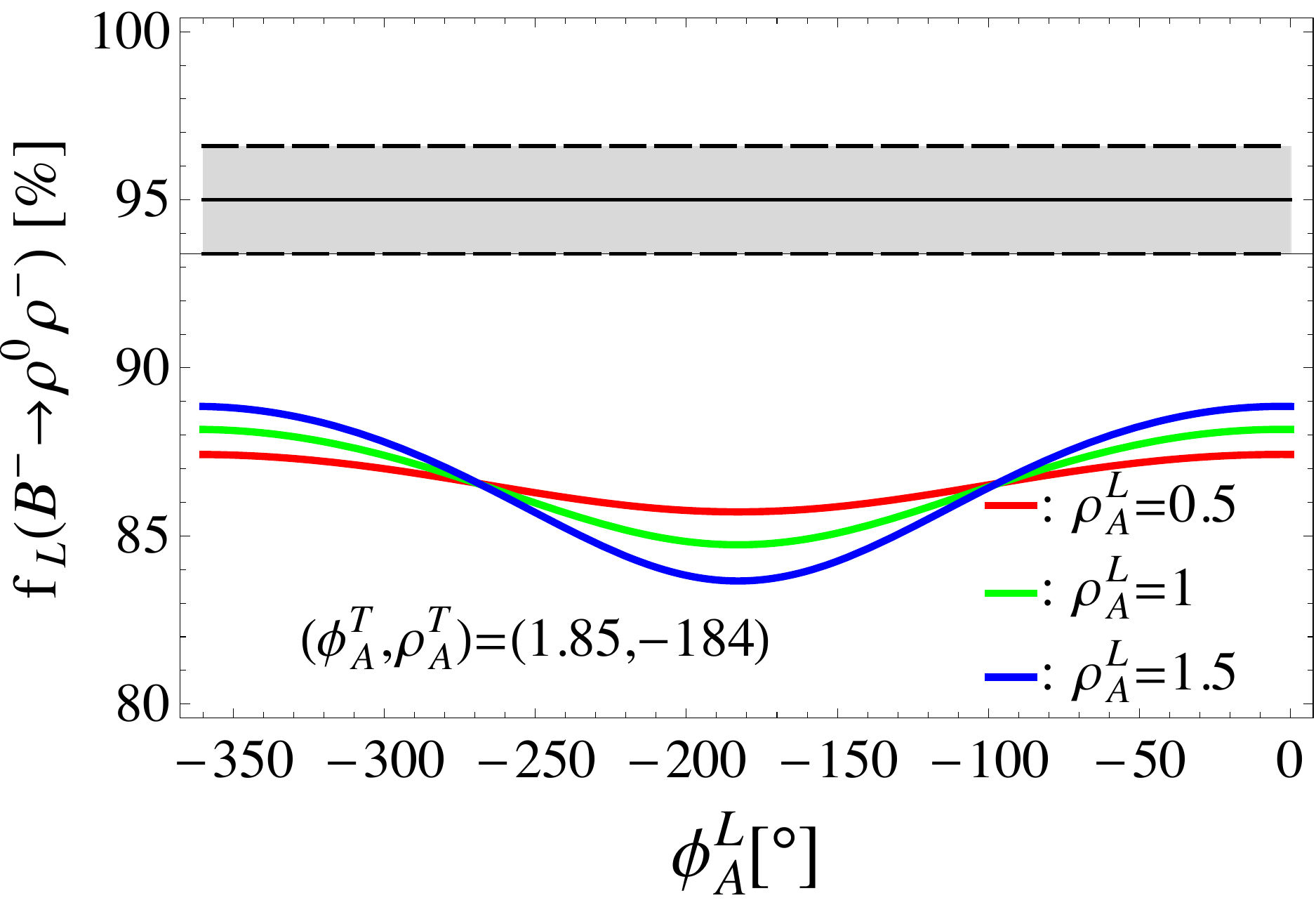}}\\
 \subfigure[]{\includegraphics[width=5.5cm]{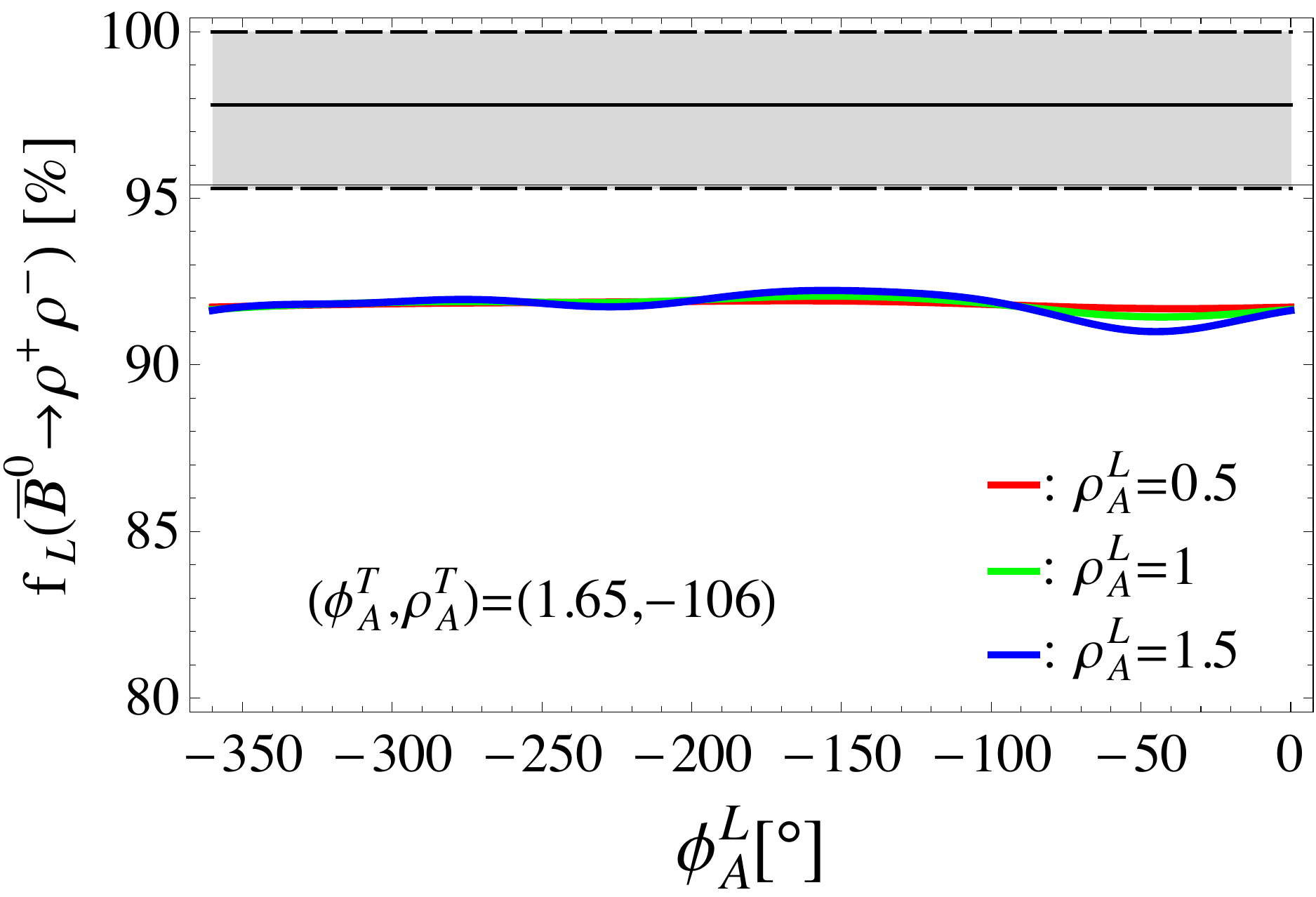}}
 \subfigure[]{\includegraphics[width=5.5cm]{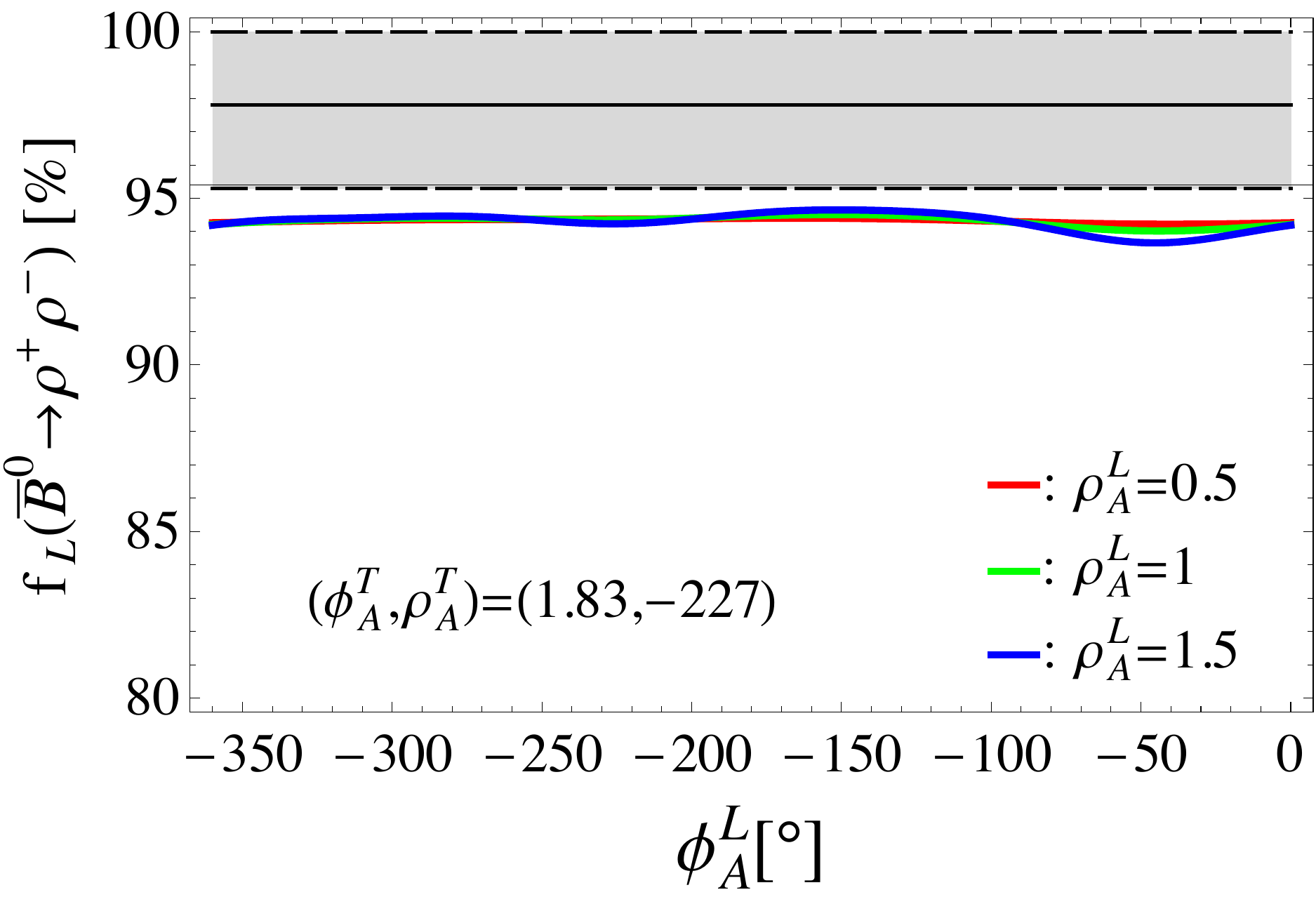}}
  \subfigure[]{\includegraphics[width=5.5cm]{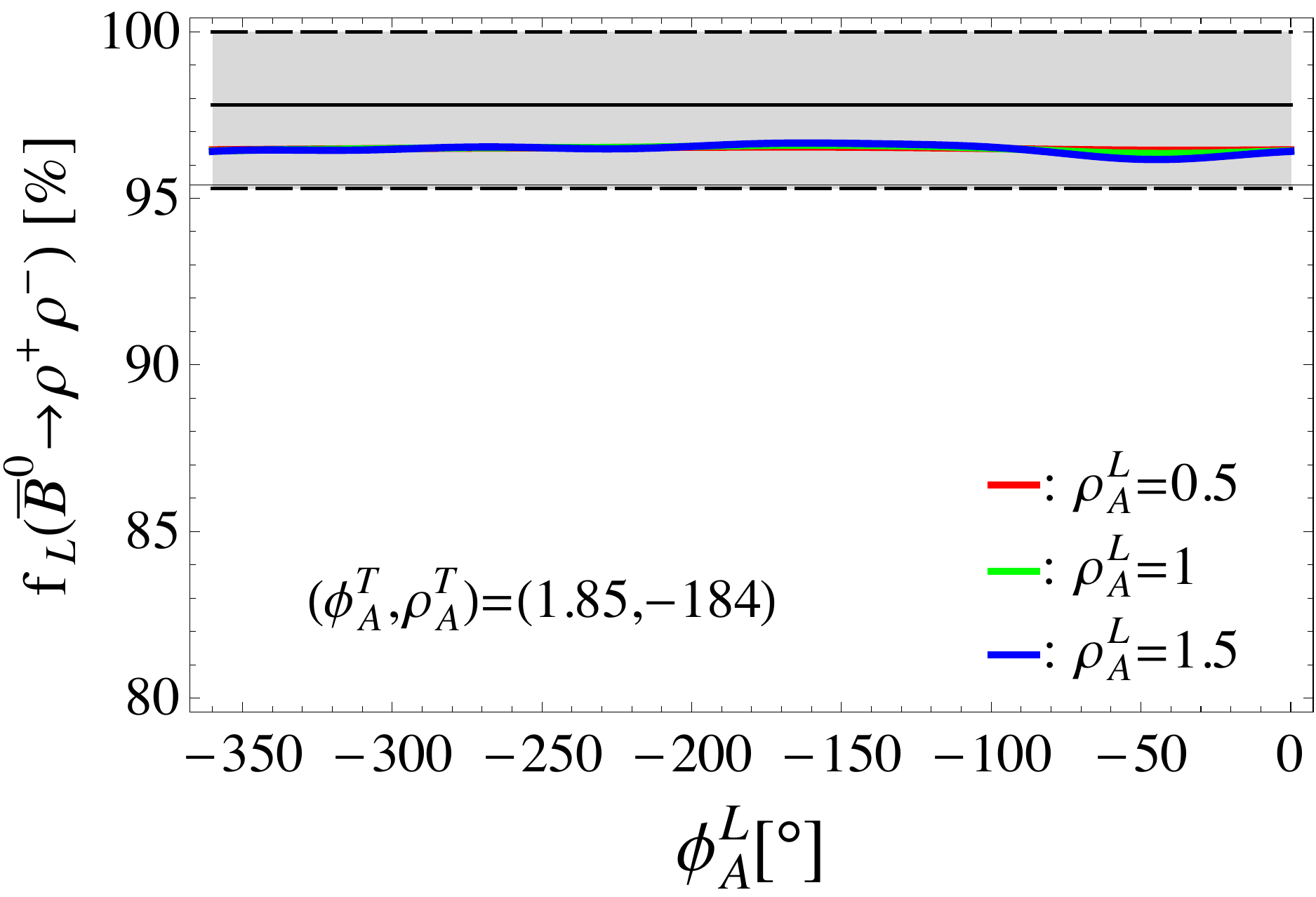}}
 \caption{\label{depc4} \small The dependences of $f_{L}(B\to \rho\rho)$ on the parameter $\phi_A^L$ with different values of $\rho_A^L$ and settled $(\rho_{A}^{T},\phi_{A}^{T})$ of solutions $\rm A'$-$\rm D'$ in Eq.~(\ref{scaseIV}), which are labeled in each figure. The shaded bands are experimental results with $1\sigma$ error. }
\end{center}
\end{figure}

Then, we would like to test such solutions further in $B\to \rho\rho$ decays. With the well settled values of $(\rho_{A}^{T},\phi_{A}^{T})$ in Eq.~(\ref{scaseIV}), the dependences of $f_{L}(B\to \rho\rho)$  on the parameter $\phi_A^L$ with different $\rho_A^L$ are shown by Fig.~\ref{depc4}. For the $\bar{B}^0\to \rho^+\rho^-$ decay, because its amplitude is dominated by the tree coefficient $\alpha_1$, the effects of HSS and WA corrections are not significant as Figs.~\ref{depc4}~(d-f) show. For the $B^-\to \rho^0\rho^-$ decay, its amplitude is irrelevant to the WA contribution but sensitive to the HSS correction through $\alpha_2$. From Figs.~\ref{depc4}~(a-c), it could be found that the solutions $\rm B'$, $\rm C'$ and $\rm D'$ are possible to be excluded by $f_{L}(B^-\to \rho^0\rho^-)$ and the solution $\rm A'$ is unwillingly acceptable. 

\begin{figure}[t]
\begin{center}
\includegraphics[width=8cm]{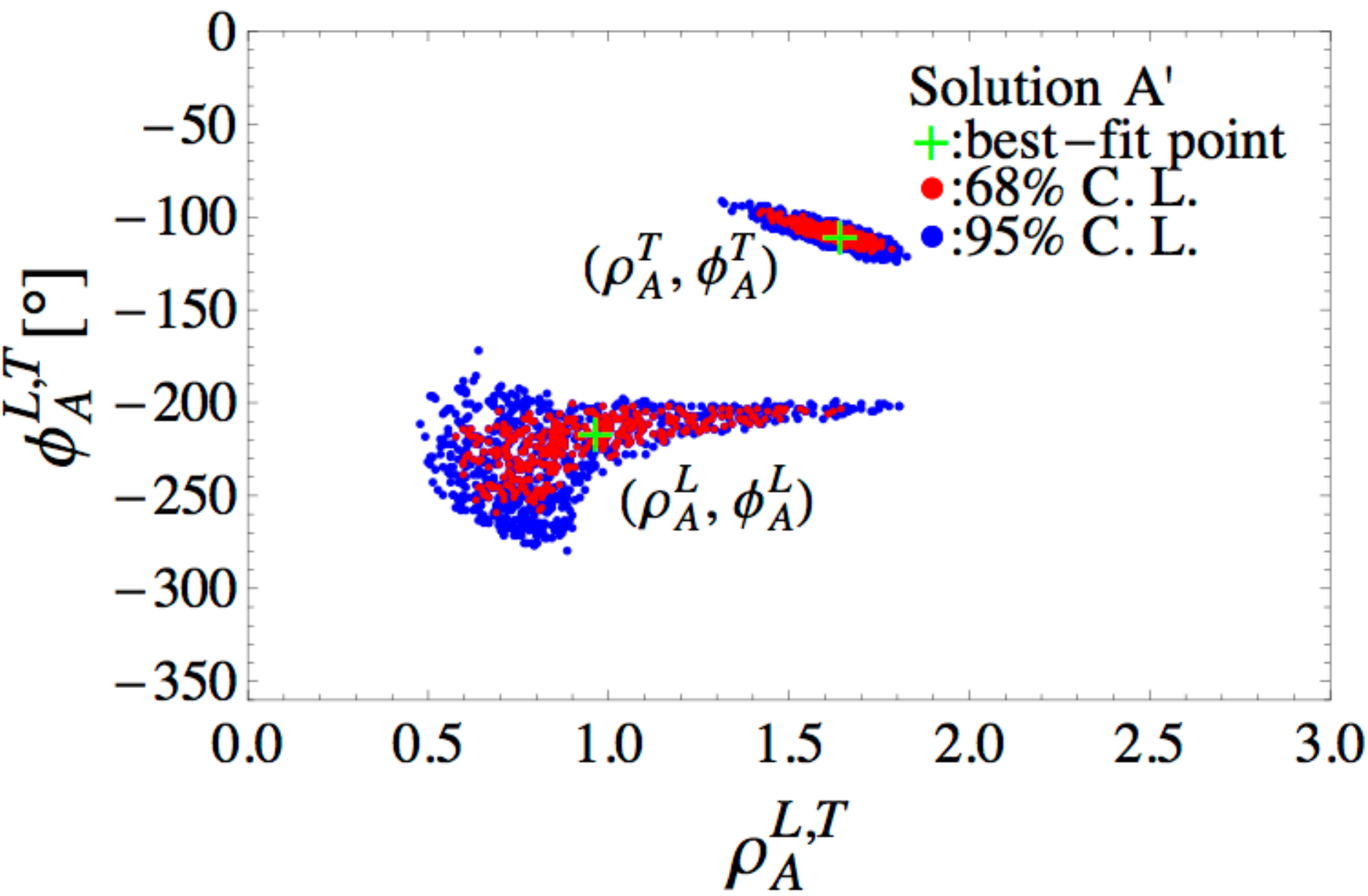}
\caption{\label{case42} \small The allowed regions of the longitudinal and transverse end-point parameters with the combined constraints from all of the decay modes considered in this paper. The best-fit point corresponds to $\chi^2_{\rm min}=16.8$. One may also see Fig.~\ref{case422} plotted in the complex plane.}
\end{center}
\end{figure}

Finally, combining the constraints from all of the decay modes considered in this paper, we present the allowed space of  longitudinal and transverse end-point parameters in Fig.~\ref{case42}. As analyses above, the solutions $\rm B'$, $\rm C'$ and $\rm D'$ gotten through $B_{u,d}\to \rho K^*, K^*\bar{K}^*$ and $\phi K^*$ decays are ruled out entirely by $B\to \rho\rho$ decays, and the parameter spaces of solution $\rm A'$ are further restricted. Numerically, we get
 \begin{equation}
\label{sol42}
 (\rho_A,\phi_A[^\circ])^{L,\,T}=
(0.97_{-0.39}^{+0.67},-214_{-44}^{+14}),~(1.65_{-0.23}^{+0.13},-108_{-10}^{+13})\,. \quad \text{solution $\rm A'$}
\end{equation}

Using the values of $(\rho_A^{L,\,T},\phi_A^{L,\,T})$ in Eq.~\eqref{sol42}, we present the theoretical results for $B\to VV$ decays in the ``case IV" columns of Tables \ref{tab:rhok}, \ref{tab:kk}, \ref{tab:phik}, \ref{tab:rhorho} and \ref{tab:phiphi}. All of the theoretical results are in consistence with the data within the errors and uncertainties. Comparing case II with case IV, one can find some significant differences, especially for the pure annihilation $\bar{B}^0\to \phi \phi$ decay. For instance, case II presents a large branching fraction  $\sim 13\times 10^{-8}$, which is in the scope of SuperKEKb/Belle-II experiment, while the prediction of case IV, $\sim 0.1\times 10^{-8}$, is very small. Moreover, case IV presents a much larger transverse polarization fraction, $\sim 31\%$, than case II. In addition, we note that the amplitude ${\cal{A}}_{\bar B^{0}\to \phi\phi}^{h}$ is only relevant to the effective coefficients $b_{4}$ and $b_{4,EW}$, and both of which involve $A_{1,2}^{i}$ only. It implies that $\bar{B}^0\to \phi \phi$ decay is very suitable for probing the non-factorizable WA contributions.  The measurement of pure annihilation decays is required for testing such results and exploring a much clearer picture of WA contributions.  
 
 \section{Conclusion}
 In summary, we have studied the effects of weak annihilation and hard spectator scattering contributions in $B_{u,d}\to VV $ decays with the QCDF approach. In order to evaluate the values of end-point parameters, comprehensive statistical $\chi^2$ analyses are preformed in four cases. Our analyses in cases I and II are based on the topology-dependent parameterization scheme, which is presented first in Ref.~\cite{Zhu:2011mm} and favored by $B\to PP\,,PV$ decays~\cite{Chang:2014yma,Sun:2014tfa}. The analyses in cases~III and IV are based on the polarization-dependent parameterization scheme ({\it i.e.}, the end-point parameters are non-universal for  longitudinal and transverse building blocks). In each of cases, a global fit of end-point parameters is performed with the data available, and the numerical results are presented. Our main conclusions and findings could be summarized as the following:
\begin{itemize}
\item The  allowed spaces of $ (\rho_A^{i},\phi_A^{i})$ at 95\% C. L. are entirely different from that of $(\rho_A^{f},\phi_A^{f})$ in $B\to VV$ decays, {\it i.e.}, $ (\rho_A^{i},\phi_A^{i})\neq (\rho_A^{f},\phi_A^{f})$, which confirms the proposal of topology-dependent scheme presented in Ref.~\cite{Zhu:2011mm}. More interestingly, the fitted result of $(\rho_A^{f},\phi_A^{f})$ in $B\to VV$ decays is very similar to the ones in $B\to PP$ and $PV$ decays, which implies possible universal end-point contributions for the factorizable annihilation topologies.  
\item The findings mentioned above are gotten mainly through penguin-dominated decays. Unfortunately, some tensions  between theoretical results and data appear when the color-suppressed tree-dominated $\bar{B}^0\to \rho^0 \rho^0$ decay is taken into account. To be exact,  a large $\rho_A^i$ and/or the simplification $ (\rho_H,\phi_H)\neq (\rho_A^{i},\phi_A^{i})$, which has been proven to be a good simplification by a global fit in $B\to PP\,,PV$ decays~\cite{Chang:2014yma,Sun:2014tfa}, are challenged especially by ${\cal B}(\bar{B}^0\to \rho^0 \rho^0)$. We further point out that any hypothesis for resolving the ``$\pi\pi$ puzzle'' through modifying HSS corrections should be carefully tested in $B\to \rho\rho$ decays.
\item For the polarization-dependent scheme, an ideal strategy is to extract the longitudinal end-point parameters through the longitudinal-polarization-dominated decay modes and further analysis their topology-dependence. However, the lack of data and large uncertainties prevent us from obtaining an exact result. Combining all of the decays considered in this paper, the fitted result at 95\% C. L. indicates that $(\rho_A^{L},\phi_A^{L})\neq (\rho_A^{T},\phi_A^{T})$. Using the fitted values of end-point parameters, the experimental data could be accommodated within QCDF Framework. 
\end{itemize}

Generally, because $B\to VV$ decays involve more observables than $B\to PP$ and $PV$ decays, more information for the WA and HSS contributions can be obtained, which surely helps us to further explore and understand the underlying mechanism. However, the measurements of $B\to VV$ decays are still very rough by now, especially for the complete angular analysis and the pure annihilation decays. With the rapid accumulation of data on B events at running LHC and forthcoming SuperKEKb/Belle-II, more refined measurements of $B\to VV$ decays are urgently expected for a much clearer picture  of  WA and HSS contributions. 

 \section*{Acknowledgments}
  The work is supported by the National Natural
  Science Foundation of China (Grant Nos. 11475055,
  11275057, U1232101 and U1332103).
  Q. Chang is also supported by the Foundation
  for the Author of National Excellent Doctoral
  Dissertation of P. R. China (Grant No. 201317),
  the Program for Science and Technology Innovation
  Talents in Universities of Henan Province
  (Grant No. 14HASTIT036) and Foundation for University Key Teacher of Henan Province (Grant No. 2013GGJS-58). 
  
\begin{appendix}
\section*{Appendix A: The decay amplitudes }
\begin{eqnarray}
\label{eq:rhok1}
{\cal{A}}_{B^{-}\to \rho^{-}\bar K^{*0}}^{h}&=&A_{\rho\bar K^{*}}^{h}
[\delta_{pu}\beta_{2}^{p,h}+\alpha_{4}^{p,h}-\half\alpha_{4,EW}^{p,h}+\beta_{3}^{p,h}+\beta_{3,EW}^{p,h}],\\
\label{eq:rhok2}
\sqrt{2}{\cal{A}}_{B^{-}\to \rho^{0}K^{*-}}^{h}&=&A_{\rho\bar K^{*}}^{h}
[\delta_{pu}(\alpha_{1}^{p,h}+\beta_{2}^{p,h})+\alpha_{4}^{p,h}+\alpha_{4,EW}^{p,h}+\beta_{3}^{p,h}+\beta_{3,EW}^{p,h}]\nonumber\\
&&+A_{\bar K^{*}\rho}^{h}[\delta_{pu}\alpha_{2}^{p,h}+\frac{3}{2}\alpha_{3,EW}^{p,h}],\\
\label{eq:rhok3}
{\cal{A}}_{\bar B^{0}\to \rho^{+} K^{*-}}^{h}&=&A_{\rho\bar K^{*}}^{h}
[\delta_{pu}\alpha_{1}^{p,h}+\alpha_{4}^{p,h}+\alpha_{4,EW}^{p,h}+\beta_{3}^{p,h}-\half\beta_{3,EW}^{p,h}],\\
\label{eq:rhok4}
\sqrt{2}{\cal{A}}_{\bar B^{0}\to \rho^{0}\bar K^{*0}}^{h}&=&A_{\rho\bar K^{*}}^{h}
[-\alpha_{4}^{p,h}+\half\alpha_{4,EW}^{p,h}-\beta_{3}^{p,h}+\half\beta_{3,EW}^{p,h}]
\nonumber\\&&+A_{\bar K^{*}\rho}^{h}[\delta_{pu}\alpha_{2}^{p,h}+\frac{3}{2}\alpha_{3,EW}^{p,h}],\\
%%%%%%%%%%%%
\label{eq:kk1}
{\cal{A}}_{B^{-}\to K^{*-}K^{*0}}^{h}&=&A_{\bar K^{*}K^{*}}^{h}
[\delta_{pu}\beta_{2}^{p,h}+\alpha_{4}^{p,h}-\half\alpha_{4,EW}^{p,h}+\beta_{3}^{p,h}+\beta_{3,EW}^{p,h}],\\
\label{eq:kk2}
{\cal{A}}_{\bar B^{0}\to K^{*-}K^{*+}}^{h}&=&A_{\bar K^{*}K^{*}}^{h}
[\delta_{pu}\beta_{1}^{p,h}+\beta_{4}^{p,h}+\beta_{4,EW}^{p,h}]+B_{K^{*}\bar K^{*}}^{h}[b_{4}^{p,h}-\half b_{4,EW}^{p,h}],\\
\label{eq:kk3}
{\cal{A}}_{\bar B^{0}\to \bar K^{*0}K^{*0}}^{h}&=&A_{\bar K^{*}K^{*}}^{h}
[\alpha_{4}^{p,h}-\half\alpha_{4,EW}^{p,h}+\beta_{3}^{p,h}+\beta_{4}^{p,h}-\half\beta_{3,EW}^{p,h}-\half\beta_{4,EW}^{p,h}]
\nonumber\\&&+B_{K^{*}\bar K^{*}}^{h}[b_{4}^{p,h}-\half b_{4,EW}^{p,h}],\\
%%%%%%%%%%%%
{\cal{A}}_{B^{-}\to K^{*-}\phi}^{h}&=&A_{\bar K^{*}\phi}^{h}
[\delta_{pu}\beta_{2}^{p,h}+\alpha_{3}^{p,h}+\alpha_{4}^{p,h}-\half\alpha_{3,EW}^{p,h}-\half\alpha_{4,EW}^{p,h}
+\beta_{3}^{p,h}+\beta_{3,EW}^{p,h}],\\
{\cal{A}}_{\bar B^{0}\to \bar K^{*0}\phi}^{h}&=&A_{\bar K^{*}\phi}^{h}
[\alpha_{3}^{p,h}+\alpha_{4}^{p,h}-\half\alpha_{3,EW}^{p,h}-\half\alpha_{4,EW}^{p,h}
+\beta_{3}^{p,h}-\half\beta_{3,EW}^{p,h}],\\
\sqrt{2}{\cal{A}}_{B^{-}\to \rho^{0}\rho^{-}}^{h}&=&A_{\rho{-}\rho{0}}^{h}
[\delta_{pu}(\alpha_{2}^{p,h}-\beta_{2}^{p,h})-\alpha_{4}^{p,h}+\frac{3}{2}\alpha_{3,EW}^{p,h}+\half\alpha_{4,EW}^{p,h}
-\beta_{3}^{p,h}-\beta_{3,EW}^{p,h}]
\nonumber\\&&+A_{\rho{0}\rho{-}}^{h}[\delta_{pu}(\alpha_{1}^{p,h}+\beta_{2}^{p,h})+\alpha_{4}^{p,h}+\alpha_{4,EW}^{p,h}
+\beta_{3}^{p,h}+\beta_{3,EW}^{p,h}],\\
{\cal{A}}_{\bar B^{0}\to \rho^{+}\rho^{-}}^{h}&=&A_{\rho\rho}^{h}
[\delta_{pu}(\alpha_{1}^{p,h}-\beta_{1}^{p,h})+\alpha_{4}^{p,h}+\alpha_{4,EW}^{p,h}\nonumber\\
&&+\beta_{3}^{p,h}+2\beta_{4}^{p,h}-\half\beta_{3,EW}^{p,h}+\half\beta_{4,EW}^{p,h}],\\
-{\cal{A}}_{\bar B^{0}\to \rho^{0}\rho^{0}}^{h}&=&A_{\rho\rho}^{h}
[\delta_{pu}(\alpha_{2}^{p,h}-\beta_{1}^{p,h})-\alpha_{4}^{p,h}+\frac{3}{2}\alpha_{3,EW}^{p,h}+\half\alpha_{4,EW}^{p,h}\nonumber\\
&&-\beta_{3}^{p,h}-2\beta_{4}^{p,h}+\half\beta_{3,EW}^{p,h}-\half\beta_{4,EW}^{p,h}],\\
{\cal{A}}_{\bar B^{0}\to \phi\phi}^{h}&=&B_{\phi\phi}^{h}
[b_{4}^{p,h}-\half b_{4,EW}^{p,h}].
\end{eqnarray}

\section*{Appendix B: The experimental data and theoretical results}
\begin{table}[!htbp]
\begin{center}
\caption{\small The observables of $B\to\rho K^*$ decays. For the theoretical results of case II and IV, the first, second and third theoretical errors are caused by the CKM parameters, the other inputs in Table.~\ref{ppvalue} and end-point parameters, respectively.  }
\label{tab:rhok}
\vspace{0.2cm}
\scriptsize \doublerulesep 0.10pt \tabcolsep 0.05in
\begin{tabular}{lcccccc}
\hline\hline
\multirow{2}{*}{Obs.} &\multirow{2}{*}{Decay modes}&\multirow{2}{*}{Exp.}&\multicolumn{2}{c}{This work} &\multicolumn{2}{c}{Previous works} \\
&&&case II&case IV&Cheng~\cite{Cheng:2009cn,Cheng:2008gxa} &Beneke~\cite{Beneke:2006hg}\\ \hline
${\cal B}[10^{-6}]$
&$B^-\to \rho^-\bar{K}^{*0}$ &$9.2\pm1.5$&$9.0_{-0.5-2.0-2.2}^{+0.4+2.8+3.0}$&$9.0_{-0.5-1.0-2.6}^{+0.4+1.1+4.6}$&$9.2^{+1.2+3.6}_{-1.1-5.4}$ &$5.9^{+0.3+6.9}_{-0.3-3.7}$\\
&$B^-\to\rho^{0}K^{*-}$
&$4.6\pm1.1$&$5.9_{-0.4-0.9-0.8}^{+0.3+1.4+1.3}$&$6.4_{-0.4-0.7-2.4}^{+0.3+0.7+1.3}$&$5.5^{+0.6+1.3}_{-0.5-2.5}$ &$4.5^{+1.5+3.0}_{-1.3-1.4}$\\
&$\bar{B}^0\to\rho^+K^{*-}$
&$10.3\pm2.6$&$9.8^{+0.4+2.3+3.0}_{-0.6-1.7-2.1}$&$8.5^{+0.4+1+4.2}_{-0.5-0.9-2.4}$&$8.9^{+1.1+4.8}_{-1.0-5.5}$ &$5.5^{+1.7+5.7}_{-1.5-2.9}$\\
&$\bar{B}^0\to\rho^0\bar{K}^{*0}$
&$3.9\pm0.8$&$5.5^{+0.3+0.7+1.6}_{-0.3-0.5-0.9}$&$3.2^{+0.1+0.4+2.0}_{-0.2-0.4-1.2}$&$4.6^{+0.6+3.5}_{-0.5-3.5}$ &$2.4^{+0.2+3.5}_{-0.1-2.0}$\\
\hline
$A_{CP}[\%]$
&$B^-\to \rho^-\bar{K}^{*0}$
&$-1\pm16$&$4_{-0-1-2}^{+0+1+3}$&$0.8^{+0+0.1+0.5}_{-0-0.1-0.3}$&$-0.3^{+0+2}_{-0-0}$ &$0^{+0+3}_{-0-1}$\\
&$B^-\to \rho^{0} K^{*-}$
&$31\pm13$&$39_{-1-4-21}^{+1+3+16}$&$14^{+0+1+15}_{-0-1-24}$&$43^{+6+12}_{-3-28}$ &$16^{+4+23}_{-4-16}$\\
&$\bar{B}^0\to\rho^+K^{*-}$
&$21\pm15$&$26^{+1+4+13}_{-1-5-18}$&$5.6^{+0.1+3.6+24}_{-0.1-3.3-13}$&$32^{+1+2}_{-3-14}$ &$5^{+1+40}_{-1-17}$\\
&$\bar{B}^0\to\rho^0\bar{K}^{*0}$
&$-6\pm9$&$-25^{+1+9+34}_{-1-8-31}$&$-20^{+1+6+15}_{-1-5-4}$&$-15^{+4+16}_{-8-14}$ &$-15^{+4+17}_{-4-32}$\\
\hline
$A_{CP}^0[\%]$
&$B^-\to \rho^-\bar{K}^{*0}$&---&$2.7^{+0.1+0.3+2.3}_{-0.1-0.3-1.8}$&$0.8^{+0+0.1+0.2}_{-0-0.1-0.4}$&---&$-1^{+0+1}_{-0-1}$\\
&$B^-\to \rho^{0}K^{*-}$&---&$5.1^{+0.3+16.6+23.7}_{-0.2-16.2-22.2}$&$3.0^{+0.3+1.2+9.1}_{-0.1-1.2-9.9}$&---&$7^{+2+12}_{-2-13}$\\
&$\bar{B}^0\to\rho^+K^{*-}$&---&$68.6^{+1.9+7.2+18.4}_{-2.1-7.6-25.4}$&$20.2^{+0.8+3.9+13.1}_{-0.5-4.3-21.7}$&---&$18^{+6+12}_{-5-29}$\\
&$\bar{B}^0\to\rho^0\bar{K}^{*0}$&---&$18.3^{+0.4+5.7+13.1}_{-0.6-7.3-10.4}$&$13.0^{+0.4+6.5+9.7}_{-0.4-7.8-5.3}$&---&$-30^{+11+60}_{-11-48}$\\
\hline
$A_{CP}^{\bot}[\%]$
&$B^-\to \rho^-\bar{K}^{*0}$&---&$-0.6^{+0+0.3+1.3}_{-0-0.5-1.5}$&$-1.2^{+0+0.1+0.5}_{-0-0.1-0.6}$&---&--- \\
&$B^-\to \rho^{0} K^{*-}$&---&$-6.6^{+0.2+30.0+45.7}_{-0.3-20.4-25.0}$&$-7.6^{+0.2+3.5+29.3}_{-0.4-4.4-20.2}$&---&--- \\
&$\bar{B}^0\to\rho^+K^{*-}$&---&$-64.3^{+1.5+9.3+16.4}_{-1.3-8.7-13.2}$&$-23.0^{+0.7+3.0+9.6}_{-0.6-2.4-24.0}$&---&--- \\
&$\bar{B}^0\to\rho^0\bar{K}^{*0}$&---&$-16.9^{+0.7+8.3+11.4}_{-0.4-7.5-12.2}$&$-6.8^{+0.3+4.5+5.7}_{-0.3-3.6-16.3}$&---&--- \\
\hline
$f_L[\%]$
&$B^-\to \rho^-\bar{K}^{*0}$
&$48\pm8$&$56_{-0-3-26}^{+0+3+20}$&$59^{+0+6+23}_{-0-6-17}$&$48^{+3+52}_{-4-40}$ &$56^{+4+48}_{-0-30}$\\
&$B^-\to \rho^{0} K^{*-}$
&$78\pm12$&$61_{-0-6-27}^{+0+6+19}$&$72^{+0+4+16}_{-1-4-15}$&$67^{+2+31}_{-3-48}$ &$84^{+2+16}_{-3-25}$\\
&$\bar{B}^0\to\rho^+K^{*-}$
&$38\pm13$&$48^{+0+1+12}_{-0-1-12}$&$53^{+1+5+21}_{-1-5-14}$&$53^{+2+45}_{-3-32}$ &$61^{+5+38}_{-7-28}$\\
&$\bar{B}^0\to\rho^0\bar{K}^{*0}$
&$40\pm14$&$52^{+0+2+7}_{-0-3-10}$&$35^{+0+7+30}_{-0-7-15}$&$39^{+0+60}_{-0-31}$ &$22^{+3+53}_{-3-14}$\\
\hline
$f_{\bot}[\%]$
&$B^-\to \rho^-\bar{K}^{*0}$&---&$20.5^{+0+2.0+11.7}_{-0-1.8-9.2}$&$20.9^{+0+3.0+8.6}_{-0-2.9-11.5}$&---&---\\
&$B^-\to \rho^{0} K^{*-}$&---&$17.9^{+0.1+3.0+12.2}_{-0.1-2.8-8.7}$&$14.4^{+0.5+2.1+7.5}_{-0.3-2.0-8.0}$&---&---\\
&$\bar{B}^0\to\rho^+K^{*-}$&---&$24.2^{+0.1+1.0+5.2}_{-0.1-1.0-5.6}$&$23.7^{+0.6+2.7+7.2}_{-0.3-2.6-10.5}$&---&---\\
&$\bar{B}^0\to\rho^0\bar{K}^{*0}$&---&$22.9^{+0.1+1.3+4.1}_{-0.1-1.1-3.2}$&$33.0^{+0.1+3.6+7.7}_{-0.1-3.6-15.1}$&---&---\\
\hline
$\phi_{\parallel}$
&$B^-\to \rho^-\bar{K}^{*0}$&---&$-1.7^{+0+0.2+0.4}_{-0-0.2-0.3}$&$3.1^{+0+0.1+5.8}_{-0-0.1-6.2}$&---&$-37^{+0+92}_{-0-59}$\\
&$B^-\to \rho^{0} K^{*-}$&---&$-1.3^{+0+0.1+0.4}_{-0-0.2-0.3}$&$3.0^{+0+0.1+5.4}_{-0-0.1-5.8}$&---&$-39^{+4+146}_{-5-88}$\\
&$\bar{B}^0\to\rho^+K^{*-}$&---&$-1.9^{+0+0.1+0.4}_{-0-0.1-0.3}$&$3.1^{+0+0+2.8}_{-0-0.1-2.7}$&---&$-36^{+4+111}_{-5-68}$\\
&$\bar{B}^0\to\rho^0\bar{K}^{*0}$&---&$-2.4^{+0+0.1+0.1}_{-0-0.1-0.1}$&$2.9^{+0+0+2.7}_{-0-0.1-3.0}$&---&$-41^{+4+63}_{-4-44}$\\
\hline
$\phi_{\perp}$
&$B^-\to \rho^-\bar{K}^{*0}$&---&$-1.8^{+0+0.2+0.3}_{-0-0.2-0.3}$&$3.0^{+0+0.1+5.8}_{-0-0.1-6.2}$&---&---\\
&$B^-\to \rho^{0} K^{*-}$&---&$-1.4^{+0+0.2+0.4}_{-0-0.2-0.4}$&$3.1^{+0+0.1+5.4}_{-0-0.1-5.8}$&---&---\\
&$\bar{B}^0\to\rho^+K^{*-}$&---&$-2.0^{+0+0.1+0.3}_{-0-0.1-0.3}$&$3.0^{+0+0+2.8}_{-0-0-2.7}$&---&---\\
&$\bar{B}^0\to\rho^0\bar{K}^{*0}$&---&$-2.5^{+0+0.1+0.1}_{-0-0.1-0.1}$&$2.9^{+0+0.1+2.7}_{-0-0.1-3.0}$&---&---\\
\hline
$\Delta\phi_{\parallel}$
&$B^-\to \rho^-\bar{K}^{*0}$&---&$0.03^{+0+0+0.03}_{-0-0-0.03}$&$-0^{+0+0+0}_{-0-0-0}$&---&$0^{+0+0}_{-0-2}$\\
&$B^-\to \rho^{0} K^{*-}$&---&$-0.14^{+0+0.03+0.30}_{-0-0.03-0.35}$&$5.6^{+0+0+0.1}_{-0-0-0.1}$&---&$-14^{+3+29}_{-4-60}$\\
&$\bar{B}^0\to\rho^+K^{*-}$&---&$-0.10^{+0.01+0.07+0.36}_{-0.01-0.07-0.22}$&$5.6^{+0+0+0.1}_{-0-0-0.1}$&---&$-19^{+5+74}_{-5-18}$\\
&$\bar{B}^0\to\rho^0\bar{K}^{*0}$&---&$-0.06^{+0+0.02+0.19}_{-0-0.03-0.18}$&$0.17^{+0+0.06+0.08}_{-0.01-0.06-0.7}$&---&$17^{+5+22}_{-5-24}$\\
\hline
$\Delta\phi_{\perp}$
&$B^-\to \rho^-\bar{K}^{*0}$&---&$0^{+0+0+0.02}_{-0-0-0.03}$&$-0^{+0+0+0}_{-0-0-0}$&---&---\\
&$B^-\to \rho^{0} K^{*-}$&---&$-0.14^{+0+0.02+0.30}_{-0-0.02-0.38}$&$5.62^{+0+0+0.15}_{-0-0-0.11}$&---&---\\
&$\bar{B}^0\to\rho^+K^{*-}$&---&$-0.12^{+0.01+0.07+0.38}_{-0.01-0.07-0.25}$&$5.62^{+0.02+6.09}_{-0.02-6.13}$&---&---\\
&$\bar{B}^0\to\rho^0\bar{K}^{*0}$&---&$-0.10^{+0+0.03+0.18}_{-0-0.04-0.16}$&$0.16^{+0+0.06+0.08}_{-0-0.06-0.07}$&---&---\\
\hline\hline
\end{tabular}
\end{center}
\end{table}

\begin{table}[!htbp]
\begin{center}
\caption{\small The observables of $B\to K^* \bar{K}^*$ decays. The other captions are the same as Table \ref{tab:rhok}. } \label{tab:kk}
\vspace{0.2cm}
\footnotesize \doublerulesep 0.10pt \tabcolsep 0.05in
\begin{tabular}{lcccccc}
\hline\hline
\multirow{2}{*}{Obs.} &\multirow{2}{*}{Decay Modes}&\multirow{2}{*}{Exp.}&\multicolumn{2}{c}{This work}&\multicolumn{2}{c}{Previous works}\\
&&&case II&case IV&Cheng~\cite{Cheng:2009cn,Cheng:2008gxa}   &Beneke~\cite{Beneke:2006hg}\\ \hline
${\cal B}[10^{-6}]$
&$B^-\to K^{*0}K^{*-}$
&$1.2\pm0.5$&$0.8^{+0+0.2+0.2}_{-0-0.1-0.1}$&$0.6^{+0+0.1+0.3}_{-0-0.1-0.2}$&$0.6^{+0.1+0.3}_{-0.1-0.3}$ &$0.5^{+0.2+0.4}_{-0.1-0.3}$\\
&$\bar{B}^0\to K^{*+}K^{*-}$
&$<2$&$1.7^{+0.1+0.1+2.6}_{-0.1-0.1-1.0}$&$0.02^{+0+0+0.02}_{-0-0-0.01}$&$0.1^{+0+0.1}_{-0-0.1}$&--- \\
&$\bar{B}^0\to K^{*0}\bar{K}^{*0}$
&$0.81\pm0.23$&$0.98^{+0.05+0.19+0.56}_{-0.06-0.14-0.40}$&$0.56^{+0.03+0.07+0.27}_{-0.03-0.07-0.14}$&$0.6^{+0.1+0.2}_{-0.1-0.3}$ &$0.6^{+0.1+0.5}_{-0.1-0.3}$\\
\hline
$A_{CP}[\%]$
&$B^-\to K^{*0}K^{*-}$&---&$-65.7^{+0.8+5.4+20.5}_{-0.9-4.0-8.9}$&$-16.6^{+0.5+1.9+5.6}_{-0.5-1.8-9.0}$&$16^{+1+17}_{-3-34}$&$0^{+0+17}_{-0-40}$\\
&$\bar{B}^0\to K^{*+}K^{*-}$&---&$0^{+0+0+0}_{-0-0-0}$&$0^{+0+0+0}_{-0-0-0}$&$0$&---\\
&$\bar{B}^0\to K^{*0}\bar{K}^{*0}$&---&$-10.2^{+0.3+1.5+5.5}_{-0.4-1.5-3.8}$&$-9.6^{+0.3+1.7+3}_{-0.3-1.6-5.2}$&$-14^{+1+6}_{-1-2}$&$-13^{+3+6}_{-4-8}$\\
\hline
$A_{CP}^0[\%]$
&$B^-\to K^{*0}K^{*-}$
&---&$-84.8^{+2.1+11.2+52.3}_{-3.4-9.5-17.2}$&$-15.2^{+0.5+2.1+7.3}_{-0.5-2.3-4.2}$&---&$9^{+3+12}_{-2-24}$\\
&$\bar{B}^0\to K^{*+}K^{*-}$&---&$0^{+0+0+0}_{-0-0-0}$&$0^{+0+0+0}_{-0-0-0}$&---&---\\
&$\bar{B}^0\to K^{*0}\bar{K}^{*0}$
&---&$-0.23^{+0.02+0.63+1.3}_{-0.02-0.67-2.3}$&$-12.3^{+0.4+1.5+8.0}_{-0.4-1.5-4.2}$&---&$0^{+0+2}_{-0-4}$\\
\hline
$A_{CP}^{\bot}[\%]$
&$B^-\to K^{*0}K^{*-}$&---&$0.69^{+0.96+7.99+32.9}_{-1.93-5.61-41.1}$&$26.5^{+0.8+3.0+12.3}_{-0.8-2.8-11.0}$&---&--- \\
&$\bar{B}^0\to K^{*+}K^{*-}$&---&$0^{+0+0+0}_{-0-0-0}$&$0^{+0+0+0}_{-0-0-0}$&---&---\\
&$\bar{B}^0\to K^{*0}\bar{K}^{*0}$&---&$0.90^{+0.05+1.36+1.57}_{-0.03-1.30-3.43}$&$23.8^{+0.8+3.0+10.6}_{-0.7-2.8-10.0}$&---&---\\
\hline
$f_L[\%]$
&$B^-\to K^{*0}K^{*-}$                      &$75^{+16}_{-26}$&$40^{+0+2+18}_{-1-2-7}$&$63^{+0+5+20}_{-0-5-16}$&$45^{+2+55}_{-4-38}$ &$62^{+1+42}_{-2-33}$\\
&$\bar{B}^0\to K^{*+}K^{*-}$&---&$76^{+0+2+5}_{-0-1-9}$&$61^{+0+1+39}_{-0-1-25}$&$\approx1$&---\\
&$\bar{B}^0\to K^{*0}\bar{K}^{*0}$
&$80^{+12}_{-13}$&$69^{+0+1+7}_{-0-1-13}$&$65^{+0+5+20}_{-0-5-16}$&$52^{+4+48}_{-7-48}$ &$69^{+1+34}_{-1-27}$\\
\hline
$f_{\bot}[\%]$
&$B^-\to K^{*0}K^{*-}$&---&$18.1^{+0.3+1.4+9.2}_{-0.5-1.1-9.5}$&$18.7^{+0.1+2.8+8.2}_{-0-2.7-10.1}$&---&---\\
&$\bar{B}^0\to K^{*+}K^{*-}$&---&$12.1^{+0+0.7+5.3}_{-0-0.8-3.1}$&$19.5^{+0+0.7+12.3}_{-0-0.6-20.8}$&---&---\\
&$\bar{B}^0\to K^{*0}\bar{K}^{*0}$&---&$26.1^{+0.1+1.3+9.4}_{-0.1-1.1-9.0}$&$14.8^{+0+2.4+7.3}_{-0-2.3-8.1}$&---&---\\
\hline
$\phi_{\parallel}$
&$B^-\to K^{*0}K^{*-}$
&---&$-1.8^{+0.1+0.2+3.3}_{-0.1-0.2-2.6}$&$3.0^{+0+0.1+6.6}_{-0-0.1-6.9}$&---&$-39^{+2+96}_{-3-57}$\\
&$\bar{B}^0\to K^{*+}K^{*-}$&---&$0.7^{+0+0.1+0.2}_{-0-0.1-0.2}$&$2.0^{+0+0+1.1}_{-0-0-0.8}$ &---&---\\
&$\bar{B}^0\to K^{*0}\bar{K}^{*0}$&---&$-3.0^{+0+0.1+3.1}_{-0-0.1-4.1}$&$3.1^{+0+0.1+5.8}_{-0-0.1-8.7}$&---
&$-32^{+0+82}_{-0-51}$\\
\hline
$\phi_{\perp}$
&$B^-\to K^{*0}K^{*-}$
&---&$-1.8^{+0+0.2+1.9}_{-0-0.2-0.5}$&$3.0^{+0+0.1+5.8}_{-0-0.1-6.9}$&---&---\\
&$\bar{B}^0\to K^{*+}K^{*-}$&---&$-2.2^{+0+0.1+0.2}_{-0-0.1-0.2}$&$-1.8^{+0+0.1+1.1}_{-0-0.1-0.8}$&---&---\\
&$\bar{B}^0\to K^{*0}\bar{K}^{*0}$&---&$-2.1^{+0+0.1+0.2}_{-0-0.1-0.2}$&$3.0^{+0+0+5.8}_{-0-0-8.7}$
&---&---\\
\hline
$\Delta\phi_{\parallel}$
&$B^-\to K^{*0}K^{*-}$&---&$-0.60^{+0.06+0.17+5.9}_{-0.06-0.19-5.7}$&$0.08^{+0+0+0.02}_{-0-0-0.05}$&---&$-5^{+1+28}_{-1-7}$\\
&$\bar{B}^0\to K^{*+}K^{*-}$&---&$0^{+0+0+0}_{-0-0-0}$&$0^{+0+0+0}_{-0-0-0}$&---&---\\
&$\bar{B}^0\to K^{*0}\bar{K}^{*0}$&---&$-0.16^{+0+0.89+0}_{-0-0.03-0}$&$0.02^{+0+0+0.03}_{-0-1-0.06}$&---&$3^{+1+14}_{-1-6}$\\
\hline
$\Delta\phi_{\perp}$
&$B^-\to K^{*0}K^{*-}$&---&$-0.02^{+0.04+0.15+0.5}_{-0.04-0.17-0.6}$&$-0.03^{+0+0+0.02}_{-0-0.01-0.05}$&---&---\\
&$\bar{B}^0\to K^{*+}K^{*-}$&---&$0^{+0+0+0}_{-0-0-0}$&$0.02^{+0+0.01+0}_{-0-0.01-0}$&---&---\\
&$\bar{B}^0\to K^{*0}\bar{K}^{*0}$&---&$-0.08^{+0+0+0.01}_{-0-0-0.02}$&$0.02^{+0+0+0.03}_{-0-0-0.07}$&---&---\\
\hline\hline
\end{tabular}
\end{center}
\end{table}

\begin{table}[p]
\begin{center}
\caption{\small The observables of $B\to \phi \bar{K}^*$ decays. The other captions  are the same as Table \ref{tab:rhok}. } \label{tab:phik}
\vspace{0.2cm}
\footnotesize \doublerulesep 0.10pt \tabcolsep 0.05in
\begin{tabular}{lcccccc}
\hline\hline
\multirow{2}{*}{Observables} &\multirow{2}{*}{Decay Modes}&\multirow{2}{*}{Exp.}&\multicolumn{2}{c}{This work}&\multicolumn{2}{c}{Previous works}\\
&&&case II&case IV&Cheng~\cite{Cheng:2009cn,Cheng:2008gxa} &Beneke~\cite{Beneke:2006hg} \\ \hline
${\cal B}\,[10^{-6}]$
&$B^-\to \phi K^{*-}$
&$10.0\pm1.1$&$6.3^{+0.3+4.2+3.4}_{-0.4-1.7-1.5}$&$12.0^{+0.5+1.3+6.8}_{-0.7-1.2-4.0}$ &$10.0^{+1.4+12.3}_{-1.3-6.1}$ &$10.1^{+0.5+12.2}_{-0.5-7.1}$\\
&$\bar{B}^0\to \phi \bar{K}^{*0}$
&$10.1^{+0.6}_{-0.5}$&$5.8^{+0.3+3.4+3.1}_{-0.3-1.5-1.3}$&$11.1^{+0.5+1.2+6.4}_{-0.6-1.1-3.8}$&$9.5^{+1.3+11.9}_{-1.2-5.9}$ &$9.3^{+0.5+11.4}_{-0.5-6.5}$\\
\hline
$A_{CP}[\%]$
&$B^-\to \phi K^{*-}$
&$-1\pm8$&$6^{+0+0+3}_{-0-1-2}$&$0^{+0+0+0.4}_{-0-0-0.3}$ &$0.05$&$0^{+0+2}_{-0-1}$\\
&$\bar{B}^0\to \phi \bar{K}^{*0}$
&$-0\pm4$&$1^{+0+0+0}_{-0-0-0}$&$0.2^{+0+0.1+0.3}_{-0-0.1-0.1}$&$0.8^{+0+0.4}_{-0-0.5}$&$1^{+0+1}_{-0-0}$\\
\hline
$A_{CP}^0[\%]$
&$B^-\to \phi K^{*-}$
&$17\pm11$&$5^{+0+4+8}_{-0-2-6}$&$1^{+0+0+0.2}_{-0-0-0.5}$&---&$-1^{+0+2}_{-0-1}$\\
&$\bar{B}^0\to \phi \bar{K}^{*0}$
&$-0.7\pm3.0$&$0.4^{+0+0.6+1.1}_{-0-0.2-0.6}$&$0.8^{+0+0.1+0.3}_{-0-0.1-0.6}$&---&$0^{+0+1}_{-0-1}$\\
\hline
$A_{CP}^{\bot}[\%]$
&$B^-\to \phi K^{*-}$
&$22\pm25$&$-3^{+0+2+9}_{-0-3-7}$&$-0.9^{+0+0.2+0.5}_{-0-0.2-0.6}$&---&---\\
&$\bar{B}^0\to \phi \bar{K}^{*0}$
&$-1.4\pm5.7$&$-0.3^{+0+0.1+0.5}_{-0-0.2-0.1}$&$-0.8^{+0+0.1+0.4}_{-0-0.1-0.4}$&---&---\\
\hline
$f_L[\%]$
&$B^-\to \phi K^{*-}$
&$50\pm5$&$50^{+0+6+46}_{-0-16-43}$&$47^{+0+7+26}_{-0-7-18}$&$49^{+4+51}_{-7-42}$&$45^{+0+58}_{-0-36}$\\
&$\bar{B}^0\to \phi \bar{K}^{*0}$
&$49.7\pm1.7$&$50.1^{+0+6.0+46.1}_{-0-15.3-43.7}$&$47^{+0+7+26}_{-0-7-18}$&$50^{+4+51}_{-6-43}$&$44^{+0+59}_{-0-36}$\\
\hline
$f_{\bot}[\%]$
&$B^-\to \phi K^{*-}$
&$20\pm5$&$21^{+0+2+20}_{-0-3-20}$&$27^{+0+4+9}_{-0-4-13}$&---&---\\
&$\bar{B}^0\to \phi \bar{K}^{*0}$
&$22.5\pm1.5$&$20.5^{+0+2.2+20.9}_{-0-2.3-19.1}$&$27^{+0+4+9}_{-0-4-13}$&---&---\\
\hline
$\phi_{\parallel}$
&$B^-\to \phi K^{*-}$
&$-0.80\pm0.17$&$-1.18^{+0+0.62+0.93}_{-0-0.55-0.64}$&$-3.0^{+0+6.1+9.4}_{-0-6.1-10.6}$&---&$-41^{+0+84}_{-0-53}$\\
&$\bar{B}^0\to \phi \bar{K}^{*0}$
&$-0.71\pm0.06$&$-1.13^{+0+0.56+0.94}_{-0-0.51-0.64}$&$-3.0^{+0+6.0+9.4}_{-0-6.1-10.6}$&---&$-42^{+0+87}_{-0-54}$\\
\hline
$\phi_{\perp}$
&$B^-\to \phi K^{*-}$
&$-0.56\pm0.17$&$-1.20^{+0+0.62+1.03}_{-0-0.54-0.72}$&$-3.0^{+0+6.2+9.4}_{-0-6.1-10.6}$&---&$-41^{+0+84}_{-0-53}$\\
&$\bar{B}^0\to \phi \bar{K}^{*0}$
&$-0.61\pm0.06$&$-1.18^{+0+0.64+1.05}_{-0-0.55-0.74}$&$-3.0^{+0+0.1-9.4}_{-0-0.1-10.6}$&---&$-42^{+0+87}_{-0-54}$\\
\hline
$\Delta\phi_{\parallel}$
&$B^-\to \phi K^{*-}$
&$0.07\pm0.21$&$0.03^{+0+0+0.05}_{-0-0-0.02}$&$0^{+0+0+0}_{-0-0-0}$&---&$0^{+0+0}_{-0-1}$\\
&$\bar{B}^0\to \phi \bar{K}^{*0}$
&$0.05\pm0.05$&$0^{+0+0+0.01}_{-0-0-0.01}$&$0^{+0+0+0}_{-0-0-0}$&---&$0^{+0+0}_{-0-0}$\\
\hline
$\Delta\phi_{\perp}$
&$B^-\to \phi K^{*-}$
&$0.19\pm0.21$&$-0.02^{+0+0+0.05}_{-0.02-0.01-0.08}$&$0^{+0+0+0}_{-0-0-0}$&---&$0^{+0+0}_{-0-1}$\\
&$\bar{B}^0\to \phi \bar{K}^{*0}$
&$0.08\pm0.05$&$0^{+0+0+0.01}_{-0-0-0.01}$&$0^{+0+0+0}_{-0-0-0}$&---&$0^{+0+0}_{-0-0}$\\
\hline\hline
\end{tabular}
\end{center}
\end{table}

\begin{table}[!htbp]
\begin{center}
\caption{\small The observables of $B\to \rho\rho$ decays.  The other captions  are the same as Table \ref{tab:rhok}. } \label{tab:rhorho}
\vspace{0.2cm}
\footnotesize \doublerulesep 0.10pt \tabcolsep 0.05in
%\begin{threeparttable}
\begin{tabular}{lcccccc}
\hline\hline
\multirow{2}{*}{Obs.} &\multirow{2}{*}{Decay Modes}&\multirow{2}{*}{Exp.}&\multicolumn{2}{c}{This work}&\multicolumn{2}{c}{Previous works} \\
&   &  &case II&case IV&Cheng~\cite{Cheng:2009cn,Cheng:2008gxa}&Beneke~\cite{Beneke:2006hg} \\\hline
${\cal B}[10^{-6}]$
&$B^-\to \rho^0 \rho^- $
&$24.0^{+1.9}_{-2.0}$&$26.2^{+2.0+6.1+5.2}_{-2.2-5.7-3.4}$&$20.0^{+1.5+3.6+2.0}_{-3.6-3.4-1.6}$&$20.0^{+4+2}_{-1.9-0.9}$ &$18.8^{+0.4+3.2}_{-0.4-3.9}$\\
&$\bar{B}^0\to\rho^+\rho^-$
&$24.2^{+3.1}_{-3.2}$&$24.4^{+1.8+4.8+0.8}_{-2.0-4.2-0.3}$&$25.7^{+1.9+1.6+0.7}_{-2-2-0.4}$&$25.5^{+1.5+2.4}_{-2.6-1.5}$ &$23.6^{+1.7+3.9}_{-1.9-3.6}$ \\
&$\bar{B}^0\to\rho^0\rho^0$
&$0.94\pm0.17$&$14.3^{+1.1+8.8+10.3}_{-1.2-7.8-4.3}$&$1.52^{+0.11+0.89+0.29}_{-0.12-0.77-0.36}$&$0.9^{+1.5+1.1}_{-0.4-0.2}$ &$0.9^{+0.6+1.9}_{-0.3-0.9}$\\
\hline
$A_{CP}[\%]$
&$B^-\to \rho^0 \rho^- $ &$-5.1\pm5.4$&$0.5^{+0+0.1+0.1}_{-0-0.2-0.1}$&$0.1^{+0+0.1+0.2}_{-0-0.1-0.1}$ &$0.06$ &$0^{+0+0}_{-0-0}$\\
&$\bar{B}^0\to\rho^+\rho^-$ &---&$-20.3^{+0.7+3.0+4.4}_{-0.6-3.2-3.1}$&$-3.4^{+0.1+0.7+3.8}_{-0.1-0.8-8.9}$&$-4^{+0+3}_{-0-3}$ &$-1^{+0+4}_{-0-8}$\\
&$\bar{B}^0\to\rho^0\rho^0$ &---&$16.1^{+0.5+8.4+8.9}_{-0.5-3.8-10.9}$&$41.5^{+1.0+19.1+10.5}_{-1.2-9.4-20.5}$ &$30^{+17+14}_{-16-26}$ &$28^{+5+53}_{-7-29}$\\
\hline
$A_{CP}^0[\%]$
&$B^-\to \rho^0 \rho^-$&---&$0.82^{+0.02+0.23+0.35}_{-0.03-0.28-0.21}$&$0.07^{+0.01+0.06+0.19}_{-0-0.05-0.11}$&---&---\\
$C_{long}[\%]$
&$\bar{B}^0\to\rho^+\rho^-$&$0\pm9$&$32^{+1+4+7}_{-1-3-8}$&$6^{+0+0+9}_{-0-0-3}$&---&---\\
&$\bar{B}^0\to\rho^0\rho^0$&$20\pm90$&$-40^{+1+12+15}_{-1-29-10}$&$-30^{+1+10+12}_{-1-22-35}$&---&---\\
$S_{long}[\%]$
&$\bar{B}^0\to\rho^+\rho^-$&$-14\pm13$&$-0^{+5+4+12}_{-7-5-6}$&$-22^{+5+0+3}_{-7-0-6}$&---&---\\
&$\bar{B}^0\to\rho^0\rho^0$&$30\pm70$&$34^{+4+8+11}_{-7-9-13}$&$48^{+4+8+19}_{-6-9-13}$&---&---\\
\hline
$A_{CP}^{\bot}[\%]$
&$B^-\to \rho^0 \rho^-$&---&$-2.4^{+0.1+0.2+0.5}_{-0.1-0.2-0.4}$&$2.0^{+0+0.2+0.6}_{-0-0.4-0.4}$&---&--- \\
&$\bar{B}^0\to\rho^+\rho^-$&---&$56.4^{+1.2+7.0+6.3}_{-1.6-7.0-8.3}$&$25.4^{+0.8+5.9+19.8}_{-0.8-7.3-17.4}$&---&--- \\
&$\bar{B}^0\to\rho^0\rho^0$&---&$2.2^{+0.1+0.6+2.8}_{-0.1-0.5-2.2}$&$6.4^{+0.3+4.1+21.7}_{-0.6-1.8-13.2}$&---&--- \\
\hline
$f_L[\%]$
&$B^-\to \rho^0 \rho^- $
&$95.0\pm1.6$&$74.5^{+0+7.8+7.7}_{-1.6-6.5-11.6}$&$91.4^{+0+1.7+1.8}_{-0-2.0-1.4}$&$96^{+1+2}_{-1-2}$ &$95.9^{+0.2+3.4}_{-0.3-6.4}$\\
&$\bar{B}^0\to\rho^+\rho^-$
&$97.8^{+2.5}_{-2.2}$&$80.7^{+0.1+5.5+7.9}_{-0.1-6.1-10.5}$&$92.1^{+0.2+1.6+1.8}_{-0.1-2.0-1.2}$&$92^{+1+1}_{-2-2}$ &$91.3^{+0.4+5.6}_{-0.3-6.4}$\\
&$\bar{B}^0\to\rho^0\rho^0$
&$59\pm13$\tnote{*}&$22.4^{+0.1+2.3+2.9}_{-0.2-3.7-4.2}$&$34.9^{+0.5+8.4+19.4}_{-0.7-4.7-14.2}$&$92^{+3+6}_{-4-37}$&$90^{+3+8}_{-4-56}$\\
\hline
$f_{\bot}[\%]$
&$B^-\to \rho^0 \rho^-$&---&$12.7^{+0+3.3+5.8}_{-0-3.9-3.8}$&$4.1^{+0+1.0+0.7}_{-0-0.9-0.9}$&---&---\\
&$\bar{B}^0\to\rho^+\rho^-$&---&$12.4^{+0.1+3.4+7.1}_{-0.1-3.0-4.8}$&$4.2^{+0+1.2+0.7}_{-0.1-1.0-0.6}$&---&---\\
&$\bar{B}^0\to\rho^0\rho^0$&---&$38.0^{+0.1+1.0+4.4}_{-0.1-1.0-3.4}$&$30.2^{+0.3+2.5+6.9}_{-0.2-4.9-9.5}$&---&---\\
\hline
$\phi_{\parallel}$
&$B^-\to \rho^0 \rho^-$&---&$-1.3^{+0+0.2+0.3}_{-0-0.2-0.3}$&$0.5^{+0+0.1+0.1}_{-0-0.2-0.1}$&---&$-5^{+0+31}_{-0-32}$\\
&$\bar{B}^0\to\rho^+\rho^-$&---&$1.2^{+0+0.2+0.4}_{-0-0.3-0.3}$&$-0.4^{+0+0.1+0}_{-0-0.1-0}$&---&$1^{+2+17}_{-2-17}$\\
&$\bar{B}^0\to\rho^0\rho^0$&---&$-2.4^{+0+0.2+0}_{-0-0.1-0}$&$1.3^{+0+0+0.3}_{-0-0-0.5}$&---&---\\
\hline
$\phi_{\perp}$
&$B^-\to \rho^0 \rho^-$&---&$-1.3^{+0+0.2+0.3}_{-0-0.2-0.3}$&$0.5^{+0+0.1+0.1}_{-0-0.2-0.1}$&---&$-5^{+0+31}_{-0-32}$\\
&$\bar{B}^0\to\rho^+\rho^-$&---&$0.9^{+0+0.2+0.3}_{-0.2-0.3-0.3}$&$-0.4^{+0+0.1+0}_{-0-0.1-0}$&---&$1^{+2+17}_{-2-17}$\\
&$\bar{B}^0\to\rho^0\rho^0$&---&$-2.6^{+0+0.1+0.1}_{-0-0-0.1}$&$1.4^{+0+0.1+0.3}_{-0-0.1-0.5}$&---&---\\
\hline
$\Delta\phi_{\parallel}$
&$B^-\to \rho^0 \rho^-$&---&$-0.02^{+0+0+0}_{-0-0-0}$&$-0^{+0+0+0}_{-0-0-0}$&---&$-6^{+2+2}_{-1-5}$\\
&$\bar{B}^0\to\rho^+\rho^-$&---&$-0.08^{+0+0.05+0.11}_{-0-0.06-0.16}$&$-0.31^{+0.01+0+0.09}_{-0.01-0-0}$&---&$4^{+1+9}_{-1-9}$\\
&$\bar{B}^0\to\rho^0\rho^0$&---&$0.16^{+0+0.07+0.03}_{-0.01-0.04-0.05}$&$0.42^{+0.01+0.18+0.26}_{-0.01-0.09-0.10}$&---&---\\
\hline
$\Delta\phi_{\perp}$
&$B^-\to \rho^0 \rho^-$&---&$-0.02^{+0+0+0}_{-0-0-0}$&$0^{+0+0+0}_{-0-0-0}$&---&$-6^{+2+2}_{-1-5}$\\
&$\bar{B}^0\to\rho^+\rho^-$&---&$0.11^{+0+0+0.12}_{-0.03-0.02-0.08}$&$-0.28^{+0.01+0.03+0.09}_{-0.01-0.03-0.04}$&---&$4^{+1+9}_{-1-9}$\\
&$\bar{B}^0\to\rho^0\rho^0$&---&$0.16^{+0+0.08+0.04}_{-0.01-0.04-0.04}$&$0.37^{+0.01+0.13+0.26}_{-0.01-0.71-0.10}$&---&---\\
\hline\hline
\end{tabular}
\end{center}
\end{table}

\begin{table}[!htbp]
\begin{center}
\caption{\small The observables of $\bar{B}^0\to \phi \phi$ decays. The CP asymmetries $A_{CP}^{(0,\perp)}$ and phases $\Delta\phi_{\parallel,\perp}$ are equal to zero, and thus not listed. The other captions  are the same as Table \ref{tab:rhok}. } \label{tab:phiphi}
\vspace{0.2cm}
\footnotesize \doublerulesep 0.10pt \tabcolsep 0.05in
\begin{tabular}{lccccc}
\hline\hline
\multirow{2}{*}{Obs.} &\multirow{2}{*}{Exp.}&\multicolumn{2}{c}{This work}&\multicolumn{2}{c}{Previous work}\\
&&case II&case IV&Beneke~\cite{Beneke:2006hg} \\ \hline
${\cal B}\,[10^{-8}]$
&$<20$&$13.4^{+0.6+1.2+23.4}_{-0.9-1.0-7.6}$&$0.10^{+0+0+0.08}_{-0-0-0.04}$&$<3$\\
\hline
$f_L[\%]$
&---&$72^{+0+2+5}_{-0-2-8}$&$38^{+0+2+51}_{-0-2-22}$&$>80$\\
\hline
$f_{\bot}[\%]$
&---&$14^{+0+1+5}_{-0-1-3}$&$31^{+0+1+10}_{-0-1-26}$&---\\
\hline
$\phi_{\parallel}$
&---&$0.49^{+0+0.06+0.12}_{-0-0.06-0.13}$&$1.86^{+0+0+1.08}_{-0-0-0.84}$&---\\
\hline
$\phi_{\perp}$
&---&$-2.44^{+0+0.06+0.14}_{-0-0.06-0.17}$&$-1.84^{+0+0-1.08}_{-0-0-0.85}$&---\\
\hline\hline
\end{tabular}
\end{center}
\end{table}

\newpage
\section*{Appendix C: The fitted results of end-point parameters in the complex plane}
\begin{figure}[ht]
\begin{center}
 \subfigure[]{\includegraphics[width=7cm]{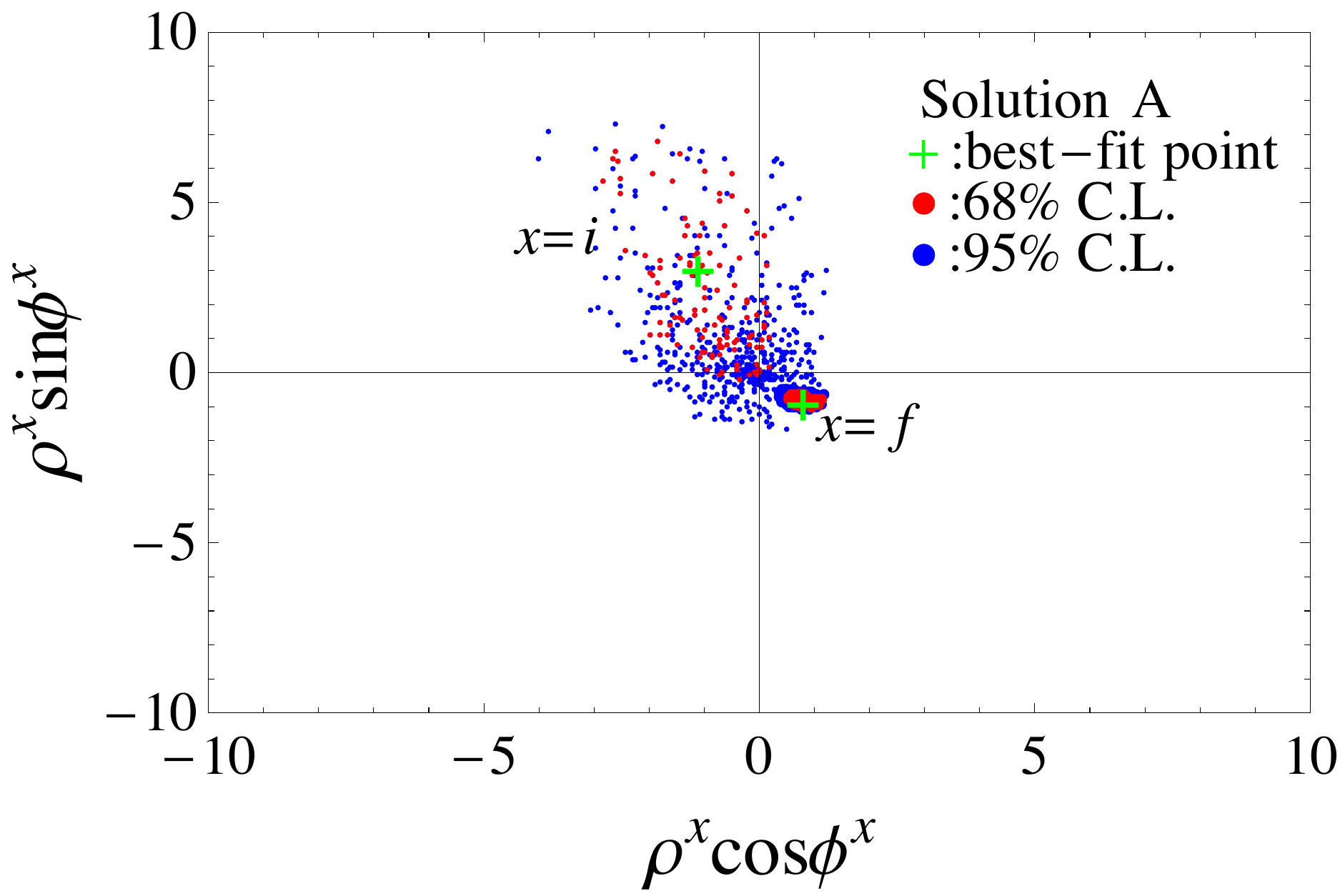}}\quad
 \subfigure[]{\includegraphics[width=7cm]{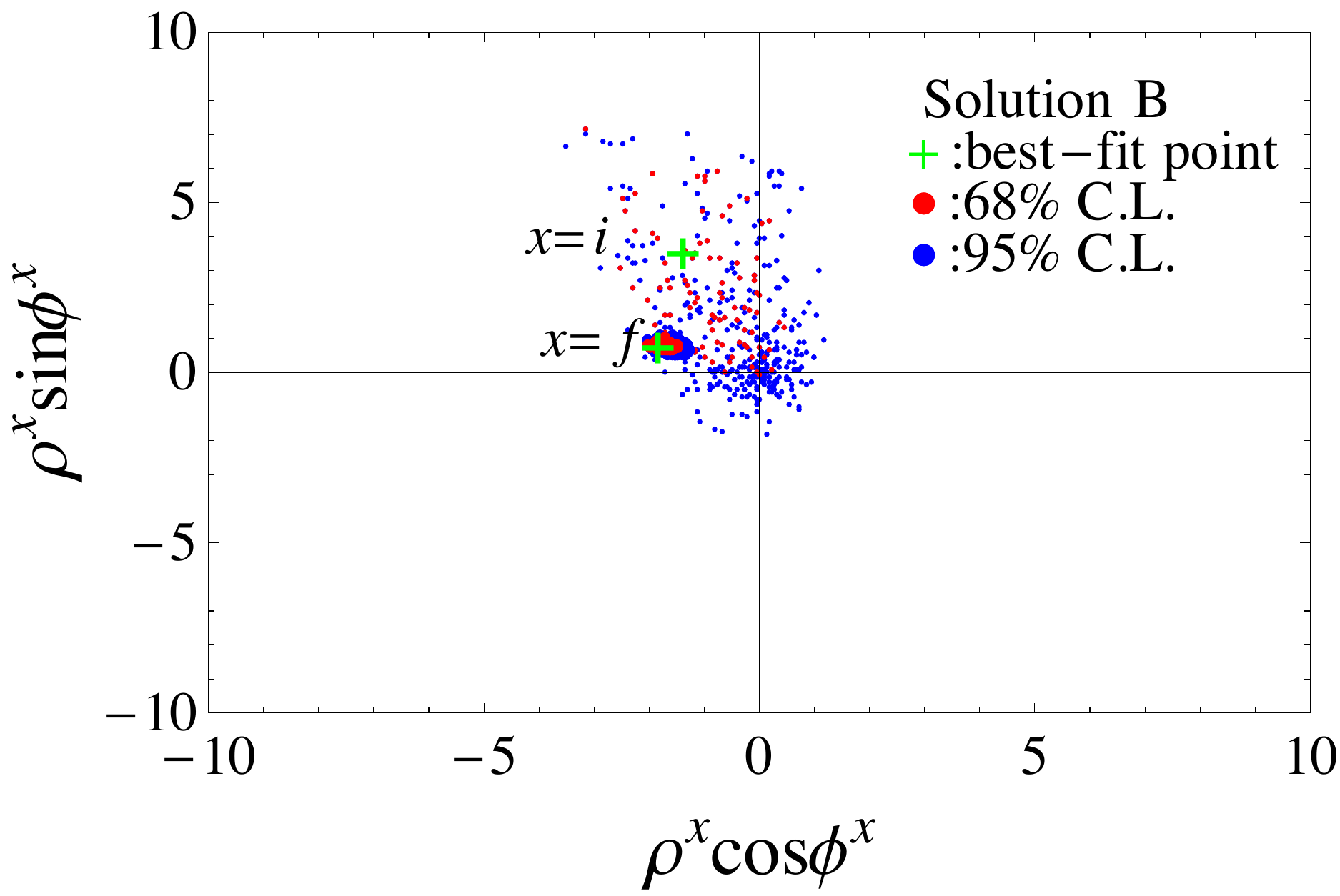}}\\
  \subfigure[]{\includegraphics[width=7cm]{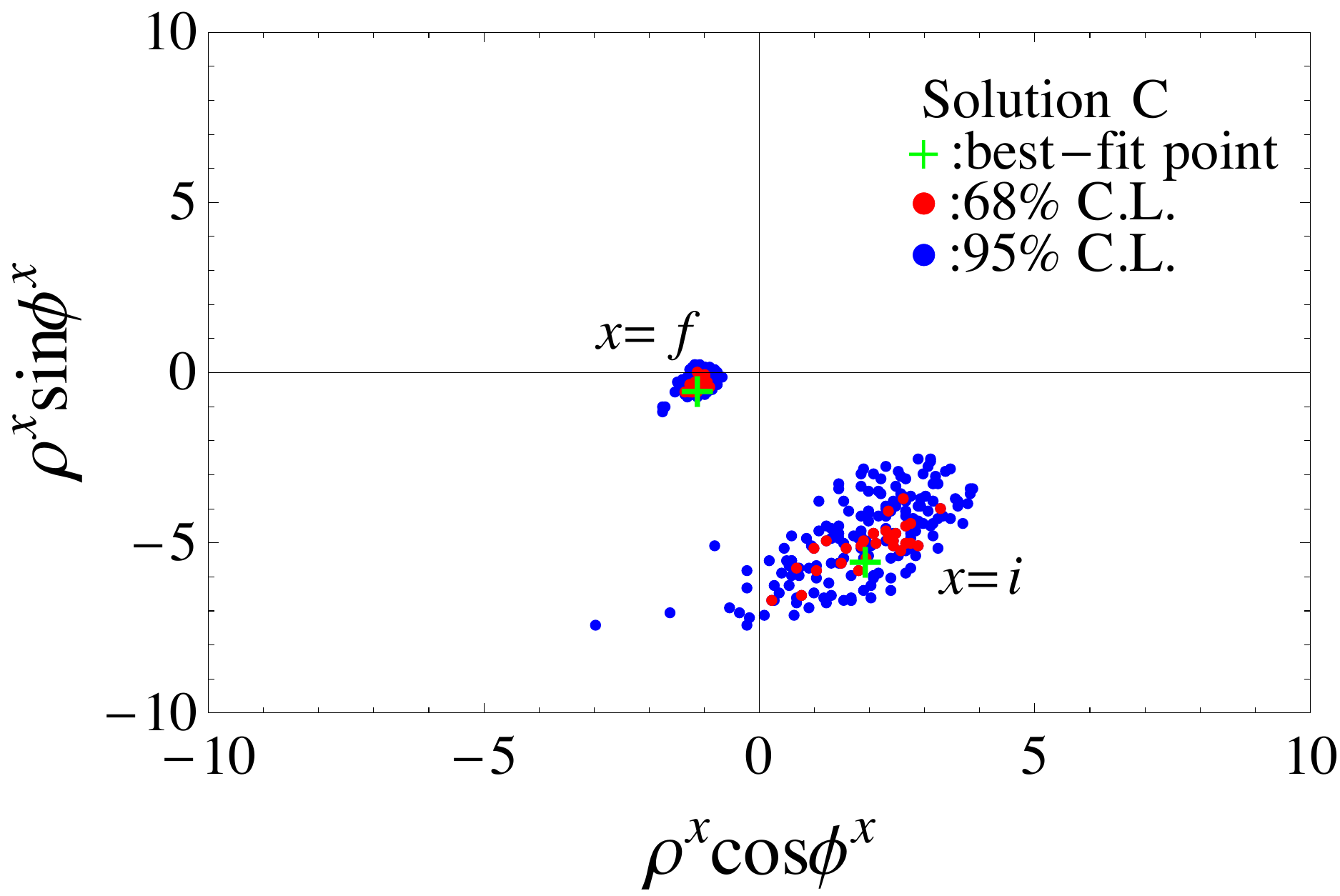}}\quad
    \subfigure[]{\includegraphics[width=7cm]{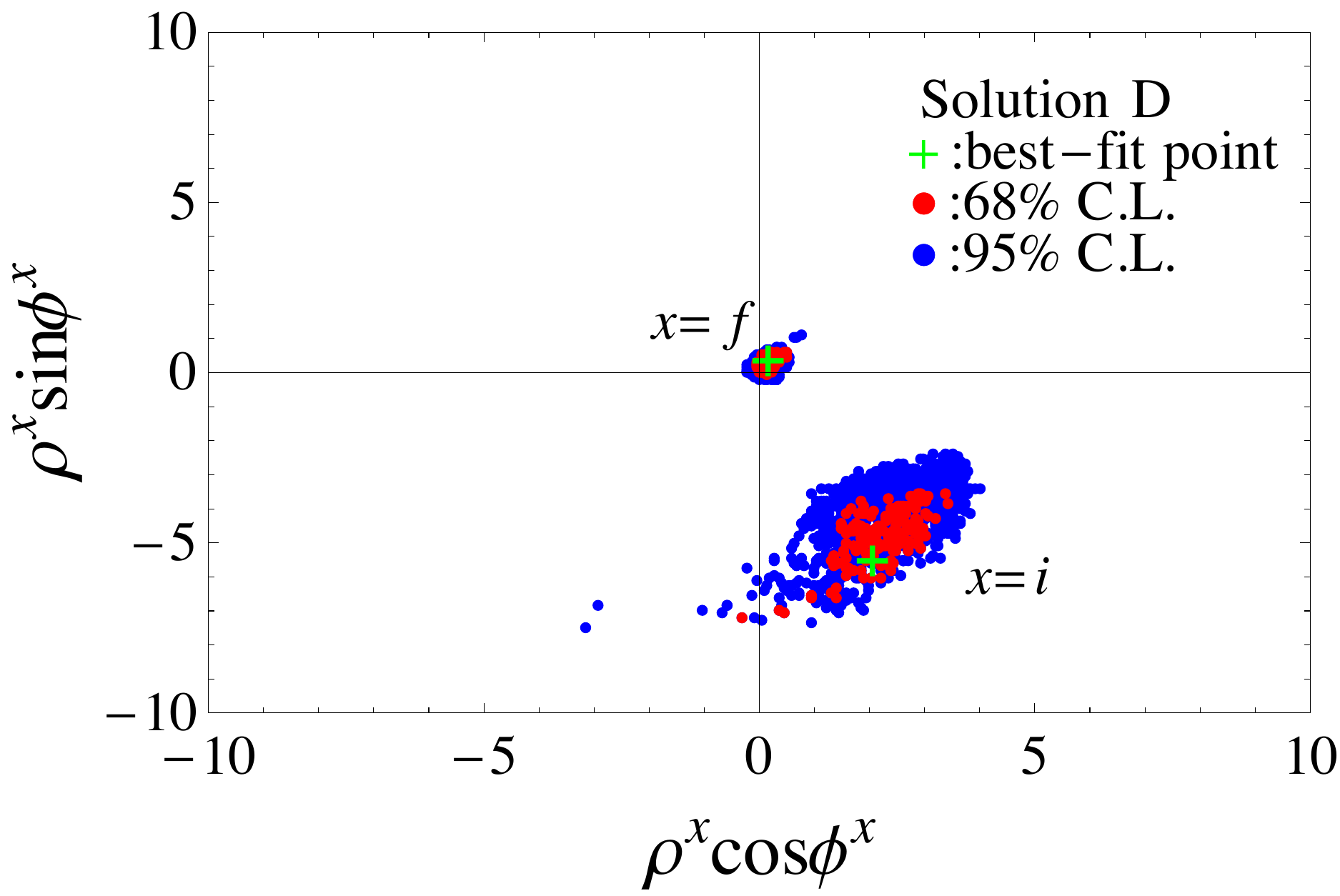}}
\caption{\label{KKrhoK2} \small Same as Fig.~\ref{KKrhoK} except for in $(\rho \cos \phi, \rho \sin \phi)$ plane.}
\end{center}
\end{figure}

\begin{figure}[ht]
\begin{center}
    \subfigure[]{\includegraphics[width=7cm]{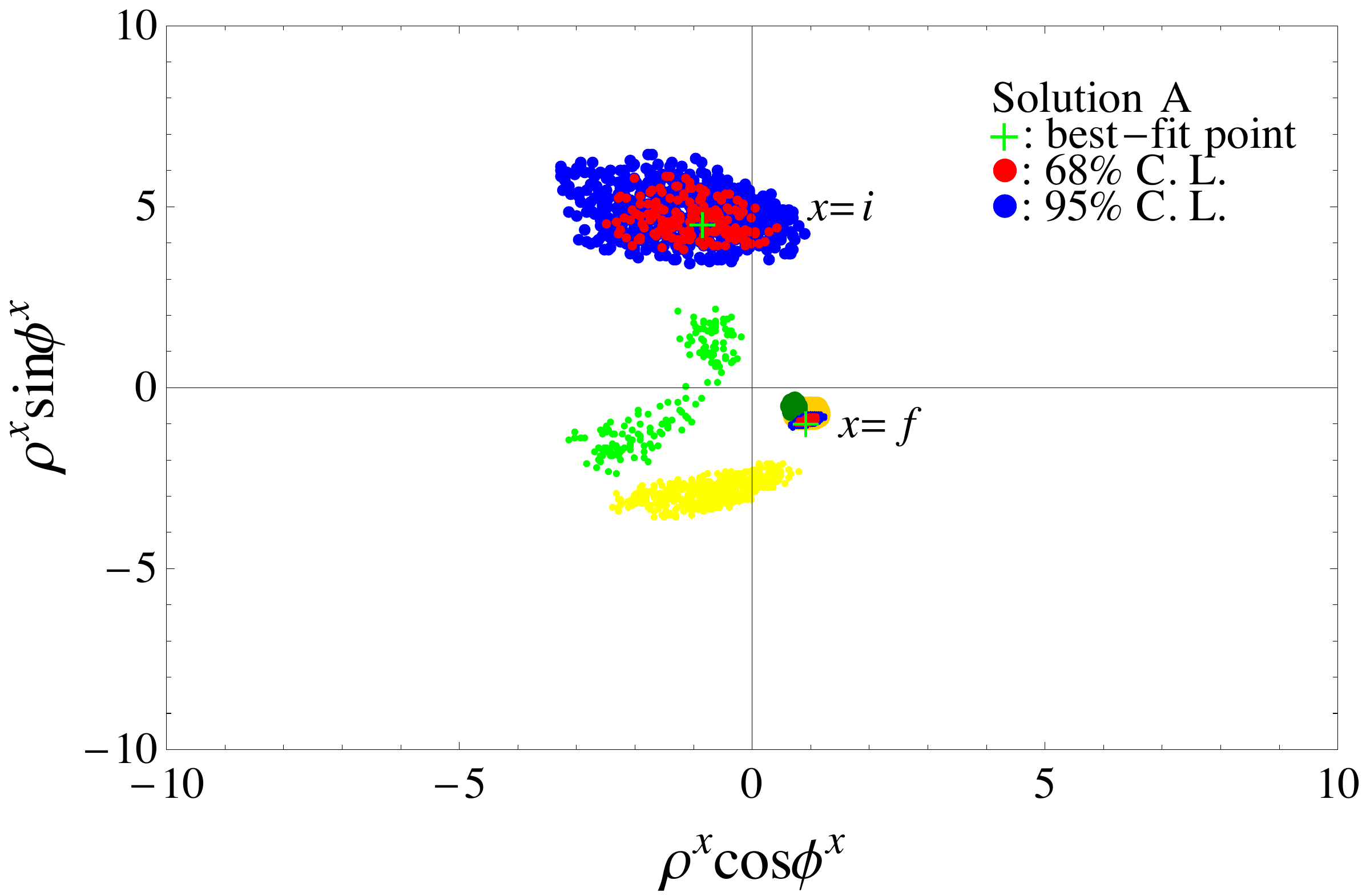}}\quad
    \subfigure[]{\includegraphics[width=7cm]{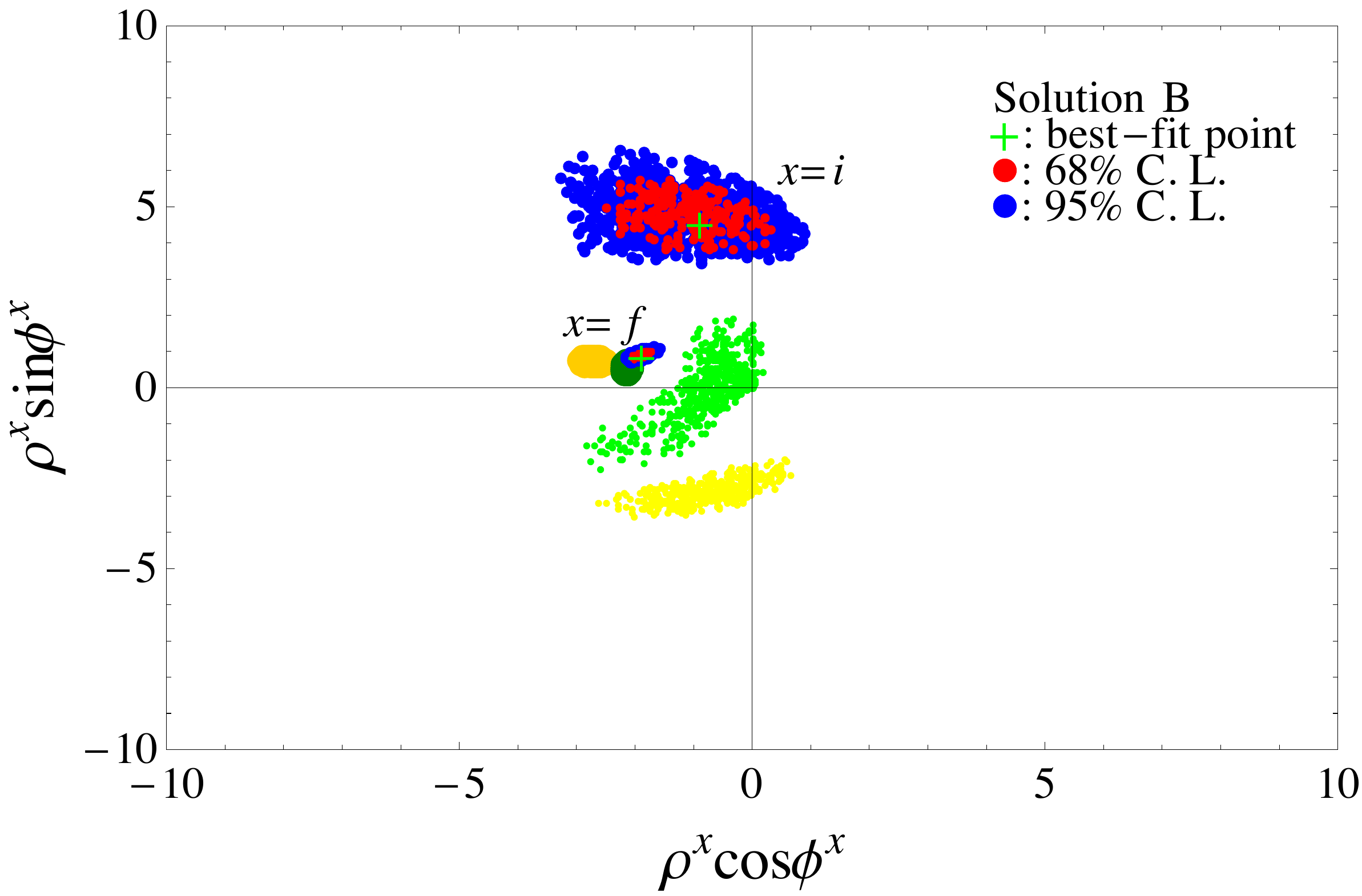}}
 \caption{\label{KKrhoKphiK2} \small Same as Fig.~\ref{KKrhoKphiK} except for in $(\rho \cos \phi, \rho \sin \phi)$ plane.}
\end{center}
\end{figure}

\vspace{2cm}
\begin{figure}[ht]
\begin{center}
 \subfigure[]{\includegraphics[width=7cm]{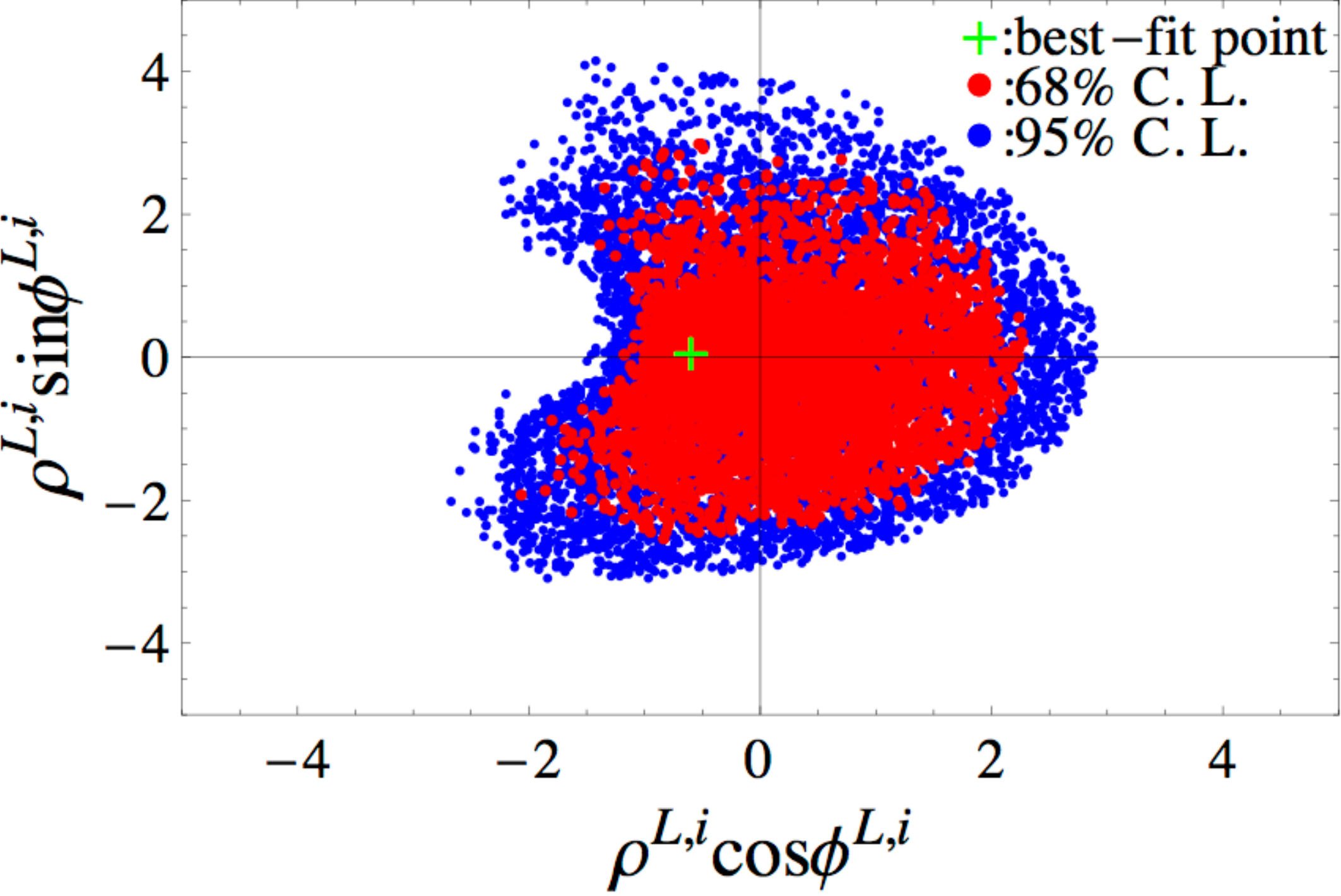}}
 \subfigure[]{\includegraphics[width=7cm]{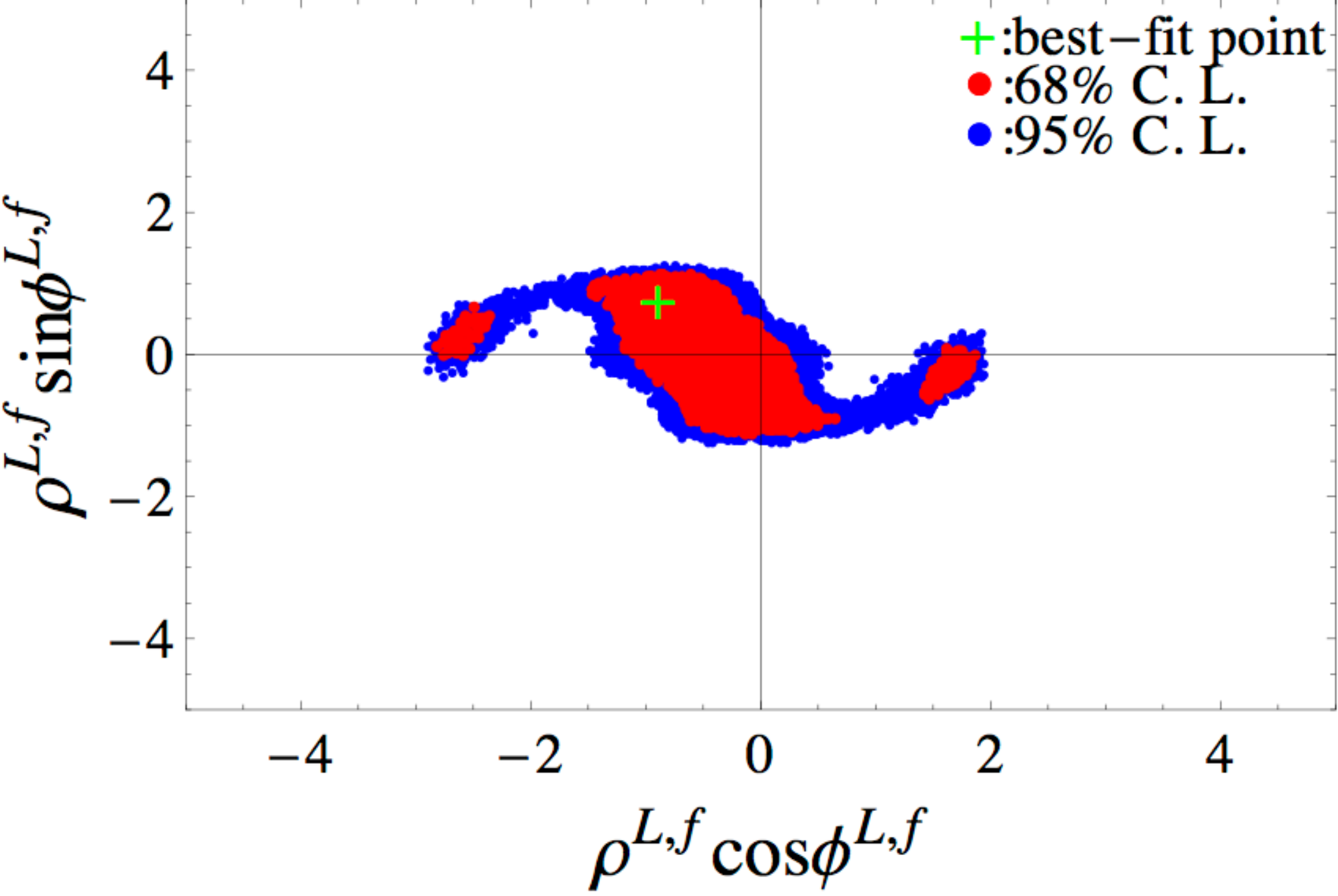}}
  \subfigure[]{\includegraphics[width=7cm]{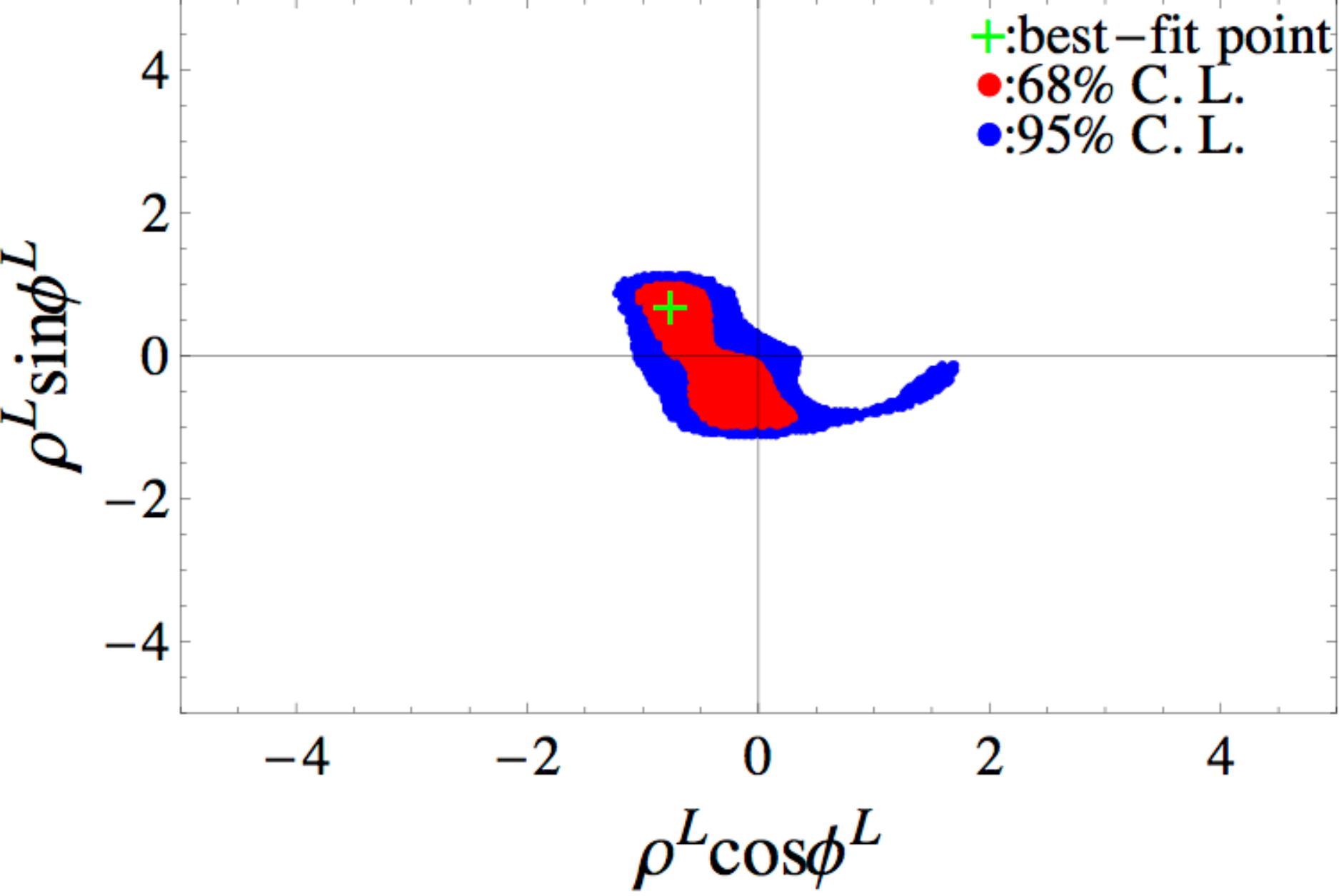}}
\caption{\label{case32} \small Same as Fig.~\ref{case3} except for in $(\rho \cos \phi, \rho \sin \phi)$ plane.}
\end{center}
\end{figure}

\begin{figure}[ht]
\begin{center}
   \subfigure[]{\includegraphics[width=7cm]{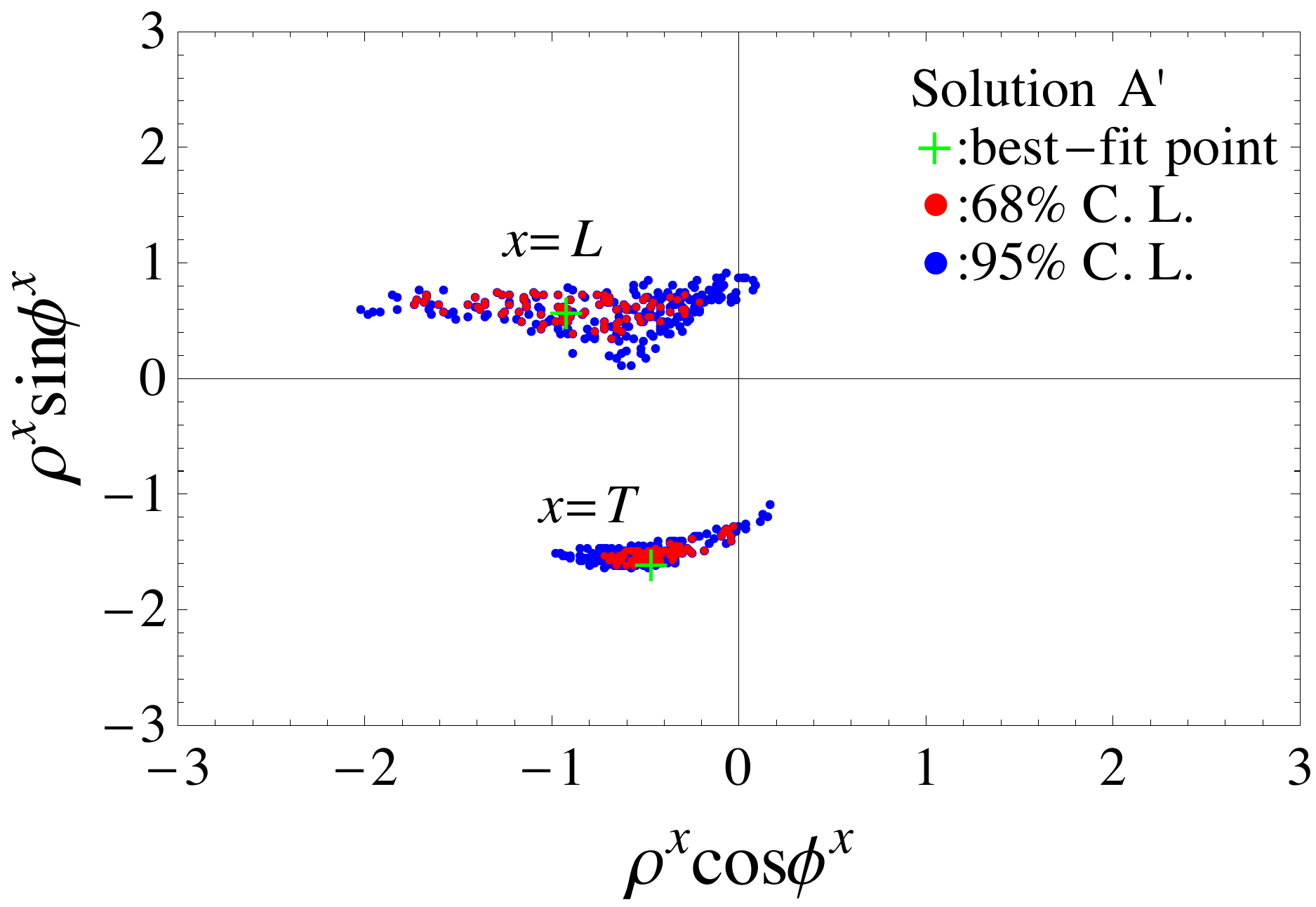}}
 \subfigure[]{\includegraphics[width=7cm]{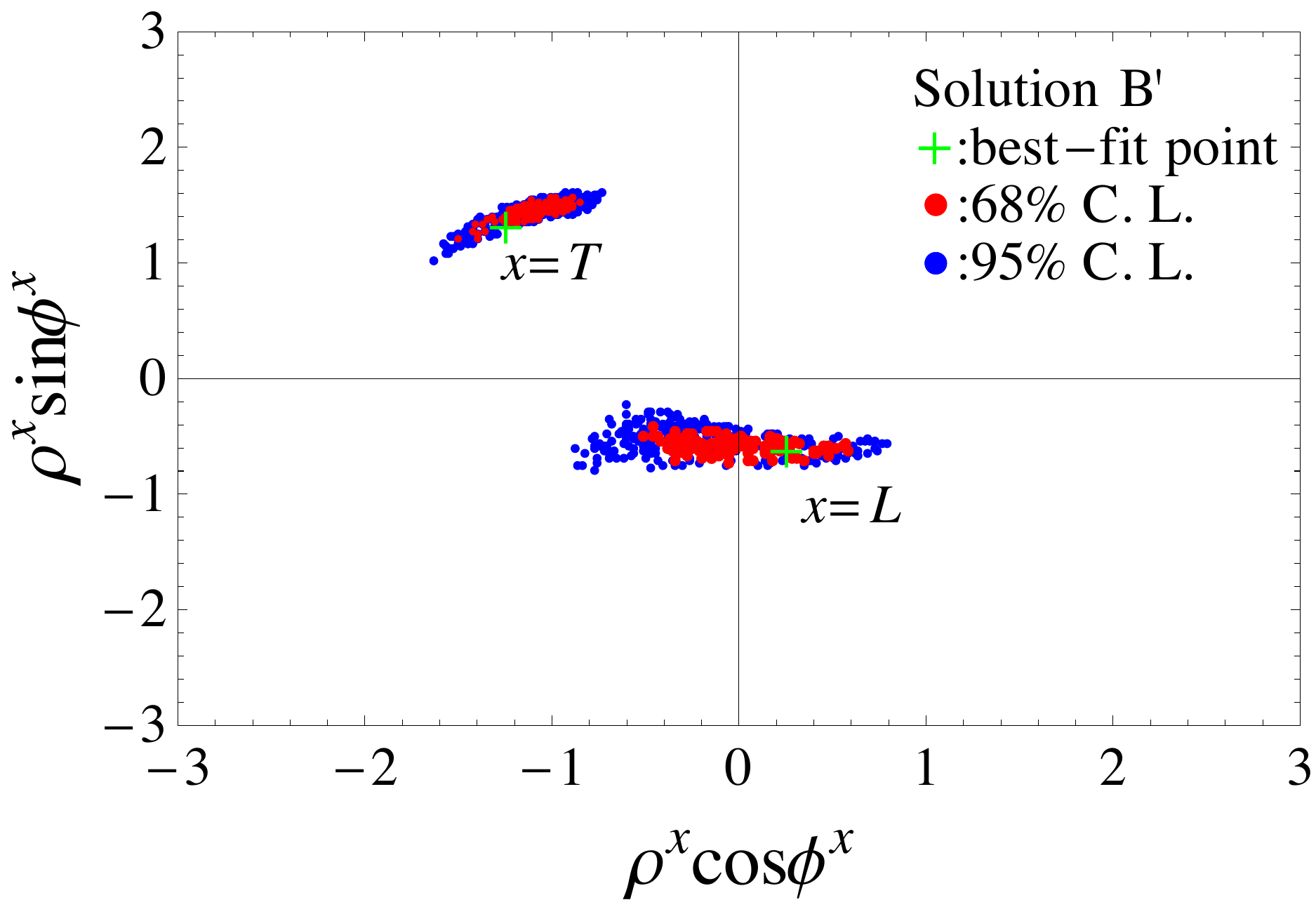}}
  \subfigure[]{\includegraphics[width=7cm]{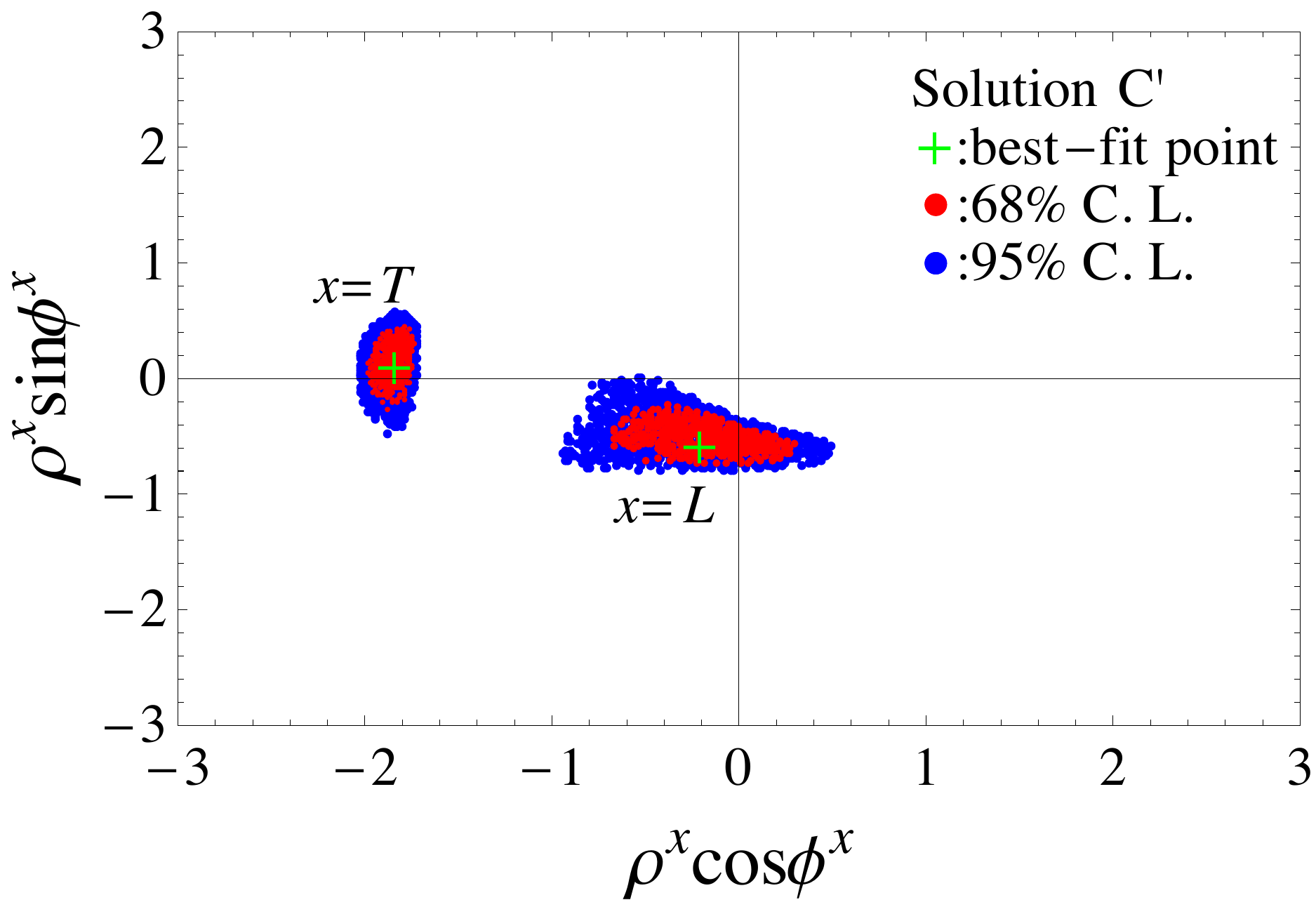}}
    \subfigure[]{\includegraphics[width=7cm]{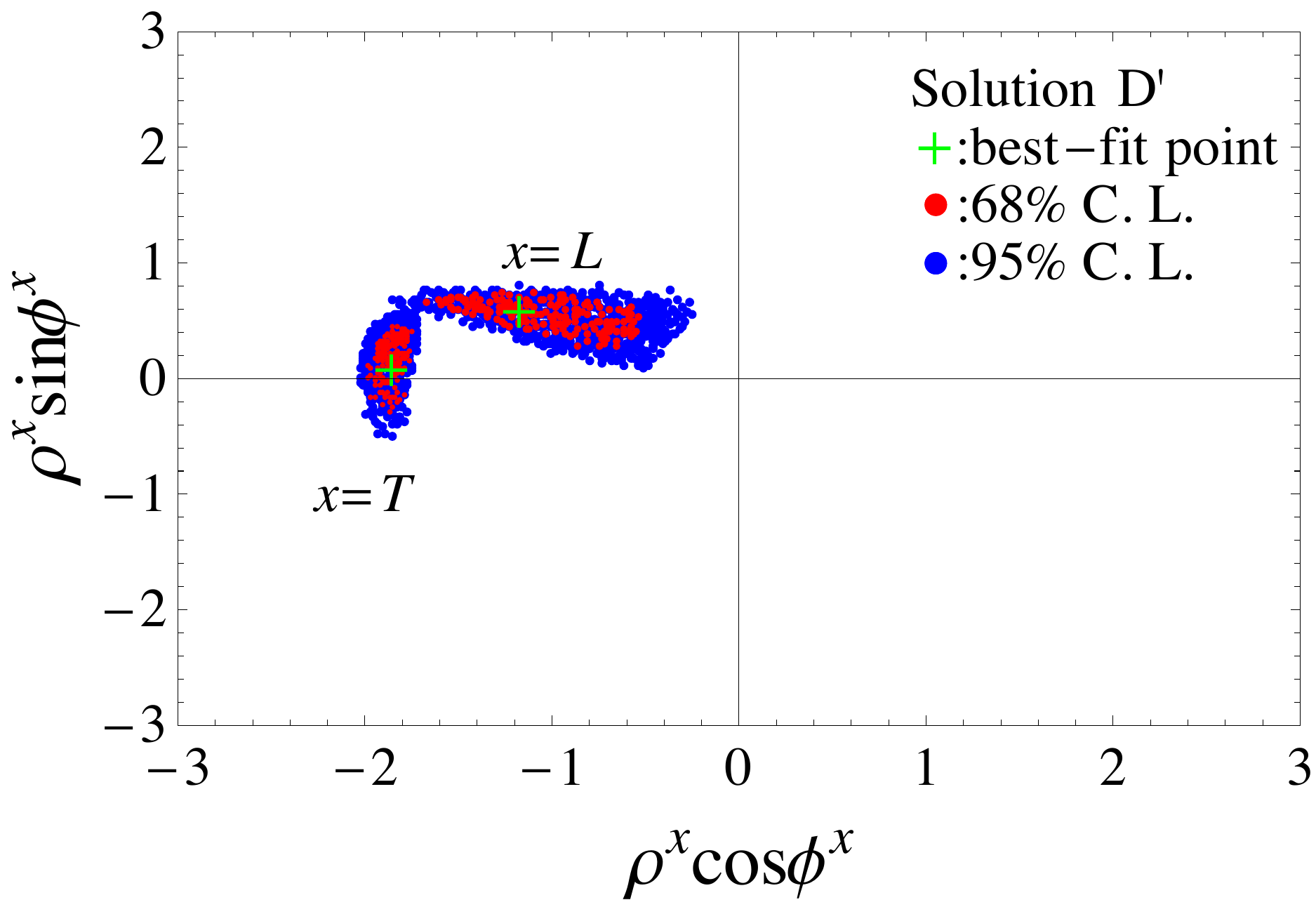}}
\caption{\label{case412} \small Same as Fig.~\ref{case41} except for in $(\rho \cos \phi, \rho \sin \phi)$ plane.}
\end{center}
\end{figure}

\begin{figure}[h]
\begin{center}
     \subfigure[]{\includegraphics[width=7cm]{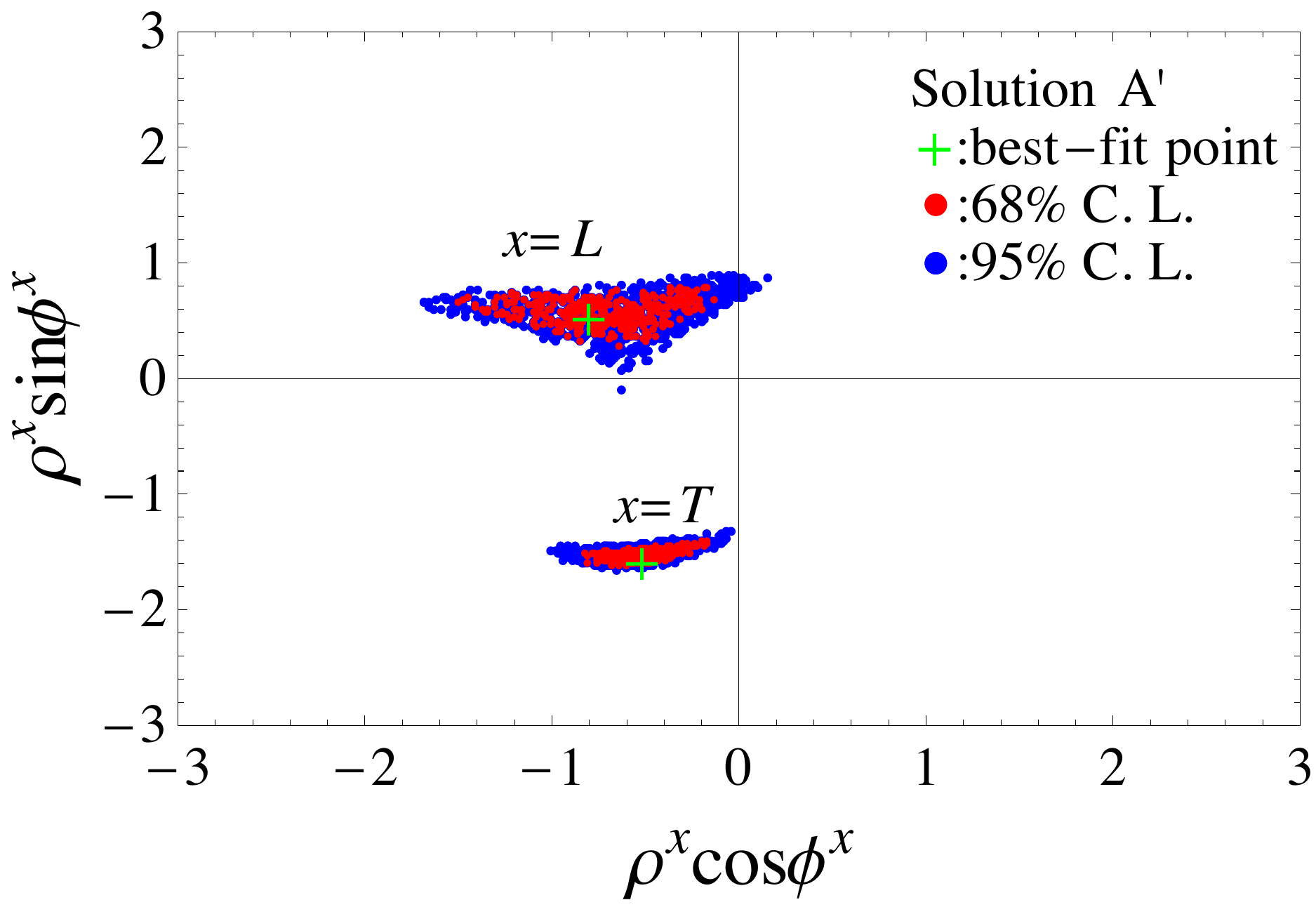}}
\caption{\label{case422} \small Same as Fig.~\ref{case42} except for in $(\rho \cos \phi, \rho \sin \phi)$ plane.}
\end{center}
\end{figure}

\end{appendix}


\newpage\newpage
\begin{thebibliography}{99}
\bibitem{Beneke:1999br}
   %``QCD factorization for B ---> pi pi decays: Strong phases and CP violation in the heavy quark limit,''  [hep-ph/9905312].
  M.~Beneke, G.~Buchalla, M.~Neubert and C.~T.~Sachrajda,
  Phys.\ Rev.\ Lett.\  {\bf 83} (1999) 1914.
  
  \bibitem{Beneke:2000ry}
  %``QCD factorization for exclusive, nonleptonic B meson decays: General arguments and the case of heavy light final states,''  [hep-ph/0006124].
  M.~Beneke, G.~Buchalla, M.~Neubert and C.~T.~Sachrajda,
  Nucl.\ Phys.\ B {\bf 591} (2000) 313.
  
  \bibitem{Keum:2000ph}
  %``Fat penguins and imaginary penguins in perturbative QCD,''  [hep-ph/0004004].
  Y.~Y.~Keum, H.~n.~Li and A.~I.~Sanda,
  Phys.\ Lett.\ B {\bf 504} (2001) 6.
  
  \bibitem{Keum:2000wi}
  %``Penguin enhancement and $B \to K \pi$ decays in perturbative QCD,''  [hep-ph/0004173].
  Y.~Y.~Keum, H.~N.~Li and A.~I.~Sanda,
  Phys.\ Rev.\ D {\bf 63} (2001) 054008.

  \bibitem{scet1}
  C. Bauer, S. Fleming and M. Luke,
  Phys. Rev. D {\bf 63} (2000) 014006.% [hep-ph/0005275];
  
 \bibitem{scet2}
  C. Bauer, S. Fleming, D. Pirjol and I. Stewart,
  Phys. Rev. D {\bf 63} (2001) 114020.% [hep-ph/0011336];
  
 \bibitem{scet3}
  C. Bauer and I. Stewart,
  Phys. Lett. B {\bf 516} (2001) 134.% [hep-ph/0107001 ];

 \bibitem{scet4}
  C. Bauer, D. Pirjol and I. Stewart,
  Phys. Rev. D {\bf 65} (2002) 054022.% [hep-ph/0109045].

\bibitem{msf}
 A. V. Manohar and I. W. Stewart, Phys. Rev. D {\bf 76} (2007) 074002. 

\bibitem{scetAnni}
 C.~M.~Arnesen, Z.~Ligeti, I.~Z. Rothstein and I.~W. Stewart, Phys. Rev. D {\bf 77} (2008) 054006.
 
 \bibitem{Beneke:2003zv}
 %``QCD factorization for B ---> PP and B ---> PV decays,''%[hep-ph/0308039].
  M.~Beneke and M.~Neubert,
 Nucl.\ Phys.\ B {\bf 675} (2003) 333. 
 
\bibitem{Beneke:2006hg}
  %``Branching fractions, polarisation and asymmetries of B ---> VV decays,''[hep-ph/0612290].
  M.~Beneke, J.~Rohrer and D.~Yang,
  Nucl.\ Phys.\ B {\bf 774} (2007) 64.
  
\bibitem{Cheng:2009cn}
 %``Revisiting Charmless Hadronic B(u,d) Decays in QCD Factorization,'' [arXiv:0909.5229 [hep-ph]].
  H.~Y.~Cheng and C.~K.~Chua,
  Phys.\ Rev.\ D {\bf 80} (2009) 114008. 

\bibitem{Cheng:2009mu}
  %``{QCD} Factorization for Charmless Hadronic $B_s$ Decays Revisited,''  [arXiv:0910.5237 [hep-ph]].
  H.~Y.~Cheng and C.~K.~Chua,
  Phys.\ Rev.\ D {\bf 80} (2009) 114026.
   
   \bibitem{Cheng:2008gxa}
  %``Branching Ratios and Polarization in B ---> VV, VA, AA Decays,''  [arXiv:0805.0329 [hep-ph]].
  H.~Y.~Cheng and K.~C.~Yang,
  Phys.\ Rev.\ D {\bf 78} (2008) 094001
   [Phys.\ Rev.\ D {\bf 79} (2009) 039903].

\bibitem{Aaltonen:2011jv}
  T.~Aaltonen {\it et al.} [CDF Collaboration],
  %``Evidence for the charmless annihilation decay mode $B^0_s \to \pi^+\pi^-$,''  [arXiv:1111.0485 [hep-ex]].
  Phys.\ Rev.\ Lett.\  {\bf 108} (2012) 211803.

  \bibitem{Aaij:2012as}
  R.~Aaij {\it et al.} [LHCb Collaboration],
  %``Measurement of $b$-hadron branching fractions for two-body decays into charmless charged hadrons,''  [arXiv:1206.2794 [hep-ex]].
  JHEP {\bf 1210} (2012) 037.
  
\bibitem{Lu:2000em}
  C.~D.~Lu, K.~Ukai and M.~Z.~Yang,
  %``Branching ratio and CP violation of B ---> pi pi decays in perturbative QCD approach,''   [hep-ph/0004213].
  Phys.\ Rev.\ D {\bf 63} (2001) 074009.

\bibitem{Li:2005vu}
  Y.~Li and C.~D.~Lu,
  %``Study pure annihilation decays B**0(s) (anti-B**0(s)) ---> D pm pi mp in PQCD approach,''   [hep-ph/0502038].
  Commun.\ Theor.\ Phys.\  {\bf 44} (2005) 659.
  
\bibitem{Ali:2007ff}
  A.~Ali, G.~Kramer, Y.~Li, C.~D.~Lu, Y.~L.~Shen, W.~Wang and Y.~M.~Wang,
  %``Charmless non-leptonic $B_s$ decays to $PP$, $PV$ and $VV$ final states in the pQCD approach,''  [hep-ph/0703162 [HEP-PH]].
  Phys.\ Rev.\ D {\bf 76} (2007) 074018.

\bibitem{Xiao:2011tx}
  Z.~J.~Xiao, W.~F.~Wang and Y.~y.~Fan, 
  %``Revisiting the pure annihilation decays $B_s\to \pi^+ \pi^-$ and $B^0 \to K^+ K^-$: the data and the pQCD predictions,''   [arXiv:1111.6264 [hep-ph]].
  Phys.\ Rev.\ D {\bf 85} (2012) 094003.

\bibitem{Chang:2012xv}
  Q.~Chang, X.~W.~Cui, L.~Han and Y.~D.~Yang,
  %``Revisiting the Annihilation Corrections in Non-leptonic $\bar{B}_s^0$ Decays within QCD Factorization,''    [arXiv:1205.4325 [hep-ph]].
  Phys.\ Rev.\ D {\bf 86} (2012) 054016.

  \bibitem{Gronau:2012gs}
  M.~Gronau, D.~London and J.~L.~Rosner,
  %``Rescattering Contributions to rare B-Meson Decays,''  [arXiv:1211.5785 [hep-ph]].
  Phys.\ Rev.\ D {\bf 87} (2013) 3,  036008.

\bibitem{Cheng:2014rfa}
  H.~Y.~Cheng, C.~W.~Chiang and A.~L.~Kuo,
  %``Updating B→PP,VP decays in the framework of flavor symmetry,''  [arXiv:1409.5026 [hep-ph]].
  Phys.\ Rev.\ D {\bf 91} (2015) 1,  014011.

\bibitem{Li:2015xna}
  Y.~Li, W.~L.~Wang, D.~S.~Du, Z.~H.~Li and H.~X.~Xu,
  %``Impact of family-non-universal $Z^\prime $ boson on pure annihilation $B_s \rightarrow \pi ^+ \pi ^-$ and $B_d \rightarrow K^+ K^-$ decays,''[arXiv:1503.00114 [hep-ph]].
  Eur.\ Phys.\ J.\ C {\bf 75} (2015) 7,  328.
  
\bibitem{Bobeth:2014rra}
  C.~Bobeth, M.~Gorbahn and S.~Vickers,
  %``Weak annihilation and new physics in charmless $B \to M M$ decays,''   [arXiv:1409.3252 [hep-ph]]
  Eur.\ Phys.\ J.\ C {\bf 75} (2015) 7,  340.

\bibitem{Zhu:2011mm}
  G.~Zhu,
  %``Implications of the recent measurement of pure annihilation $B_s \to \pi^+ \pi^-$ decays in QCD factorization,''   [arXiv:1106.4709 [hep-ph]].
  Phys.\ Lett.\ B {\bf 702} (2011) 408. 

\bibitem{Wang:2013fya}
  K.~Wang and G.~Zhu,
  %``Flavor dependence of annihilation parameters in QCD factorization,''  [arXiv:1304.7438 [hep-ph]].
  Phys.\ Rev.\ D {\bf 88} (2013) 014043.

\bibitem{Chang:2014yma}
  Q.~Chang, J.~Sun, Y.~Yang and X.~Li,
  %``A combined fit on the annihilation corrections in B$_{u,d,s}$ → PP decays within QCDF,''    [arXiv:1409.2995 [hep-ph]].
   Phys.\ Lett.\ B {\bf 740} (2015) 56.
  
\bibitem{Sun:2014tfa}
  J.~Sun, Q.~Chang, X.~Hu and Y.~Yang,
  %``Constraints on hard spectator scattering and annihilation corrections in $B_{u,d}$ ${\to}$ $PV$ decays within QCD factorization,''  [arXiv:1412.2334 [hep-ph]].
  Phys.\ Lett.\ B {\bf 743} (2015) 444.

\bibitem{HFAG}
Y. Amhis  {\it et al.} (HFAG), arXiv:1207.1158; updated results and plots available at http://www.slac.stanford.edu/xorg/hfag/.

\bibitem{Buchalla:1995vs}
  G.~Buchalla, A.~J.~Buras and M.~E.~Lautenbacher,
  %``Weak decays beyond leading logarithms,'' [hep-ph/9512380].
  Rev.\ Mod.\ Phys.\  {\bf 68} (1996) 1125.
  
  \bibitem{Buras:1998raa}
  A.~J.~Buras,
  %``Weak Hamiltonian, CP violation and rare decays,''
  hep-ph/9806471.
  
\bibitem{Bell:2009fm}
  G.~Bell and V.~Pilipp,
  %``B- ---> pi- pi0/rho- rho0 to NNLO in QCD factorization,'' [arXiv:0907.1016 [hep-ph]].
  Phys.\ Rev.\ D {\bf 80} (2009) 054024. 
  
\bibitem{Beneke:2009ek}
  M.~Beneke, T.~Huber and X.~Q.~Li,
  %``NNLO vertex corrections to non-leptonic B decays: Tree amplitudes,'' [arXiv:0911.3655 [hep-ph]].
  Nucl.\ Phys.\ B {\bf 832} (2010) 109.
  
  \bibitem{Bell:2015koa}
 G.~Bell, M.~Beneke, T.~Huber and X.~Q.~Li,
  %``Two-loop current–current operator contribution to the non-leptonic QCD penguin amplitude,''  [arXiv:1507.03700 [hep-ph]].
  Phys.\ Lett.\ B {\bf 750} (2015) 348. 

   \bibitem{XQLi}
 M. Beneke, X. Q. Li and L. Vernazza, Eur. Phys. J. C {\bf 61} (2009) 429. 
 
\bibitem{Vanhoefer:2015ijw}
  P.~Vanhoefer {\it et al.} [Belle Collaboration],
  %``Study of $\mathbf{B^{0}\rightarrow\rho^{+}\rho^{-}}$ decays and implications for the CKM angle $\mathbf{\phi_2}$,''
  arXiv:1510.01245 [hep-ex].

  \bibitem{PDG}
K. Olive {\it et al.} (Particle Data Group), Chin. Phys. C 38 (2014) 090001.

\bibitem{CKMfitter}
J. Charles  {\it et al.} (CKMfitter~Group). Eur. Phys. J. C {\bf 41} (2005) 1; updated results and plots available
at http://ckmfitter.in2p3.fr.

\bibitem{lattice}
J. Laiho, E. Lunghi and R. Van de Water, Phys. Rev. D {\bf 81} (2010) 034503 ; updated results and plots available
at http://www.latticeaverages.org.

\bibitem{Ball:2006eu}
  P.~Ball, G.~W.~Jones and R.~Zwicky,
  %``B ---> V gamma beyond QCD factorisation,''  [hep-ph/0612081].
  Phys.\ Rev.\ D {\bf 75} (2007) 054004.
  
%\cite{Ball:2004rg}
\bibitem{Ball:2004rg}
  P.~Ball and R.~Zwicky,
  %``B(D,S) ---> rho, omega, K*, phi decay form-factors from light-cone sum rules revisited,''   [hep-ph/0412079].
  Phys.\ Rev.\ D {\bf 71} (2005) 014029.

%\cite{Ball:2007rt}
\bibitem{Ball:2007rt} 
  P.~Ball and G.~W.~Jones,
  %``Twist-3 distribution amplitudes of K* and phi mesons,''   [hep-ph/0702100 [HEP-PH]].
  JHEP {\bf 0703} (2007) 069. 

\bibitem{Chang:2014rla}
  Q.~Chang, J.~Sun, Y.~Yang and X.~Li,
  %``Spectator scattering and annihilation contributions as a solution to the $\pi K$ and $\pi \pi$ puzzles within QCD factorization approach,''  [arXiv:1409.1322 [hep-ph]].
  Phys.\ Rev.\ D {\bf 90} (2014) 5,  054019.

 \bibitem{Hofer:2010ee}
  L.~Hofer, D.~Scherer and L.~Vernazza,
  %``$B_s \to \phi \rho^0$ and $B_s \to \phi \pi^0$ as a handle on isospin-violating New physics,''  [arXiv:1011.6319 [hep-ph]].
  JHEP {\bf 1102} (2011) 080.

\bibitem{Kagan:2004uw}
  A.~L.~Kagan,
  %``Polarization in B ---> VV decays,''   [hep-ph/0405134].
  Phys.\ Lett.\ B {\bf 601} (2004) 151.
  
\bibitem{Zou:2015iwa}
  Z.~T.~Zou, A.~Ali, C.~D.~Lu, X.~Liu and Y.~Li,
  %``Improved Estimates of The $B_{(s)}\to V V$ Decays in Perturbative QCD Approach,'' doi:10.1103/PhysRevD.91.054033   [arXiv:1501.00784 [hep-ph]].
  Phys.\ Rev.\ D {\bf 91} (2015) 054033.

\end{thebibliography}
 \end{document}